\documentclass{amsart}

\usepackage{amsmath}%
\usepackage{amsfonts}%
\usepackage{amssymb}%
\usepackage{graphicx}

\newtheorem{theo}{Theorem}[section]
\newtheorem{lemme}[theo]{Lemma}

\newtheorem{defi}[theo]{Definition}

\newtheorem{prop}[theo]{Proposition}

\numberwithin{equation}{section}

\DeclareMathAlphabet{\mathonebb}{U}{bbold}{m}{n}
\newcommand{\one}{\ensuremath{\mathonebb{1}}}
\begin{document}
\title[Grazing collision limit of Boltzmann's equation]{Asymptotic of grazing collisions for the spatially homogeneous Boltzmann equation for soft and Coulomb potentials}
\author{David Godinho}

\begin{abstract}
We give an explicit bound for the Wasserstein distance with quadratic cost between the solutions of Boltzmann's and Landau's equations in the case of soft and Coulomb potentials. This gives an explicit rate of convergence for the grazing collisions limit. Our result is local in time for very soft and Coulomb potentials and global in time for moderately soft potentials.
\end{abstract}
\maketitle


\vskip0.5cm

David Godinho: Laboratoire d'Analyse et de Math\'ematiques Appliqu\'ees,
CNRS UMR 8050, Universit\'e Paris-Est, 61 avenue du G\'en\'eral de Gaulle, 94010
Cr\'eteil Cedex, France.

\textit{E-mail address:} david.godinho-pereira@u-pec.fr

\section{Introduction and main result}

\subsection{The Boltzmann equation}
If we denote by $f_t(v)$ the density of particles which move with velocity $v\in\mathbb R^3$ at time $t\geq0$ in a spatially homogeneous dilute gas, then, under some assumptions, $f$ solves the Boltzmann equation
\begin{align} \label{boltz}
\partial_tf_t(v)=\int_{\mathbb R^3}dv_*\int_{\mathbb S^2}d\sigma B(|v-v_*|,\theta)\big[f_t(v')f_t(v_*')-f_t(v)f_t(v_*)\big],
\end{align}
where the pre-collisional velocities are given by
\begin{align} \label{v'}
v'=\frac{v+v_*}{2}+\frac{|v-v_*|}{2}\sigma,\quad v_*'=\frac{v+v_*}{2}-\frac{|v-v_*|}{2}\sigma,
\end{align}
and $\theta$ is the so-called \textit{deviation angle} defined by $\cos\theta=\frac{(v-v_*)}{|v-v_*|}.\sigma$. The function $B=B(|v-v_*|,\theta)=B(|v'-v_*'|,\theta)$ is called the collision kernel and depends on the nature of the interactions between particles.

Let us interpret this equation: for each $v\in\mathbb R^3$, new particles with velocity $v$ appear due to a collision between two particles with velocities $v'$ and $v_*'$, at rate $B(|v'-v_*'|,\theta)$, while particles with velocity $v$ disappear because they collide with another particle with velocity $v_*$, at rate $B(|v-v_*|,\theta)$. See Cercignani \cite{CER}, Desvillettes \cite{DES}, Villani \cite{VIL3} and Alexandre \cite{ALE} for much more details.  

Since the collisions are assumed to be elastic, conservation of mass, momentum and kinetic energy hold at least formally for solutions to (\ref{boltz}) and we will assume without loss of generality that $\int_{\mathbb R^3}f_0(v)dv=1$.
\vskip0.5cm

We will first assume that the collision kernel $B$ has the following form
\renewcommand\theequation{{\bf A1($\gamma$)}}
\begin{align} \label{A1}
B(|v-v_*|,\theta)\sin\theta=|v-v_*|^\gamma\beta(\theta),
\end{align} 
\renewcommand\theequation{\thesection.\arabic{equation}}\\
\addtocounter{equation}{-1}
where $\beta:(0,\pi]\mapsto[0,\infty)$ is a function and $\gamma\in\mathbb R$. We consider the case of particles which interact through repulsive forces following an inverse power law, which means that two particles apart from a distance $r$ exert on each other a force proportional to $1/r^s$, with $s\in(2,\infty)$. In this case, we have
\begin{align}
\beta(\theta)\overset{0}{\sim}cst\theta^{-1-\nu}\quad {\rm with}\ \nu=\frac{2}{s-1}\in(0,2),\quad {\rm and}\  \gamma=\frac{s-5}{s-1}\in(-3,1).
\end{align}
One classically names \textit{hard potentials} the case where $\gamma\in(0,1)$ (i.e. $s>5$), \textit{Maxwellian molecules} the case where $\gamma=0$ (i.e. $s=5$), \textit{moderately soft potentials} the case where $\gamma\in(-1,0)$ (i.e. $s\in(3,5)$), \textit{very soft potentials} the case where $\gamma\in(-3,-1]$ (i.e. $s\in(2,3]$). We will study in this paper all soft potentials.

In all these cases, we have $\int_0^\pi\beta(\theta)d\theta=+\infty$, which means that there is an infinite number of \textit{grazing collisions} (collisions with a very small deviation) for each particle during any time interval. We will consider the Boltzmann equation without cutoff where we assume
\renewcommand\theequation{{\bf A2}}
\begin{align} \label{A2}
\int_0^\pi\theta^2\beta(\theta)d\theta=\frac{4}{\pi},
\end{align}
\renewcommand\theequation{\thesection.\arabic{equation}}\\
\addtocounter{equation}{-1}
which corresponds to the real physical situation. The classical assumption is only $\int_0^\pi\theta^2\beta(\theta)d\theta<\infty$ but we can assume without loss of generality that it is equal to $\frac{4}{\pi}$ (it suffices to make a change of time).

In the case of soft potentials, we will suppose that for some $\nu\in(0,2)$ and $0<c_1<c_2$,
\renewcommand\theequation{{\bf A3($\nu$)}}
\begin{align} \label{A3}
c_1\theta^{-1-\nu}\leq\beta(\theta)\leq c_2\theta^{-1-\nu}\quad \text{for all} \quad \theta\in(0,\pi].
\end{align} 
\renewcommand\theequation{\thesection.\arabic{equation}}\\
\addtocounter{equation}{-1}
In order to focus on grazing collisions for soft potentials, we also set, for $0<\epsilon\leq\pi$, 
\begin{align} \label{betaeps}
B_\epsilon(|v-v_*|,\theta)\sin\theta=|v-v_*|^\gamma\beta_\epsilon(\theta) \quad{\rm with} \quad \beta_{\epsilon}(\theta)=\frac{\pi^3}{\epsilon^3}\beta\Big(\frac{\pi\theta}{\epsilon}\Big)\one_{|\theta|<\epsilon}.
\end{align}
Observe that $\beta_\epsilon$ is concentrated on small deviation angles, but for all $\epsilon\in(0,\pi)$,
\begin{align} \label{theta2}
	\int_0^\pi\theta^2\beta_\epsilon(\theta)d\theta=\frac{4}{\pi}.
\end{align}

When the particles exert on each other a force proportional to $1/r^2$, we talk about Coulomb potential. As explained in Villani \cite[Section 7]{VIL}, the Boltzmann equation does not make sense in this case because grazing collisions become preponderant over all other collisions. To treat the Coulomb case, we will consider the following collision kernel
\renewcommand\theequation{{\bf AC}}
\begin{align} \label{AC}
B_\epsilon(|v-v_*|,\theta)\sin\theta=(|v-v_*|+h_\epsilon)^{-3}\beta_\epsilon(\theta),
\end{align} 
\renewcommand\theequation{\thesection.\arabic{equation}}\\
\addtocounter{equation}{-1}
where $\epsilon\in(0,1)$, $h_\epsilon\in(0,1)$ decreases to 0 as $\epsilon$ tends to 0  and for $\theta\in(0,\pi]$,
\begin{align} \label{betaepsc}
\beta_\epsilon(\theta)=\frac{c_\epsilon}{\log\frac{1}{\epsilon}}\frac{\cos\theta/2}{\sin^3\theta/2}\one_{\epsilon\leq\theta\leq\pi/2},
\end{align}
where $c_\epsilon$ is such that (\ref{theta2}) is satisfied. We can compute explicitly $c_\epsilon$ and we get
\begin{align*}
c_\epsilon=\frac{4}{\pi}\frac{\log\frac{1}{\epsilon}}{\frac{\epsilon^2}{\sin^2\epsilon/2}+\frac{4\epsilon\cos\epsilon/2}{\sin\epsilon/2}+8\log\frac{1}{\sqrt2\sin\epsilon/2}-\pi^2/2-2\pi},
\end{align*} 
which tends to $\frac{1}{2\pi}$ as $\epsilon\rightarrow0$. 

We thus take the same collision kernel as in Villani \cite[Section 7]{VIL} with two small modifications. We add $h_\epsilon$ in the velocity part only to get easily existence and uniqueness of solutions to (\ref{boltz}). Indeed, we do not need it for the calculus of the rate of convergence in Theorem \ref{colrascoul} (observe that we only ask  to $h_\epsilon$ to decrease to 0 without asking any rate for this convergence). We use $c_\epsilon$ to get (\ref{theta2}) for our convenience, but it does not change the nature of the cross section since $c_\epsilon$ is close to $\frac{1}{2\pi}$ when $\epsilon$ is small.

Since we have (\ref{theta2}) for each $\epsilon>0$ and since $\int_0^\pi\theta^4\beta_\epsilon(\theta)d\theta\leq \frac{C}{\log1/\epsilon}\rightarrow0$, this cross section indeed concentrates on grazing collisions.    

\subsection{The Landau equation}
We consider the spatially homogeneous Landau equation in dimension 3 for soft and Coulomb potentials. This equation of kinetic physics, also called Fokker-Planck-Landau equation, has been derived from the Boltzmann equation by Landau in 1936 when the grazing collisions prevail in the gas. It describes the density $g_t(v)$ of particles having the velocity $v\in\mathbb R^3$ at time $t\geq0$:
\begin{align} \label{landau}
\partial_tg_t(v)=\frac{1}{2}\sum_{i,j=1}^3\partial_i\Big\{\int_{\mathbb R^3}l_{ij}(v-v_*)\Big[g_t(v_*)\partial_jg_t(v)-g_t(v)\partial_jg_t(v_*)\Big]dv_*\Big\},
\end{align}
where $l(z)$ is a symmetric nonnegative $3\times3$ matrix for each $z\in\mathbb R^3$, depending on a parameter $\gamma\in[-3,0)$, defined by
\begin{align} \label{alandau}
l_{ij}(z)=|z|^\gamma(|z|^2\delta_{ij}-z_iz_j).
\end{align}
As for the Boltzmann equation, we can observe that the solutions to (\ref{landau}) conserve at least formally the mass, the momentum and the kinetic energy and we assume without loss of generality that $\int_{\mathbb R_3}g_0(v)dv=1$.

We refer to Villani \cite{VIL,VIL3} for more details on this equation, especially its physical meaning and its derivation from the Boltzmann equation.

\subsection{Notation}
We denote by $C_b^2(\mathbb R^3)$ the set of real bounded functions which are in $C^2(\mathbb R^3)$ with first and second derivatives bounded and by $L^p(\mathbb R^3)$ the space of measurable functions $f$ with $||f||_{L^p}:=(\int_{\mathbb R^3}|f(v)|^pdv)^{1/p}<+\infty$.

For $k\geq0$, we denote by $\mathcal P_k(\mathbb R^3)$ the set of probability measures on $\mathbb R^3$ admitting a moment of order $k$ (i.e. such that $m_k(f):=\int_{\mathbb R^3}|v|^kf(dv)<\infty)$ and for $\alpha\in(-3,0]$, we introduce the space $\mathcal J_\alpha(\mathbb R^3)$ of probability measures $f$ on $\mathbb R^3$ such that
\begin{align} \label{Jalphadef}
J_\alpha(f):=\sup_{v\in\mathbb R^3}\int_{\mathbb R^3}|v-v_*|^\alpha f(dv_*)<\infty.
\end{align}

For any $T>0$, we finally denote by $L^\infty([0,T],\mathcal P_2(\mathbb R^3))$, $L^\infty([0,T],L^p(\mathbb R^3))$, $L^1([0,T],\mathcal J_{\alpha}(\mathbb R^3))$ and $L^1([0,T],L^p(\mathbb R^3))$ the set of measurable families $(f_t)_{t\in[0,T]}$ of probability measures on $\mathbb R^3$ with $\sup_{[0,T]}m_2(f_t)<+\infty$, $\sup_{[0,T]}||f_t||_{L^p}<+\infty$, $\int_0^TJ_\alpha(f_t)dt<+\infty$ and $\int_0^T||f_t||_{L^p}dt<+\infty$ respectively. We finally denote the entropy of a nonnegative function $f\in L^1(\mathbb R^3)$ by
\begin{align*}
H(f):=\int_{\mathbb R^3}f(v)\log\big(f(v)\big)dv.
\end{align*}

In this article, we will use the Wasserstein distance with quadratic cost for our results of convergence: if $f,g\in\mathcal P_2(\mathbb R^3)$, 
\begin{align*}
\mathcal W_2(f,g)=\inf\Big\{\mathbb E[|U-V|^2]^{1/2}, U\sim f, V\sim g\Big\},
\end{align*}
where the infimum is taken over all $\mathbb R^3$-valued random variables $U$ with law $f$ and $V$ with law $g$. It is known that the infimum is reached and more precisely if we fix $U\sim f$, then there exists $V\sim g$ such that $\mathcal W_2^2(f,g)=\mathbb E[|U-V|^2]$. See e.g. Villani \cite{VIL2} for many details on the subject. 

\subsection{The main results}
We first give an explicit rate of convergence for the asymptotic of grazing collisions for soft potentials. Observe that the existence and the uniqueness of solutions to (\ref{boltz}) and (\ref{landau}) that we state in the following result are direct consequences of the papers of Fournier-Mouhot \cite{FOUMOU} and Fournier-Gu\'erin \cite{FOUGUE1}-\cite{FOUGUE2}. The precise notion of weak solutions that we use is given in the next section.

\begin{theo} \label{colras}
Let $\gamma\in(-3,0)$, $\nu\in(0,2)$ and $B$ be a collision kernel which satisfies (\ref{A1}-\ref{A2}-\ref{A3}). For $\epsilon\in(0,\pi]$, we consider $B_\epsilon$ as in (\ref{betaeps}).\\
(i) If $\gamma\in(-1,0)$ and $\nu\in(-\gamma,1)$, let $f_0\in\mathcal P_{p+2}(\mathbb R^3)$ for some $p>\max(5,\gamma^2/(\nu+\gamma))$ such that $H(f_0)<\infty$. Then there exists a unique weak solution $(g_t)_{t\in[0,\infty)}$ to (\ref{landau}) with $g_0=f_0$, and for any $\epsilon\in(0,\pi]$, there exists a unique weak solution $(f_t^\epsilon)_{t\in[0,\infty)}$ to (\ref{boltz}) with collision kernel $B_\epsilon$ and initial condition $f_0^\epsilon=f_0$. Moreover, for any $T>0$ and $\epsilon\in(0,1)$,
\begin{align*}
\sup_{[0,T]}\mathcal W_2(f_t^\epsilon,g_t)\leq C\epsilon^\frac{p}{2p+3},
\end{align*}
where $C$ is a constant depending on $T,p,\gamma,f_0$.\\
(ii) If $\gamma\in(-3,0)$, let $f_0\in\mathcal P_{p+2}(\mathbb R^3)$ for some $p\geq5$ such that $f_0\in L^q(\mathbb R^3)$ for some $q>\frac{3}{3+\gamma}$. Then there exists $T_*=T_*(q,||f_0||_{L^q})>0$ such that there exists a unique weak solution $(g_t)_{t\in[0,T_*]}$ to (\ref{landau}) with $g_0=f_0$, and for any $\epsilon\in(0,\pi]$, there exists a unique weak solution $(f_t^\epsilon)_{t\in[0,T_*]}$ to (\ref{boltz}) with collision kernel $B_\epsilon$ and initial condition $f_0^\epsilon=f_0$. Moreover, for any $\epsilon\in(0,1)$,
\begin{align*}
\sup_{[0,T_*]}\mathcal W_2(f_t^\epsilon,g_t)\leq \epsilon^\frac{p}{2p+3},
\end{align*}
where $C$ is a constant depending on $p,q,\gamma,f_0$.
\end{theo}

Point $(i)$ applies to the case of moderately soft potentials ($s\in(3,5)$) and $(ii)$ applies to the case of very soft potentials ($s\in(2,3]$). The proof of this result is based on a more general inequality, see Theorem \ref{dist}.

We now treat the case of Coulomb potential. The existence and the uniqueness of solutions to (\ref{boltz}) and (\ref{landau}) stated below are direct consequences of the papers of Fournier-Gu\'erin \cite{FOUGUE1} for (\ref{boltz}) and Arsen'ev-Peskov \cite{ARSPES} and Fournier \cite{FOU2} for (\ref{landau}).
\begin{theo}\label{colrascoul}
Let $\gamma =-3$, $B_\epsilon$ be given by (\ref{AC}) and let $f_0\in\mathcal P_{p}(\mathbb R^3)\cap L^\infty(\mathbb R^3)$ for some $p\geq7$. Then there exists $T_*=T_*(||f_0||_{L^\infty})$ such that there exists a unique weak solution $(g_t)_{t\in[0,T_*]}$ to (\ref{landau}) with $g_0=f_0$, and for any $\epsilon\in(0,1)$, there exists a unique weak solution $(f_t^{\epsilon})_{t\in[0,T_*]}$ to (\ref{boltz}) with collision kernel $B_\epsilon$ and initial condition $f_0^\epsilon=f_0$. Moreover, for any $\epsilon\in(0,1)$,
\begin{align*}
\sup_{[0,T_*]}\mathcal W_2(f_t^{\epsilon},g_t)\leq C\Big(h_\epsilon^{a}+\Big(\frac{1}{\log\frac{1}{\epsilon}}\Big)^{a}\Big),
\end{align*}
where $C$ and $a>0$ depend on $p$ and $f_0$.
\end{theo}
\vskip0.5cm
The constant $a$ can be made explicit from the proof. We have two error terms. The first one ($h_\epsilon^{a}$) comes from the fact that we introduce a parameter in the collision kernel in order to get easily existence and uniqueness of solutions to (\ref{boltz}). The second one \Big($\Big(\frac{1}{\log\frac{1}{\epsilon}}\Big)^{a}$\Big) is the true rate of convergence that we get for the asymptotic of grazing collisions in the Coulomb case. These two terms are not linked, so that assuming existence and uniqueness for (\ref{boltz}), we could take $h_\epsilon=0$ (which still makes our proofs valid). Anyway, since we allow $h_\epsilon$ to decrease to 0 as fast as one wants, we believe that this is not really a limitation.

\subsection{Comments and main difficulties}
It was already known that in the limit of grazing collisions, the solution to Boltzmann's equation converges to the solution of the Landau equation. To be more precise, Degond and Lucquin-Desreux \cite{DEG} and Desvillettes \cite{DES2} have shown the convergence of the operators (not of the solutions) and Villani \cite{VIL} has shown some compactness results and the convergence of subsequences. The uniqueness results of Fournier-Gu\'erin \cite{FOUGUE1} and Fournier \cite{FOU2} show the true convergence (under some more restrictive assumptions). In this article, we give an explicite rate for this convergence and we thus justify the fact that the Landau equation is a good approximation of the Boltzmann equation in the limit of grazing collisions.   
\vskip0.5cm

In all cases (soft or Coulomb potentials), we expect to get a bound for $\mathcal W_2(f_t^\epsilon,g_t)$ of order $\sqrt{\int_0^\pi\theta^4\beta_\epsilon(\theta)d\theta}$ as for the Kac equation (see \cite{FOUGOD}). For soft potentials, the rate of convergence that we get is $\epsilon^{1/2-}$ (if $f_0$ is nice) instead of $\epsilon$. For the Coulomb potential (which is the only case which has a real physical interest), we get a rate of order $\Big(\frac{1}{\log\frac{1}{\epsilon}}\Big)^{a}$ with $a>0$ very small (if $f_0$ is nice) instead of $\sqrt\frac{1}{\log\frac{1}{\epsilon}}$. This last case is very complicated because of the huge singularity, and there may be underlying reasons for the slow convergence.

The results are local in time, except for moderately soft potentials, but this was expected since the uniqueness results for the Boltzmann and Landau equations are also local in time.

To our knowledge, the present paper is the first, with the one of He \cite{HE}, which states an explicit rate of convergence. He obtains a better rate ($\epsilon$ instead of $\epsilon^{1/2-}$ for soft potentials) but considers much more regular solutions (lying in $\mathcal P_p(\mathbb R^3)\cap H_l^N(\mathbb R^3)$ for some $N\geq6$, $l>0$ and $p$ which depends on $N$). Furthermore, for the Coulomb case, He uses a cross section which does not seem to correspond to the physical situation (it resembles more at the case of soft potentials).
\vskip0.5cm

Our result has two main interests. A physical one, since it gives a justification for the Landau equation, and a numerical one. Indeed, in a recent paper about the Kac equation \cite{FOUGOD}, using the same kind of result for grazing collisions, we have shown numerically and theoretically that it is much more efficient to replace small collisions (which cannot be simulated) by a Landau-type term than to neglect them. Theorem 1.1 shows that this should also be the case for the Boltzmann equation for soft potentials.
\vskip0.5cm

Our proofs use probabilistic methods. The first who used probabilistic methods 
to study a Boltzmann-type equation (the Kac equation) is McKean \cite{McK,McK2}. He was investigating the 
convergence to equilibrium and he proposed some probabilistic representation of Wild's sums, using some tools
now known as the McKean graphs. The present article is strongly inspired by Tanaka \cite{TAN2}. He proved that the Wasserstein distance with quadratic cost between two solutions of Kac's equation is non-increasing. He extended the same ideas in \cite{TAN} to the Boltzmann equation for Maxwell molecules. His study was based on
the use of some nonlinear stochastic processes related to the Kac and Boltzmann equations. The same kind of ideas is also used in Desvillettes-Graham-M\'el\'eard \cite{MEL}. 

In this article, we will also use a result of Zaitsev \cite{ZAI} in order to obtain a bound for the Wasserstein distance between a compensated Poisson integral and a Gaussian random variable. Such an idea comes from the paper of Fournier \cite{FOU} about the approximation of L\'evy-driven stochastic differential equations in one dimension, see also \cite{FOUGOD}. Since we work here in dimension 3, such a result is much more difficult to obtain. 

If we compare the present work to our similar result for the Kac equation, another difficulty is the fact that we treat the case of soft and Coulomb potentials ($\gamma\in[-3,0))$ instead of the Maxwell case ($\gamma=0$) where the velocity part of the collision kernel is constant. These reasons explain why we are not able to obtain an optimal rate of convergence.

\subsection{Plan of the paper}
In the next section, we precise the notion of weak solutions that we shall use, we give well-posedness results and some properties of the solutions to Boltzmann's and Landau's equations. In Section 3, we give a general result about the Wasserstein distance between solutions of Boltzmann's and Landau's equations for soft potentials and we deduce Theorem \ref{colras}. In Section 4  we give a probabilistic interpretation of the equations (\ref{boltz}) and (\ref{landau}). Section 5 is devoted to the proof of our general result for soft potentials. In Section 6, we study the Coulomb case. We end the paper with an appendix where we give a result about the distance between a compensated Poisson integral and a centered Gaussian law with the same variance, a result about the ellipticity of the diffusion matrix $l$ (recall (\ref{alandau})), a generalized Gr\"onwall Lemma and another technical result . 

\section{Weak solutions}

\subsection{Preliminary observations}
\subsubsection{Soft potentials}
We consider a collision kernel which satisfies (\ref{A1}-\ref{A2}-\ref{A3}) and we set, for $\theta\in(0,\pi]$,
\begin{align} \label{HetG}
H(\theta):=\int_\theta^\pi\beta(x)dx\quad {\rm and}\quad G(z):=H^{-1}(z).
\end{align}
The function $H$ is a continuous decreasing bijection from $(0,\pi]$ into $[0,+\infty)$ and $G:[0,+\infty)\rightarrow(0,\pi]$ is its inverse function. By Fournier-Gu\'erin \cite[Lemma 1.1, (i)]{FOUGUE1}, Assumption (\ref{A3}) implies that there exists $\kappa_1>0$ such that for all $x,y\in\mathbb R_+$,
\renewcommand\theequation{{\bf A4}}
\begin{align} \label{A4}
\int_0^\infty\big(G(z/x)-G(z/y)\big)^2dz\leq\kappa_1\frac{(x-y)^2}{x+y}.
\end{align}
\renewcommand\theequation{\thesection.\arabic{equation}}\\
\addtocounter{equation}{-1}

\begin{lemme} \label{verifA4}
For $\epsilon\in(0,\pi]$, we consider $\beta_\epsilon$ as in (\ref{betaeps}), and we set for $\theta\in(0,\epsilon]$
\begin{align*} 
H_\epsilon(\theta):=\int_\theta^\epsilon\beta_\epsilon(x)dx\quad {\rm and}\quad G_\epsilon(z):=H_\epsilon^{-1}(z).
\end{align*}
The function $H_\epsilon$ is a continuous decreasing bijection from $(0,\epsilon]$ into $[0,+\infty)$ and $G_\epsilon:[0,+\infty)\rightarrow(0,\epsilon]$ is its inverse function. Then for all $\epsilon\in(0,\pi]$, $G_\epsilon$ satisfies (\ref{A4}) with the same $\kappa_1>0$ as $G$.
\end{lemme}

\textbf{Proof.} Observing that $H_\epsilon(\theta)=\frac{\pi^2}{\epsilon^2}H(\frac{\pi\theta}{\epsilon})$ and $G_\epsilon(z)=\frac{\epsilon}{\pi}G(\frac{\epsilon^2z}{\pi^2})$, we have, for all $x,y>0$ and all $\epsilon\in(0,\pi]$,
\begin{align*}
\int_0^\infty\Big(G_\epsilon\big(\frac{z}{x})-G_\epsilon\big(\frac{z}{y}\big)\Big)^2dz&=\int_0^\infty\frac{\epsilon^2}{\pi^2}\Big(G\big(\frac{\epsilon^2z}{\pi^2x})-G\big(\frac{\epsilon^2z}{\pi^2y}\big)\Big)^2dz\\
&=\int_0^\infty\Big(G\big(\frac{u}{x})-G\big(\frac{u}{y}\big)\Big)^2du.
\end{align*}
That concludes the proof. \hfill$\square$
\vskip0.5cm

To deal with soft potentials, we will use that for $\alpha\in(-3,0)$ and for $q\in(3/(3+\alpha),\infty]$, there exists a constant $C_{\alpha,q}$ such that for any $h\in\mathcal P(\mathbb R^3)\cap L^q(\mathbb R^3)$,
\begin{align}
\label{Jalpha}
J_\alpha(h)&=\sup_{v\in\mathbb R^3}\int_{\mathbb R^3}h(v_*)|v-v_*|^\alpha dv_*\\
\nonumber
&\leq \sup_{v\in\mathbb R^3}\int_{|v_*-v|<1}h(v_*)|v-v_*|^\alpha dv_*+\sup_{v\in\mathbb R^3}\int_{|v_*-v|\geq1}h(v_*)dv_*\\
\nonumber
&\leq C_{\alpha,q}||h||_{L^q(\mathbb R^3)}+1,
\end{align}
where 
\begin{align*}
C_{\alpha,q}=\Big[\int_{|v_*|\leq1}|v_*|^{\alpha q/(q-1)}dv_*\Big]^{(q-1)/q}<\infty,
\end{align*}
since by assumption $\alpha q/(q-1)>-3$. This computation will be useful in many proofs of this article.

\subsubsection{Coulomb potential}
We consider the collision kernel $B_\epsilon$ given by (\ref{AC}) and we set, for $\epsilon\in(0,1)$ and $\theta\in[\epsilon,\pi/2]$,
\begin{align} \label{HetGcoul}
H_\epsilon(\theta):=\int_\theta^{\pi/2}\beta_\epsilon(x)dx\quad {\rm and}\quad G_\epsilon(z):=H_\epsilon^{-1}(z).
\end{align}
The function $H_\epsilon$ is a continuous decreasing bijection from $[\epsilon,\pi/2]$ into $[0,H_\epsilon(\epsilon)]$ and we extend its inverse function $G_\epsilon:[0,H_\epsilon(\epsilon)]\rightarrow[\epsilon,\pi/2]$ on $[0,\infty)$ by setting $G_\epsilon(z)=0$ for $z>H_\epsilon(\epsilon)$.
\begin{lemme}
There exists $\kappa_2>0$ such that for all $x,y\in\mathbb R_+$, for all $\epsilon\in(0,1)$,
\renewcommand\theequation{{\bf A5}}
\begin{align} \label{A5}
\int_0^\infty\big(G_\epsilon(z/x)-G_\epsilon(z/y)\big)^2dz\leq\kappa_2\Big(\frac{(x-y)^2}{x+y}+\frac{\max(x,y)}{\log\frac{1}{\epsilon}}\log\frac{\max(x,y)}{\min(x,y)}\Big).
\end{align}
\renewcommand\theequation{\thesection.\arabic{equation}}\\
\addtocounter{equation}{-1}
\end{lemme}

\textbf{Proof.} We have, for $\theta\in[\epsilon,\pi/2]$ and $z\in[0,\infty)$,
\begin{align*}
H_\epsilon(\theta)=\frac{c_\epsilon}{\log\frac{1}{\epsilon}}(\sin^{-2}\frac{\theta}{2}-2)\quad {\rm and}\quad G_\epsilon(z)=2\arcsin\Big(\frac{\log\frac{1}{\epsilon}}{c_\epsilon}z+2\Big)^{-\frac{1}{2}}\one_{\{z<H_\epsilon(\epsilon)\}}.
\end{align*}
We consider $0<x<y$. We have
\begin{align*}
\int_0^\infty\big(G_\epsilon(z/x)-G_\epsilon(z/y)\big)^2dz&= \int_0^{xH_\epsilon(\epsilon)}\big(G_\epsilon(z/x)-G_\epsilon(z/y)\big)^2dz\\
&\quad+\int_{xH_\epsilon(\epsilon)}^{yH_\epsilon(\epsilon)}G_\epsilon^2(z/y)dz\\
&=:A+B.
\end{align*} 
Using that for any $a,b>2$, 
\begin{align*}
\Big(\arcsin\frac{1}{\sqrt a}-\arcsin\frac{1}{\sqrt b}\Big)^2\leq 2\Big(\frac{1}{\sqrt{a}}
-\frac{1}{\sqrt{b}}\Big)^2=2\Big(\frac{b-a}{\sqrt{ab}(\sqrt{a}+\sqrt{b})}\Big)^2\leq 2\frac{(b-a)^2}{ab(a+b)},
\end{align*}
and setting $K_\epsilon:=\frac{\log\frac{1}{\epsilon}}{c_\epsilon}$, we have, recalling that $0<x<y$,
\begin{align*}
A&\leq C\int_0^{\frac{x}{K_\epsilon\sin^2\frac{\epsilon}{2}}}K_\epsilon^2\Big|\frac{1}{x}-\frac{1}{y}\Big|^2\frac{z^2dz}{\Big(\frac{z}{x}K_\epsilon+1\Big)\Big(\frac{z}{y}K_\epsilon+1\Big)\Big(\frac{z}{x}K_\epsilon+\frac{z}{y}K_\epsilon+1\Big)}\\
&\leq C\frac{(x-y)^2}{y}K_\epsilon^2\int_0^{\frac{x}{K_\epsilon\sin^2\frac{\epsilon}{2}}}\frac{z^2dz}{\Big(zK_\epsilon+x\Big)^2\Big(zK_\epsilon+y\Big)}\\
&\leq C\frac{(x-y)^2}{x+y}K_\epsilon^2\int_0^{\frac{x}{K_\epsilon\sin^2\frac{\epsilon}{2}}}\frac{z^2dz}{\Big(zK_\epsilon+x\Big)^3}\\
&\leq C\frac{(x-y)^2}{x+y}\Big(\int_0^{\frac{x}{K_\epsilon}}\frac{z^2K_\epsilon^2dz}{x^3}+\int_{\frac{x}{K_\epsilon}}^{\frac{x}{K_\epsilon\sin^2\frac{\epsilon}{2}}}\frac{dz}{zK_\epsilon}\Big)\\
&\leq C\frac{(x-y)^2}{x+y}\Big(\frac{1}{K_\epsilon}+\frac{\log\frac{1}{\sin^2\frac{\epsilon}{2}}}{K_\epsilon}\Big)\\
&\leq C\frac{(x-y)^2}{x+y}.
\end{align*}
We finally used that $K_\epsilon\sim\frac{1}{2\pi\log\frac{1}{\epsilon}}$ as $\epsilon\rightarrow0$. Using that $\arcsin\frac{1}{\sqrt a}\leq \sqrt2\frac{1}{\sqrt a}$ for any $a>2$, we get for $B$,
\begin{align*}
B\leq 8\int_{xH_\epsilon(\epsilon)}^{yH_\epsilon(\epsilon)}\frac{ydz}{K_\epsilon z+y}=8\frac{y}{K_\epsilon}\log\frac{K_\epsilon yH_\epsilon(\epsilon)+y}{K_\epsilon xH_\epsilon(\epsilon)+y}&\leq8\frac{y}{K_\epsilon}\log\frac{K_\epsilon yH_\epsilon(\epsilon)+y}{K_\epsilon xH_\epsilon(\epsilon)+x}\\
&=8\frac{c_\epsilon y}{\log\frac{1}{\epsilon}}\log\frac{y}{x},
\end{align*}
which ends the proof since $\sup_{\epsilon\in(0,1)}c_\epsilon<\infty$ (recall that $c_\epsilon\rightarrow\frac{1}{2\pi}$). \hfill$\square$

\subsection{The Landau equation}
We consider the operator $L$ defined, for any $\phi\in C_b^2(\mathbb R^3)$, by
\begin{equation} \label{L}
L\phi(v,v_*)=\frac{1}{2}\sum_{i,j=1}^3l_{ij}(v-v_*)\partial_{ij}^2\phi(v)+\sum_{i=1}^3b_i(v-v_*)\partial_i\phi(v),
\end{equation}
where $l_{ij}$ is defined in (\ref{alandau}) and
\begin{equation} \label{blandau}
b_i(z)=\sum_{j=1}^3\partial_jl_{ij}(z)=-2|z|^\gamma z_i,\quad \rm{for}\ i=1,2,3.
\end{equation}
For any $\phi\in C_b^2$, we have
\begin{align*}
|L\phi(v,v_*)|&\leq C_\phi(|v-v_*|^{\gamma+1}+|v-v_*|^{\gamma+2})\\
&\leq C_\phi\Big(1+|v|^2+|v_*|^2+|v-v_*|^{\gamma+1}\one_{\gamma\in[-3,-1)}\Big).
\end{align*}
We can thus observe that all the terms in the following definition are well-defined.
\begin{defi}
Let $\gamma\in[-3,0)$. We say that $(g_t)_{t\in[0,T]}\in L^\infty([0,T],\mathcal P_2(\mathbb R^3))$ is a weak solution to (\ref{landau}) if 
\begin{align} 
\int_0^T\int_{\mathbb R^3}\int_{\mathbb R^3}|v-v_*|^{\gamma+1}g_t(dv)g_t(dv_*)dt<\infty,
\end{align}
(which is automatically satisfied if $\gamma\in[-1,0)$) and if for any $\phi\in C_b^2(\mathbb R^3)$ and any $t\in[0,T]$,
\begin{align}
\int_{\mathbb R^3}\phi(v)g_t(dv)=\int_{\mathbb R^3}\phi(v)g_0(dv)+\int_0^t\int_{\mathbb R^3}\int_{\mathbb R^3}L\phi(v,v_*)g_s(dv)g_s(dv_*)ds.
\end{align}
\end{defi}

We now recall a result of Fournier and Gu\'erin \cite{FOUGUE2} which gives existence and uniqueness of a weak solution for the Landau equation.
\begin{theo} \label{unilandau}
(i) Assume that $\gamma\in(-2,0)$. Let $p(\gamma):=\gamma^2/(2+\gamma)$. Let $g_0\in\mathcal P_2(\mathbb R^3)\cap\mathcal P_p(\mathbb R^3)$ for some $p>p(\gamma)$ satisfy also $H(g_0)<\infty$. Consider $q\in(3/(3+\gamma),(3p-3\gamma)/(p-3\gamma))\subset(3/(3+\gamma),3)$. Then the Landau equation (\ref{landau}) has a unique weak solution $(g_t)_{t\geq0}$ in $L_{loc}^\infty([0,\infty),\mathcal P_2(\mathbb R^3))\cap L_{loc}^1([0,\infty),L^q(\mathbb R^3))$.\\
(ii) Assume that $\gamma\in(-3,0)$, and let $q>3/(3+\gamma)$. Let $g_0\in\mathcal P_2(\mathbb R^3)\cap L^q(\mathbb R^3)$. Then there exists $T_*>0$ depending on $q,||g_0||_{L^q}$ such that there exists a unique weak solution $(g_t)_{t\in[0,T_*]}$ to (\ref{landau}) lying in $L^\infty([0,T_*],\mathcal P_2(\mathbb R^3)\cap L^q(\mathbb R^3))$.\\  
(iii) Assume that $\gamma=-3$. Let $g_0\in\mathcal P_2(\mathbb R^3)\cap L^\infty(\mathbb R^3)$. Then there exists $T_*>0$ depending on $||g_0||_{L^\infty}$ such that there exists a unique weak solution $(g_t)_{t\in[0,T_*]}$ to (\ref{landau}) lying in $L^\infty([0,T_*],\mathcal P_2(\mathbb R^3)\cap L^\infty(\mathbb R^3))$.\\
(iv) For any $t\geq0$ (case (i)) or $t\in[0,T_*]$ (case (ii) and (iii)), we have
\begin{align} \label{conservationl}
\int_{\mathbb R^3}g_t(v)\phi(v)dv=\int_{\mathbb R^3}g_0(v)\phi(v)dv,\qquad \phi(v)=1,v,|v|^2.
\end{align}
We also have the decay of entropy: for all $t\geq0$ (case (i)) or $t\in[0,T_*]$ (case (ii) and (iii)),
\begin{align} \label{entropie}
\int_{\mathbb R^3} g_t(v)\log g_t(v)dv\leq \int_{\mathbb R^3} g_0(v)\log g_0(v)dv.
\end{align}
Furthermore, if $m_p(g_0)<\infty$ for some $p\geq2$, then $\sup_{[0,T]}m_p(g_s)<\infty$ for all $T\geq0$ (case (i)) or all $T\in[0,T_*]$ (case (ii) and (iii)).
\end{theo}

For Points $(i)$ and $(ii)$, one can see Fournier-Gu\'erin \cite[Corollary 1.4]{FOUGUE2}. For Point $(iii)$, one can see Arsen'ev-Peskov \cite{ARSPES} for the existence and Fournier \cite{FOU2} for the uniqueness of $(g_t)_{t\in[0,T_*]}$. The conservation of mass, momentum and energy and the decay of entropy are classical in Point $(iv)$. For the propagation of moments, one can see Villani \cite[Section 2.4 p 73]{VIL3} for $\gamma\in(-2,0)$ and \cite[Appendix B p 193]{VIL4} for $\gamma\in[-3,-2]$.

\subsection{The Boltzmann equation}
We take here the notation of Fournier-M\'el\'eard \cite{FOUMEL}. For each $X\in\mathbb R^3$, we introduce $I(X),J(X)\in\mathbb R^3$ such that $(\frac{X}{|X|},\frac{I(X)}{|X|},\frac{J(X)}{|X|})$ is an orthonormal basis of $\mathbb R^3$. We also require that $I(-X)=-I(X)$ and $J(-X)=-J(X)$ for convenience. For $X,v,v_*\in\mathbb R^3$, for $\theta\in[0,\pi]$ and $\varphi\in[0,2\pi)$, we set
\begin{align} \label{a(v)}
\begin{cases}
\Gamma(X,\varphi):=(\cos\varphi)I(X)+(\sin\varphi)J(X),\\
v':=v'(v,v_*,\theta,\varphi):=v-\frac{1-\cos\theta}{2}(v-v_*)+\frac{\sin\theta}{2}\Gamma(v-v_*,\varphi),\\
v_*':=v_*'(v,v_*,\theta,\varphi):=v_*+\frac{1-\cos\theta}{2}(v-v_*)-\frac{\sin\theta}{2}\Gamma(v-v_*,\varphi),\\
a:=a(v,v_*,\theta,\varphi):=(v'-v)=-(v_*'-v_*),
\end{cases}
\end{align}
which is nothing but a suitable spherical parametrization of (\ref{v'}): we write $\sigma\in\mathbb S^2$ as $\sigma=\frac{v-v_*}{|v-v_*|}\cos\theta+\frac{I(v-v_*)}{|v-v_*|}\sin\theta\cos\varphi+\frac{J(v-v_*)}{|v-v_*|}\sin\theta\sin\varphi$. We can now give the notion of weak solution of Boltzmann's equation.
\begin{defi}
Consider a collision kernel $B(|v-v_*|,\theta)\sin\theta=\Phi(|v-v_*|)\beta(\theta)$ with $\beta$ satisfying (\ref{A2}). We say that a family $(f_t)_{t\in[0,T]}\in L^\infty([0,T],\mathcal P_2(\mathbb R^3))$ is a weak solution to (\ref{boltz}) if
\begin{align} \label{boltz1}
\int_0^T\int_{\mathbb R^3}\int_{\mathbb R^3}|v-v_*|^{2}\Phi(|v-v_*|)f_t(dv)f_t(dv_*)dt<\infty,
\end{align}
and if for any $\phi\in C_b^2(\mathbb R^3)$ and any $t\in[0,T]$,
\begin{align} \label{boltz2}
\int_{\mathbb R^3}\phi(v)f_t(dv)=\int_{\mathbb R^3}\phi(v)f_0(dv)+\int_0^t\int_{\mathbb R^3}\int_{\mathbb R^3} A\phi(v,v_*)f_s(dv)f_s(dv_*)ds,
\end{align}
where
\begin{align} \label{Aboltz}
A\phi(v,v_*)=\frac{\Phi(|v-v_*|)}{2}\int_0^\pi\int_0^{2\pi}[\phi(v')+\phi(v_*')-\phi(v)-\phi(v_*)]d\varphi \beta(\theta) d\theta.
\end{align}
\end{defi}
\vskip0.5cm

For any $v,v_*\in\mathbb R^3$, $\theta\in[0,\pi]$ and $\phi\in C_b^2(\mathbb R^3)$, we have (see Villani \cite[p 291]{VIL})
\begin{align}
\Big|\int_0^{2\pi}[\phi(v')+\phi(v_*')-\phi(v)-\phi(v_*)]d\varphi\Big|\leq C||\phi''||_\infty\theta^2|v-v_*|^2,
\end{align}
so that (\ref{A2}) and (\ref{boltz1}) ensure that all the terms in (\ref{boltz2}) are well-defined.

We now give a result of existence and uniqueness for the Boltzmann equation with soft potentials.
\begin{theo} \label{uniboltz}
Let $\gamma\in(-3,0)$, $\nu\in(0,2)$ and $B$ be a collision kernel which satisfies (\ref{A1}-\ref{A2}-\ref{A3}). For $\epsilon\in(0,\pi]$, we consider $B_\epsilon$ as in (\ref{betaeps}).\\
(i) We assume that $\gamma\in(-1,0)$ and $\nu\in(-\gamma,1)$. For some $p>\gamma^2/(\nu+\gamma)$, let $f_0\in\mathcal P_2(\mathbb R^3)\cap\mathcal P_p(\mathbb R^3)$ with $H(f_0)<\infty$. Then for any $\epsilon\in(0,\pi]$, there exists a unique weak solution $(f_t^\epsilon)_{t\in[0,\infty)}$ to (\ref{boltz}) with collision kernel $B_\epsilon$ starting from $f_0$ lying in $L_{loc}^\infty\big([0,\infty),\mathcal P_2(\mathbb R^3)\big)\cap L_{loc}^1\big([0,\infty), L^q(\mathbb R^3)\big)$ for some (explicit) $q\in(3/(3+\gamma),3/(3-\nu))$ with estimates uniform in $\epsilon$.\\
(ii) We next consider the general case. Let $q\in(3/(3+\gamma),\infty)$. For any $f_0\in\mathcal P_2(\mathbb R^3)\cap L^q(\mathbb R^3)$, there exists $T_*=T_*(||f_0||_{L^q},q)>0$ such that for any $\epsilon\in(0,\pi]$, there exists a unique weak solution $(f_t^\epsilon)_{t\in[0,T_*]}$ to (\ref{boltz}) with collision kernel $B_\epsilon$ starting from $f_0$ lying in $L^\infty\big([0,T_*],\mathcal P_2(\mathbb R^3)\cap L^q(\mathbb R^3)\big)$, with estimates uniform in $\epsilon$. \\
(iii) For any $t\geq0$ (case (i)) or $t\in[0,T_*]$ (case (ii)), any $\epsilon\in(0,\pi]$, 
\begin{align} \label{conservationb}
\int_{\mathbb R^3}f_t^\epsilon(v)\phi(v)dv=\int_{\mathbb R^3}f_0(v)\phi(v)dv,\qquad \phi(v)=1,v,|v|^2,
\end{align} 
and
\begin{align} 
\label{entroboltz}
\int_{\mathbb R^3} f_t^\epsilon(v)\log f_t^\epsilon(v)dv\leq \int_{\mathbb R^3} f_0(v)\log f_0(v)dv.
\end{align}
Furthermore, if $\gamma\in(-2,0)$ and $f_0\in\mathcal P_p(\mathbb R^3)$ for some $p\geq4$, then for any $\epsilon\in(0,\pi]$, any $T\geq0$ (case (i)) or any $T\in[0,T_*]$ (case (ii)),
\begin{align*}
\sup_{[0,T]}m_p(f_t^\epsilon)\leq C_{p,T}m_p(f_0),
\end{align*}
where $C_{p,T}$ is a constant which does not depend on $\epsilon$.
\end{theo}
To prove $(i)$ and $(ii)$, we follow the line of some proofs in Fournier-Mouhot \cite{FOUMOU} and Fournier-Gu\'erin \cite{FOUGUE1}. 

\textbf{Proof.} Point $(ii)$ is a consequence of \cite[Proof of Corollary 1.5, Step 2]{FOUGUE1} (recall (\ref{A4})). More precisely, we only need to check in their proof that $T_*$ does not depend on $\epsilon$. For this, it suffices to prove that for any $\epsilon\in(0,\pi]$, there exists a constant $C$ which does not depend on $\epsilon$ such that any weak solution to (\ref{boltz}) (with cross section $B_\epsilon$) \textit{a priori} satisfies
\begin{align} \label{fepslq}
\frac{d}{dt} ||f_t^\epsilon||_{L^q}\leq C(1+||f_t^\epsilon||_{L^q}^2).
\end{align}
This will guarantee that for $0\leq t\leq T_*:=\frac{1}{2C}(\pi/2-\arctan||f_0||_{L^q})$, we  have
\begin{align*}
||f_t^\epsilon||_{L^q}\leq \tan(\arctan||f_0||_{L^q}+Ct)\leq\tan\Big(\frac{\pi}{4}+\frac{1}{2}\arctan||f_0||_{L^q}\Big).
\end{align*}

We classically may replace in $A\phi$ (recall (\ref{Aboltz})) $\beta_\epsilon(\theta)$ by $\hat\beta_\epsilon(\theta)=[\beta_\epsilon(\theta)+\beta_\epsilon(\pi-\theta)]\one_{\theta\in(0,\pi/2]}$, see e.g. Desvillettes-Mouhot \cite[Section 2]{DESMOU}. Following the line of \cite[proof of Proposition 3.2]{DESMOU}, we get
\begin{align*}
&\frac{d}{dt}\int_{\mathbb R^3}|f_t^\epsilon(v)|^qdv\\
&\leq(q-1)\int_{\mathbb R^3}f_t^\epsilon(v_*)dv_*\int_{\mathbb R^3}dv|v-v_*|^\gamma\int_0^{\pi/2}\hat\beta_\epsilon(\theta)d\theta\int_0^{2\pi}d\varphi[(f_t^\epsilon)^q(v')-(f_t^\epsilon)^q(v)].
\end{align*}
Using now the cancellation Lemma of Alexandre-Desvillettes-Villani-Wennberg \cite[Lemma 1]{ADVW} (with $N=3$, $f$ given by $(f_t^\epsilon)^q$, and $B(|v-v_*|,\cos\theta)\sin\theta=\hat\beta_\epsilon(\theta)|v-v_*|^\gamma$), we obtain
\begin{align*}
\frac{d}{dt}\int_{\mathbb R^3}|f_t^\epsilon(v)|^qdv\leq2\pi(q-1)\int_{\mathbb R^3}&f_t^\epsilon(v_*)dv_*\int_{\mathbb R^3}(f_t^\epsilon)^qdv\int_0^{\pi/2}\hat\beta_\epsilon(\theta)d\theta\\
&\big|\cos^{-3}(\theta/2)(|v-v_*|\cos^{-1}(\theta/2))^\gamma-|v-v_*|^\gamma\big|.
\end{align*}
One easily checks that $\big|\cos^{-3}(\theta/2)(|v-v_*|\cos^{-1}(\theta/2))^\gamma-|v-v_*|^\gamma\big|\leq C|v-v_*|^\gamma\theta^2$ for all $\theta\in(0,\pi/2]$ (where $C$ depends only on $\gamma$). Since $\int_0^{\pi/2}\theta^2\hat\beta_\epsilon(\theta)d\theta\leq\int_0^\pi\theta^2\beta_\epsilon(\theta)d\theta=\frac{4}{\pi}$, we finally get with $C=C(\gamma,q)$,
\begin{align*}
\frac{d}{dt}\int_{\mathbb R^3}|f_t^\epsilon(v)|^qdv&\leq C\int_{\mathbb R^3}(f_t^\epsilon)^q(v)dv\int_{\mathbb R^3}|v-v_*|^\gamma f_t^\epsilon(v_*)dv_*\\
&\leq C\int_{\mathbb R^3}(f_t^\epsilon)^q(v)dv+C_{\gamma,q}\Big[\int_{\mathbb R^3}(f_t^\epsilon)^q(v)dv\Big]^{1+1/q},
\end{align*}
by (\ref{Jalpha}) and since $q>3/(3+\gamma)$. This yields
\begin{align*}
\frac{d}{dt}||f_t^\epsilon||_{L^q}=\frac{1}{q}||f_t^\epsilon||_{L^q}^{1-q}\frac{d}{dt}\int_{\mathbb R^3}|f_t^\epsilon(v)|^qdv\leq C||f_t^\epsilon||_{L^q}+C_{\gamma,q}||f_t^\epsilon||^2_{L^q},
\end{align*}
from which (\ref{fepslq}) immediately follows. 
\vskip0.5cm

We now prove $(iii)$. First observe that the conservation of mass, momentum and kinetic energy and the decay of entropy are classical. 

Next let $\gamma\in(-2,0)$ and $p\geq4$. We want to apply (\ref{boltz2}) with $\phi(v)=|v|^p$. We set $\Delta=|v'|^p+|v_*'|^p-|v|^p-|v_*|^p$ (see (\ref{a(v)})). Observing that $v'=v+a$, $v_*'=v_*-a$, and $\nabla\phi(v)=p|v|^{p-2}v$, $\phi''(v)=p|v|^{p-2}I_3+p(p-2)|v|^{p-4}vv^*$ (where $\phi''$ is the Hessian matrix of $\phi$) and using Taylor's formula, we have
\begin{align*}
\Delta&=a.(p|v|^{p-2}v-p|v_*|^{p-2}v_*)\\
&\quad+\frac{1}{2}a.\Big[p(|w_1|^{p-2}+|w_2|^{p-2})a+p(p-2)\Big(|w_1|^{p-4}(w_1w_1^*)a+|w_2|^{p-4}(w_2w_2^*)a\Big)\Big]\\
&=pa.\big(|v|^{p-2}(v-v_*)+(|v|^{p-2}-|v_*|^{p-2})v_*\big)\\
&\quad+\frac{p}{2}\Big[(|w_1|^{p-2}+|w_2|^{p-2})|a|^2+(p-2)\Big(|w_1|^{p-4}(a.w_1)^2+|w_2|^{p-4}(a.w_2)^2\Big)\Big],
\end{align*}
where $w_1=v+\lambda_1a$ for some $\lambda_1\in[0,1]$ and $w_2=v_*+\lambda_2a$ for some $\lambda_2\in[0,1]$. We have $|w_1|^{p-2}+|w_2|^{p-2}\leq C_p(|v|^{p-2}+|v_*|^{p-2})$ where $C_p$ is a constant which only depends on $p$. Observing that
\begin{align*}
\big||v|^{p-2}-|v_*|^{p-2}\big||v_*|&\leq C_p|v-v_*|(|v|^{p-3}+|v_*|^{p-3})|v_*|\\
&\leq C_p|v-v_*|(|v|^{p-2}+|v_*|^{p-2}),
\end{align*}
that $|a|^2=\frac{1-\cos\theta}{2}|v-v_*|^2$, $\int_0^{2\pi}ad\varphi=-\frac{1-\cos\theta}{2}(v-v_*)$, $\int_0^\pi\frac{1-\cos\theta}{2}\beta_\epsilon(\theta)d\theta\leq\frac{4}{\pi}$ by (\ref{theta2}) and using (\ref{boltz2}) with $\phi$, we get
\begin{align*}
\frac{d}{dt}m_p(f_t^\epsilon)&\leq C_p\int_{\mathbb R^3}\int_{\mathbb R^3}|v-v_*|^{\gamma+2}(|v|^{p-2}+|v_*|^{p-2})f_t^\epsilon(dv)f_t^\epsilon(dv_*)\\
&\leq C_p\Big(1+m_p(f_t^\epsilon)+m_2(f_t^\epsilon)m_{p-2}(f_t^\epsilon)\Big)\\
&\leq C_p(1+m_p(f_t^\epsilon)),
\end{align*} 
with $C$ depending on $p,\gamma,m_2(f_0)$ (we used that $x^{\gamma+2}\leq C_\gamma(1+x^2)$ for any $x\geq0$). Point $(iii)$ immediately follows.
\vskip0.5cm

The existence and the uniqueness in $(i)$ are already proved in Fournier-Gu\'erin \cite{FOUGUE1}. We only have to check that the estimates are uniform in $\epsilon$. For that, it suffices to show that for any $\alpha\in(0,\gamma)$
\begin{align} \label{desmou}
\int_0^T	||(1+|v|^{\gamma-\alpha})f_t^\epsilon||_{L^{\frac{3}{3-\nu}}}dt\leq C(1+T),
\end{align}
with $C$ independent of $\epsilon$. Indeed, since we have $\sup_{[0,T]}m_p(f_t^\epsilon)\leq C$ for some $p>\frac{\gamma^2}{\gamma+\nu}$ (with $C$ independent of $\epsilon$) by $(iii)$, we will get
\begin{align*}
||f^\epsilon||_{L^1\big([0,T],L^q(\mathbb R^3)\big)}\leq C_{T,q},	
\end{align*}
for some $q\in(3/(3+\gamma),3/(3-\nu))$ by Fournier-Mouhot \cite[Step 3 of the proof of Corollary 2.4]{FOUGUE1}. Looking at Desvillettes-Mouhot \cite[paragraph before Equation (3.2)]{DESMOU}, we see that to prove (\ref{desmou}), it suffices to check that
\begin{align} \label{advw}
	\int_0^T||\sqrt{f_t^\epsilon}||_{H^{\nu/2}(|v|\leq R)}^2dt\leq CR^{|\gamma|}(1+T),
\end{align}
for some constant $C$ which does not depend on $\epsilon$. It remains to follow the line of Alexandre-Desvillettes-Villani-Wennberg \cite[Theorem 1]{ADVW} to get (\ref{advw}). More precisely, we have to check that the constants which appear in the following inequality \cite[Theorem 1]{ADVW} (observe that here $\Phi(|v|)=|v|^\gamma$ does not vanish at 0)
\begin{align} \label{fepsHnu}
||\sqrt{f_t^\epsilon}||_{H^{\nu/2}(|v|<R)}^2\leq 2c_{f^\epsilon}^{-1} R^{|\gamma|}\Big(D(f_t^\epsilon) + (C_1 + C_2)||(1+|v|^2)f_t^\epsilon||_{L^1}^2\Big),
\end{align}
do not depend on $\epsilon$, where $D(f_t^\epsilon)$ is the functional of dissipation of entropy (see (\ref{dissentrop}) below). The constant $C_1$ comes from \cite[Corollary 2]{ADVW}. This constant is such that (observe that there is a misprint in the corollary)
\begin{align*}
\Lambda(|v-v_*|)+|v-v_*|\Lambda'(|v-v_*|)\leq C_1(|v-v_*|^\gamma+|v-v_*|^2),
\end{align*}
where
\begin{align*}
\Lambda(|v-v_*|)=\int_0^\pi |v-v_*|^\gamma(1-\cos\theta)\beta_\epsilon(\theta)d\theta,
\end{align*}
and
\begin{align*}
\Lambda'(|v-v_*|)=\int_0^\pi \sup_{1<\lambda\leq\sqrt2}\frac{|v-v_*|^\gamma(\lambda^\gamma-1)}{|v-v_*|(\lambda-1)}(1-\cos\theta)\beta_\epsilon(\theta)d\theta.
\end{align*}
We can thus take $C_1=\frac{|\gamma|+1}{2}\int_0^\pi\theta^2\beta_\epsilon(\theta)d\theta=2\frac{|\gamma|+1}{\pi}$. Then we deal with the constant $C_2$ which comes from \cite[Lemma 2]{ADVW}. This constant depends on
\begin{align*}
\int_{0}^{\pi/2}\cos^{-4}\frac{\theta}{2}\sin^2\frac{\theta}{2}\beta_\epsilon(\theta)d\theta\leq C\int_{0}^{\pi}\theta^2\beta_\epsilon(\theta)d\theta
\end{align*}
and since this last integral is equal to $\frac{4}{\pi}$, the constant $C_2$ does not depend on $\epsilon$.

The constant $c_{f^\epsilon}$ comes from \cite[Proposition 2]{ADVW}. It is of the form $C'_{f^\epsilon}K$. First $C'_{f^\epsilon}>0$ is controled (from below) by upperbounds of $m_1(f_t^\epsilon)$ and $\int_{\mathbb R^3}f_t^\epsilon\log(1+f_t^\epsilon(v))dv$, which are both classically controled (uniformly in $\epsilon$) by $m_2(f_0)$ and $H(f_0)$. Next, $K>0$ is such that for all $|\xi|\geq1$,
\begin{align*}
\int_0^{\pi/2}\Big(\frac{|\xi|^2}{2}(1-\cos\theta)\wedge1\Big)\beta_\epsilon(\theta)d\theta\geq K|\xi|^\nu.
\end{align*}
One easily deduces from (\ref{A3}) that such an inequality holds uniformly in $\epsilon\in(0,\pi]$.

Hence (\ref{fepsHnu}) holds uniformly in $\epsilon\in(0,\pi]$, and we find that
\begin{align} \label{fepsHnu2}
	||\sqrt{f_t^\epsilon}||_{H^{\nu/2}(|v|<R)}^2\leq CR^{|\gamma|}\Big[D(f_t^\epsilon)+\big(1+m_2(f_t^\epsilon)\big)^2\Big],
\end{align}
for some constant $C$ depending only on $f_0$ (and on $\gamma$, $\beta$ but not on $\epsilon$). Integrating (\ref{fepsHnu2}) in time and using that  
\begin{align} \label{dissentrop}
	\int_0^TD(f_t^\epsilon)dt=H(f_0)-H(f_t^\epsilon)\leq H(f_0)+Cm_2(f_0),
\end{align}
(because classically, $H(f)\geq-Cm_2(f)$), we finally deduce (\ref{advw}) and that concludes the proof. \hfill$\square$
\vskip0.5cm

We finally treat the Coulomb case.

\begin{theo} \label{uniboltzcoul} 
Assume (\ref{AC}) and let $f_0\in\mathcal P_2(\mathbb R^3)$. Then there exists a unique weak solution $(f_t^{\epsilon})_{t\in[0,\infty)}$ to (\ref{boltz}). Furthermore, if $f_0\in L^\infty(\mathbb R^3)$, then there exists $T_*=T_*(||f_0||_{L^\infty})>0$ such that $\sup_{\epsilon\in(0,1)}\sup_{[0,T_*]}||f_t^{\epsilon}||_{L^\infty}<\infty$.
\end{theo}

\textbf{Proof.}
We observe that for $\epsilon\in(0,1)$ fixed, we consider a cutoff case with a bounded cross section: for any $v,v_*\in\mathbb R^3$ and $\theta\in[0,\pi/2]$, $B_\epsilon(|v-v_*|,\theta)\leq C_\epsilon$. The existence and the uniqueness of $(f_t^{\epsilon})_{t\in[0,\infty)}$ are thus classical.

For the stability in $L^\infty(\mathbb R^3)$, like in the previous proof (there is no need to introduce $\hat\beta_\epsilon$ here since $\beta_\epsilon$ is supported in $[0,\pi/2]$), we have for all $q\geq1$, all $\epsilon\in(0,1)$,
\begin{align*}
&\frac{d}{dt}\int_{\mathbb R^3}|f_t^{\epsilon}(v)|^qdv\\
&\leq(q-1)\int_{\mathbb R^3}f_t^{\epsilon}(v_*)dv_*\int_{\mathbb R^3}dv(|v-v_*|+h_\epsilon)^{-3}\int_0^{\pi/2}\beta_\epsilon(\theta)d\theta\\
&\qquad\qquad\qquad\qquad\qquad\qquad\qquad\qquad\qquad\qquad\int_0^{2\pi}d\varphi[(f_t^{\epsilon})^q(v')-(f_t^{\epsilon})^q(v)]\\
&\leq(q-1)\int_{\mathbb R^3}f_t^{\epsilon}(v_*)dv_*\int_{\mathbb R^3}dv|v-v_*|^{-3}\int_0^{\pi/2}\beta_\epsilon(\theta)d\theta\\
&\qquad\qquad\qquad\qquad\qquad\qquad\qquad\qquad\qquad\qquad\int_0^{2\pi}d\varphi[(f_t^{\epsilon})^q(v')-(f_t^{\epsilon})^q(v)].
\end{align*}
Using now the cancellation Lemma of Alexandre-Villani \cite[Proposition 3]{ALEVIL} (with $N=3$, $f$ given by $(f_t^{\epsilon})^q$, and $B(|v-v_*|,\cos\theta)\sin\theta=\beta_\epsilon(\theta)|v-v_*|^{-3}$), we obtain
\begin{align*}
\frac{d}{dt}\int_{\mathbb R^3}|f_t^{\epsilon}(v)|^qdv\leq \lambda_\epsilon(q-1)\int_{\mathbb R^3}(f_t^{\epsilon}(v_*))^{q+1}dv_*\leq C(q-1)||f_t^{\epsilon}||_{L^\infty}||f_t^{\epsilon}||_{L^q}^q,
\end{align*}
since
\begin{align*}
\lambda_\epsilon:&=\frac{4\pi^2}{3}\int_0^{\pi/2}\log\frac{1}{\cos\theta/2}\beta_\epsilon(\theta)d\theta\leq \frac{4\pi^2}{3}\int_0^{\pi/2}\frac{1}{\cos\theta/2}(1-\cos\theta/2)\beta_\epsilon(\theta)d\theta\\
&\leq \frac{\sqrt 2 \pi^2}{3}\int_0^{\pi/2}\theta^2\beta_\epsilon(\theta)d\theta=\frac{4\sqrt 2 \pi}{3}.
\end{align*}
We thus get
\begin{align*}
\frac{d}{dt}||f_t^\epsilon||_{L^q}&\leq \frac{1}{q}\Big(\int_{\mathbb R^3}|f_t^{\epsilon}(v)|^qdv\Big)^{1/q-1}C(q-1)||f_t^{\epsilon}||_{L^\infty}||f_t^{\epsilon}||_{L^q}^q\\
&\leq C||f_t^{\epsilon}||_{L^\infty}||f_t^{\epsilon}||_{L^q}.
\end{align*}
Making $q$ tend to infinity, we get
\begin{align*}
\frac{d}{dt}||f_t^{\epsilon}||_{L^\infty}\leq C||f_t^{\epsilon}||_{L^\infty}^2,
\end{align*}
and thus taking $T_*<\frac{1}{C||f_0||_{L^\infty}}$, we have for any $t<T_*$
\begin{align*}
||f_t^{\epsilon}||_{L^\infty}\leq \frac{||f_0||_{L^\infty}}{1-C||f_0||_{L^\infty} t}.
\end{align*}
This concludes the proof. \hfill$\square$

\section{A general estimate for soft potentials}

In this section, we give a general estimate for the distance between a solution of Boltzmann's equation and a solution of Landau's equation (for soft potentials) from which Theorem \ref{colras} follows.

\begin{theo} \label{dist}
Let $\gamma\in(-3,0)$ and let $B$ be a collision kernel which satisfies (\ref{A1}-\ref{A2}-\ref{A4}). Let $T>0$ and $p\geq5$. Let $f=(f_t)_{t\in[0,T]}$ be a weak solution of (\ref{boltz}) with collision kernel $B$ and $g=(g_t)_{t\in[0,T]}$ be a weak solution of (\ref{landau}) with $H(g_0)<\infty$. We assume that $f\in L^1([0,T],J_\gamma)$, $g\in L^1([0,T],J_\gamma)\cap L^\infty([0,T],\mathcal P_{p+2}(\mathbb R^3))$ and if $\gamma\in(-3,-1)$, that $f$ and $g$ belong to $L^\infty([0,T],J_{\gamma+1})$. Assume furthermore that $\int_0^\pi\theta^4\beta(\theta)d\theta\leq1$. Then for any $n\geq1$, $\eta\in(0,\pi)$ and $M>\sqrt{2m_2(g_0)}$, 
\begin{align*}
\sup_{[0,T]}\mathcal W_2^2(f_t,g_t)\leq C&\Big[\mathcal W_2^2(f_0,g_0)+\frac{1}{n}+\int_0^\pi\theta^4\beta(\theta)d\theta\\
&+\int_\eta^\pi\theta^2\beta(\theta)d\theta+\eta^2M^2n\Big(\log^2(r_\eta)+\log^2(n\eta^2)+M\Big)+\frac{1}{M^p}\Big],
\end{align*}
where
\begin{align} \label{reta}
r_\eta=\frac{\pi}{4}\int_0^\eta\theta^2\beta(\theta)d\theta
\end{align}
and where $C$ depends on $p,T,\kappa_1,\gamma$, $\int_0^TJ_\gamma(f_s+g_s)ds$, $\sup_{[0,T]}m_{p+2}(g_s)$, $H(g_0)$, and additionally on $\sup_{[0,T]}J_{\gamma+1}(f_s+g_s)$ if $\gamma\in(-3,-1)$.
\end{theo}
\vskip0.5cm

This result is proved in Section \ref{general}. We can now deduce Theorem \ref{colras}.
\vskip0.5cm

\textbf{Proof of Theorem \ref{colras}.} We consider a collision kernel which satisfies (\ref{A1}-\ref{A2}-\ref{A3}) and we set $\beta_\epsilon=\frac{\pi^3}{\epsilon^3}\beta\Big(\frac{\pi\theta}{\epsilon}\Big)\one_{|\theta|<\epsilon}$ and $B_\epsilon(|v-v_*|,\theta)\sin\theta=|v-v_*|^\gamma\beta_\epsilon(\theta)$. We first note that (\ref{A2}) is satisfied by $B_\epsilon$ (see (\ref{theta2})) and that (\ref{A3}) implies (\ref{A4}) (see Lemma \ref{verifA4}).
 
We now prove point $(i)$. We thus assume that $\gamma \in(-1,0)$, $\nu\in(-\gamma,1)$ and fix $T>0$. Since $f_0\in\mathcal P_{p+2}(\mathbb R^3)$ for some $p>\max(5,\gamma^2/(\nu+\gamma))$ and since $H(f_0)<\infty$, by Theorems \ref{uniboltz} and \ref{unilandau}, there exists $(f_t^\epsilon)_{t\in[0,T]}$ solution to (\ref{boltz}) with collision kernel $B_\epsilon$ and $(g_t)_{t\in[0,T]}$ solution to (\ref{landau}) both starting from $f_0$ and lying in $L^\infty([0,T],\mathcal P_{p+2}(\mathbb R^3))\cap L^1([0,T],L^q(\mathbb R^3))$ for some $q\in(3/(3+\gamma),3/(3-\nu))$ (uniformly in $\epsilon\in(0,1)$). Now using $(\ref{Jalpha})$, we get that $(f_t^\epsilon)_{t\in[0,T]}$ and $(g_t)_{t\in[0,T]}$ belong to $L^1([0,T],J_\gamma(\mathbb R^3))$ (uniformly in $\epsilon\in(0,1)$). We thus can use Theorem \ref{dist} with $\beta=\beta_\epsilon$, $\eta=\epsilon$, $n\approx\epsilon^\frac{-2p}{2p+3}$ and $M=\sqrt{2m_2(f_0)}\epsilon^\frac{-2}{2p+3}$ and we get (observe that $\int_\eta^\pi\theta^2\beta_\epsilon(\theta)d\theta=0$, $r_\eta=1$ and that $\int_0^\pi\theta^4\beta_\epsilon(\theta)d\theta\leq C\epsilon^2$)
\begin{align*}
\sup_{[0,T]}\mathcal W_2^2(f_t,g_t)&\leq C\Big[\epsilon^{\frac{2p}{2p+3}}+\epsilon^2+\epsilon^{\frac{2p+2}{2p+3}}(\log^2\epsilon+\epsilon^{\frac{-2}{2p+3}})+\epsilon^{\frac{2p}{2p+3}}\Big]\\
&\leq C \epsilon^\frac{2p}{2p+3},
\end{align*}
since $\log^2\epsilon\leq C\epsilon^{\frac{-2}{2p+3}}$ for any $\epsilon\in(0,1)$. Point $(i)$ is proved.

For point $(ii)$, we consider $f_0\in\mathcal P_{p+2}(\mathbb R^3)\cap L^q(\mathbb R^3)$ for some $p\geq5$ and $q>\frac{3}{3+\gamma}$ with $H(f_0)<\infty$. By Theorems \ref{uniboltz} and \ref{unilandau}, there exists $T_*>0$, $(f_t^\epsilon)_{t\in[0,T_*]}$ solution to (\ref{boltz}) with collision kernel $B_\epsilon$ and $(g_t)_{t\in[0,T_*]}$ solution to (\ref{landau}) both starting from $f_0$ and lying in $L^\infty([0,T_*],\mathcal P_2(\mathbb R^3)\cap L^q(\mathbb R^3))$ (uniformly in $\epsilon\in(0,1)$). We also have that $(g_t)_{t\in[0,T_*]}$ belongs to $L^\infty([0,T_*],\mathcal P_{p+2}(\mathbb R^3))$. Using again $(\ref{Jalpha})$, we get that $(f_t^\epsilon)_{t\in[0,T_*]}$ and $(g_t)_{t\in[0,T_*]}$ belong to $L^1([0,T_*],J_\gamma(\mathbb R^3))$ and to $L^\infty([0,T_*],J_{\gamma+1}(\mathbb R^3))$ if $\gamma\in(-3,-1)$, all this uniformly in $\epsilon\in(0,1)$. We conclude the proof as previously.  \hfill$\square$

\section{Probabilistic interpretation of the equations}

We will use probabilistic tools in order to prove Theorems \ref{colras} and \ref{colrascoul}, like in the paper of Tanaka \cite{TAN}. Until the end of the article,  $(\Omega,\mathcal{F},(\mathcal{F}_t)_{t\geq0},\mathbb P)$ will designate a Polish filtered probability space satisfying the usual conditions. Such a space is Borel isomorphic to the Lebesgue space $([0,1],\mathcal{B}([0,1]),d\alpha)$ which we will use as an auxiliary space. To be as clear as possible, we will use the notation $\mathbb E$ for the expectation and $\mathcal{L}$ for the law of a random variable or process defined on $(\Omega,\mathcal{F},\mathbb P)$, and we will use the notation $\mathbb E_{\alpha}$ and $\mathcal{L}_{\alpha}$ for the expectation and law of random variables or processes on $([0,1],\mathcal{B}([0,1]),d\alpha)$. The processes on $([0,1],\mathcal{B}([0,1]),d\alpha)$ will be called $\alpha$-processes.

\subsection{The Boltzmann equation}
We first need to rewrite the collision operator $A$ defined in (\ref{Aboltz}) as in Fournier-Gu\'erin \cite{FOUGUE1}. The goal of this operation is to make disappear the velocity-dependance $|v-v_*|^\gamma$ in the rate. One can find the following lemma and its proof in \cite[Lemma 2.1]{FOUGUE1}. 
\begin{lemme}
Let $B(|v-v_*|,\theta)\sin\theta=\Phi(|v-v_*|)\beta(\theta)$ with $\beta$ satisfying (\ref{A2}). We set 
\begin{align}  \label{b}
k:=\pi\int_0^\pi(1-\cos\theta)\beta(\theta)d\theta.
\end{align}
Recalling (\ref{HetG}) and (\ref{a(v)}), we define for $z\in(0,\infty)$, $\varphi\in[0,2\pi)$, $v,v_*\in\mathbb R^3$,
\begin{align} \label{cboltz}
c(v,v_*,z,\varphi):=a[v,v_*,G(z/\Phi(|v-v_*|)),\varphi].
\end{align}
We have $A\phi(v,v_*)=\frac{1}{2}[A_1\phi(v,v_*)+A_1\phi(v_*,v)]$ for all $v,v_*\in\mathbb R^3$ and $\phi\in C_b^2(\mathbb R^3)$, where
\begin{align}
\nonumber
A_1\phi(v,v_*)&=\int_0^\infty \int_0^{2\pi}\Big(\phi[v+c(v,v_*,z,\varphi)]-\phi[v]-c[v,v_*,z,\varphi].\nabla\phi[v]\Big)d\varphi dz\\
\nonumber
&\qquad -k\Phi(|v-v_*|)\nabla\phi(v).(v-v_*)\\
\nonumber
&=\int_0^\infty \int_0^{2\pi}\Big(\phi[v+c(v,v_*,z,\varphi+\varphi_0)]-\phi[v]\\
\label{+phi0}
&\qquad\quad-c[v,v_*,z,\varphi+\varphi_0].\nabla\phi[v]\Big)d\varphi dz -k\Phi(|v-v_*|)\nabla\phi(v).(v-v_*),
\end{align}
the second equality holding for any $\varphi_0\in[0,2\pi)$ (which may depend on $v$, $v_*$, $z$). As a consequence, we may replace $A$ by $A_1$ in (\ref{boltz2}).
\end{lemme}
We now recall a fundamental  remark by Tanaka \cite{TAN}, slighlty precised in Fournier-M\'el\'eard \cite[Lemma 2.6]{FOUMEL}.
\begin{lemme} \label{tanaka}
There exists a measurable function $\varphi_0:\mathbb R^3\times\mathbb R^3\rightarrow[0,2\pi)$, such that for all $X,Y\in\mathbb R^3$, all $\varphi\in[0,2\pi)$,
\begin{align}
|\Gamma(X,\varphi)-\Gamma(Y,\varphi+\varphi_0(X,Y))|\leq3|X-Y|,
\end{align}
where $\Gamma(X,Y)$ is defined in (\ref{a(v)}). 
\end{lemme}
We now introduce a nonlinear stochastic differential equation linked with (\ref{boltz}).
\begin{prop} \label{probaboltz}
Let $B(|v-v_*|,\theta)\sin\theta=\Phi(|v-v_*|)\beta(\theta)$ satisfying (i) (\ref{A1}) for some $\gamma\in(-3,0)$, (\ref{A2}) and (\ref{A4}) or (ii) (\ref{AC}). For some $T>0$, let $f=(f_t)_{t\in[0,T]}$ be a solution to (\ref{boltz}) lying in (i) $L^1([0,T],J_\gamma(\mathbb R^3))\cap L^\infty([0,T],\mathcal P_2(\mathbb R^3))$ or in (ii) $L^\infty([0,T],L^\infty(\mathbb R^3))\cap L^\infty([0,T],\mathcal P_2(\mathbb R^3))$. Consider any $\alpha$-process $(\tilde V_t)_{t\in[0,T]}$ such that $\mathcal L_\alpha(\tilde V_t)=f_t$ for all $t\in[0,T]$. Let also $N$ be a $(\mathcal F_t)_{t\in[0,T]}$-Poisson measure on $[0,T]\times[0,\infty)\times[0,2\pi]\times[0,1]$ with intensity measure $dsdzd\varphi d\alpha$, and $V_0$ a $\mathcal F_0$-measurable random variable with law $f_0$. Then there exists a unique process $(V_t)_{t\in[0,T]}$ such that for all $t\in[0,T]$,
\begin{align}
\nonumber
V_t=&V_0+\int_0^t\int_0^\infty\int_0^{2\pi}\int_0^1 c(V_{s-},\tilde V_s(\alpha),z,\varphi)\tilde N(ds,dz,d\varphi,d\alpha)\\
\label{Vt}
&-k\int_0^t\int_0^1\Phi(|V_s-\tilde V_s(\alpha)|)\big(V_s-\tilde V_s(\alpha)\big)dsd\alpha,
\end{align}
with $k$ given by (\ref{b}) and $c$ given by (\ref{cboltz}). Furthermore, $\mathcal L(V_t)=f_t$ for all $t\in[0,T]$.
\end{prop}

\textbf{Proof.}
We start with case $(i)$. In this case, the existence and the uniqueness of $(V_t)_{t\in[0,T]}$ are already proved in Fournier-Gu\'erin \cite[proof of Lemma 4.6, Steps 3 to 6]{FOUGUE1}. We set $\mu_t=\mathcal L(V_t)$. Using It\^o's formula for jump processes (see e.g. Ikeda-Watanabe \cite[Theorem 5.1]{IKE}) and taking expectations, we have for any $\phi\in C_b^2(\mathbb R^3)$
\begin{align*}
\int_{\mathbb R^3}\phi(v)\mu_t(dv)=\int_{\mathbb R^3}\phi(v)f_0(dv)+\int_0^t\int_{\mathbb R^3}\int_{\mathbb R^3}A_1\phi(v,v_*)\mu_s(dv)f_s(dv_*)ds.
\end{align*}
We thus have $\mu_t=f_t$ for any $t\in[0,T]$ by \cite[Lemma 4.6]{FOUGUE1}.

The case $(ii)$ is easier since it is a cutoff case with bounded collision kernel and we leave it to the reader. \hfill $\square$

\subsection{The Landau equation}
To give a probabilistic interpretation of (\ref{landau}), we need to use a three-dimensional space-time white noise $W(ds,d\alpha)$ on $[0,T]\times[0,1]$ with covariance measure $dsd\alpha$ (in the sense of Walsh \cite{WAL}). Recall that $W$ is an orthogonal martingale measure with covariance $dsd\alpha$.
\begin{prop} \label{probalandau}
(i) Let $\gamma\in[-3,0)$. For some $T>0$, let $g=(g_t)_{t\in[0,T]}$ be a solution to (\ref{landau}) lying in $L^1([0,T],J_\gamma(\mathbb R^3))\cap L^\infty([0,T],\mathcal P_2(\mathbb R^3))$ if $\gamma\in(-3,0)$ and in $L^\infty([0,T],L^\infty(\mathbb R^3))\cap L^\infty([0,T],\mathcal P_2(\mathbb R^3))$ if $\gamma=-3$. Consider any $\alpha$-process $(\tilde Y_t)_{t\in[0,T]}$ such that $\mathcal L_\alpha(\tilde Y_t)=g_t$ for all $t\in[0,T]$. Let also $W$ be a three-dimensional space-time white noise on $[0,T]\times[0,1]$ with covariance measure $dsd\alpha$, and $Y_0$ a $\mathcal F_0$-measurable random variable with law $g_0$. Then there exists a unique process $(Y_t)_{t\in[0,T]}$ such that for all $t\in[0,T]$,
\begin{align} \label{EDSlandau}
Y_t=Y_0+\int_0^t\int_0^1\sigma\big(Y_s-\tilde Y_s(\alpha)\big)W(ds,d\alpha)+\int_0^t\int_0^1b\big(Y_s-\tilde Y_s(\alpha)\big)dsd\alpha,
\end{align}
with for any $z\in\mathbb R^3$, $b(z)$ given in (\ref{blandau}) and 
\begin{align} \label{sigmalandau}
\sigma(z)=|z|^{\gamma/2}\begin{pmatrix}
z_2 & -z_3 & 0 \\
-z_1 & 0 & z_3\\
0 & z_1 & -z_2
\end{pmatrix}.
\end{align}
We observe that $\sigma(z)\sigma^*(z)=l(z)$ with $l(z)$ given by (\ref{alandau}). Furthermore, $\mathcal L(Y_t)=g_t$ for all $t\in[0,T]$.\\
(ii) It is possible to handle this construction in such a way that $\mathcal L_\alpha((\tilde Y_t)_{t\in[0,T]})=\mathcal L((Y_t)_{t\in[0,T]})$.
\end{prop}
One can see Fournier-Gu\'erin \cite[Proposition 2.1]{FOUGUE2} for the proof of point (i) when $\gamma\in(-3,0)$ and Fournier \cite[Proposition 10]{FOU2} when $\gamma=-3$.

\textbf{Proof of point (ii).} We first observe that the law of $(Y_t)_{t\in[0,T]}$ does not depend on the choice of $(\tilde Y_t)_{t\in[0,T]}$. To get convinced, use a substitution to rewrite $\int_0^t\int_0^1\sigma\big(Y_s-\tilde Y_s(\alpha)\big)W(ds,d\alpha)$ as $\int_0^t\int_{\mathbb R^3}\sigma\big(Y_s-z\big)\hat W(ds,dz)$ where $\hat W(ds,dz)$ is a white noise with covariance $g_s(dz)ds$.

We thus consider some $\alpha$-process $(\tilde Z_t)_{t\in[0,T]}$ such that $\mathcal L_\alpha(\tilde Z_t)=g_t$ for any $t\in{[0,T]}$ from which we build $(Z_t)_{t\in[0,T]}$ solution to (\ref{EDSlandau}). Next we consider an $\alpha$-process $(\tilde Y_t)_{t\in[0,T]}$ such that $\mathcal L_\alpha((\tilde Y_t)_{t\in[0,T]})=\mathcal L((Z_t)_{t\in[0,T]})$ from which we build $(Y_t)_{t\in[0,T]}$. Due to the previous observation, we have $\mathcal L((Y_t)_{t\in[0,T]})=\mathcal L((Z_t)_{t\in[0,T]})$ and thus $\mathcal L((Y_t)_{t\in[0,T]})=\mathcal L_\alpha((\tilde Y_t)_{t\in[0,T]})$. \hfill$\square$

\section{Soft potentials} \label{general}

This section is devoted to the proof of Theorem \ref{dist}. We fix $\gamma\in(-3,0)$, $T>0$ and we consider a collision kernel satisfying (\ref{A1}-\ref{A2}-\ref{A4}). We consider $(f_t)_{t\in[0,T]}$ and $(g_t)_{t\in[0,T]}$ solutions of (\ref{boltz}) and (\ref{landau}) respectively. 

\subsection{Definition of the processes}
We consider two random variables $V_0$ and $Y_0$ with law $f_0$ and $g_0$ respectively such that $\mathbb E[|V_0-Y_0|^2]=\mathcal W_2^2(f_0,g_0)$. We fix a white noise $W$ on $[0,T]\times[0,1]$ with covariance measure $dsd\alpha$ and we consider a process $(Y_t)_{t\in[0,T]}$ and an $\alpha$-process $(\tilde Y_t)_{t\in[0,T]}$ such that for any $t\in[0,T]$, $\mathcal L(Y_t)=\mathcal L_\alpha(\tilde Y_t)=g_t$, such that $\mathcal L_\alpha\big((\tilde Y_t)_{t\in[0,T]}\big)=\mathcal L\big((Y_t)_{t\in[0,T]}\big)$ and such that (\ref{EDSlandau}) is satisfied. 
For any $t\in[0,T]$, we consider an $\alpha$-random variable $\tilde V_t$ with law $f_t$ such that $\mathcal W_2^2(f_t,g_t)=\mathbb E_\alpha[|\tilde V_t-\tilde Y_t|^2]$ and we consider the solution $(V_t)_{t\in[0,T]}$ to (\ref{Vt}) for some $(\mathcal F_t)_{t\in[0,T]}$-Poisson measure $N$ as in Proposition \ref{probaboltz}. We will precise later the dependence of $N$ with the white noise $W$. We recall the equations satisfied by $(V_t)_{t\in[0,T]}$ and $(Y_t)_{t\in[0,T]}$, and we introduce some intermediate processes (here $n\in\mathbb N^*$ is fixed) 
\begin{align}
\nonumber
V_t=&V_0+\int_0^t\int_0^\infty\int_0^{2\pi}\int_0^1 c(V_{s-},\tilde V_s(\alpha),z,\varphi)\tilde N(ds,dz,d\varphi,d\alpha)\\
\nonumber
&-k\int_0^t\int_0^1|V_s-\tilde V_s(\alpha)|^\gamma\big(V_s-\tilde V_s(\alpha)\big)dsd\alpha,\\
\nonumber
V_t^{n}=&V_0+\int_{a_0}^t\int_0^\infty\int_0^{2\pi}\int_0^1 c(Y_{\rho_n(s)},\tilde Y_{\rho_n(s)}(\alpha),z,\varphi+\Phi_n(s,\alpha))\tilde N(ds,dz,d\varphi,d\alpha)\\
\nonumber
&-k\int_0^t\int_0^1|V_s-\tilde V_s(\alpha)|^\gamma\big(V_s-\tilde V_s(\alpha)\big)dsd\alpha,\\
\nonumber
I_t^n=&V_0+\int_{a_0}^t\int_0^\infty\int_0^{2\pi}\int_0^1 d(Y_{\rho_n(s)},\tilde Y_{\rho_n(s)}(\alpha),z,\varphi+\Phi_n(s,\alpha))\tilde N(ds,dz,d\varphi,d\alpha)\\
\nonumber
&+\int_0^t\int_0^1b\big(V_s-\tilde V_s(\alpha)\big)dsd\alpha,\\
\nonumber
J_t^n=&V_0+\int_{a_0}^t\int_0^1 \sigma\big(Y_{\rho_n(s)}-\tilde Y_{\rho_n(s)}(\alpha)\big)W(ds,d\alpha)+\int_0^t\int_0^1b\big(V_s-\tilde V_s(\alpha)\big)dsd\alpha,\\
\nonumber
Y_t=&Y_0+\int_0^t\int_0^1\sigma\big(Y_s-\tilde Y_s(\alpha)\big)W(ds,d\alpha)+\int_0^t\int_0^1b\big(Y_s-\tilde Y_s(\alpha)\big)dsd\alpha,
\end{align}
where (recall Lemma \ref{tanaka})
\begin{align}
\Phi_n(s,\alpha)&=\varphi_0(V_s- \tilde V_s(\alpha),Y_{\rho_n(s)}-\tilde Y_{\rho_n(s)}(\alpha))\\
\label{dboltz}
d(v,w,z,\varphi)&=\frac{1}{2}G\Big(\frac{z}{|v-w|^\gamma}\Big)\Gamma(v-w,\varphi),
\end{align}
and where $a_0$ and $\rho_n$ are defined as follows.

We consider the subdivision $0<a_0^{n}<...<a_{\lfloor 2nT\rfloor-1}^{n}<a_{\lfloor 2nT\rfloor}^{n}=T$ obtained using Proposition \ref{subdivision} with $h(s)=J_\gamma(g_s)$ on $[0,T]$. In order to lighten notation, we write $a_i=a_i^{n}$. For $s\in[0,T]$, we set 
\begin{align*}
\rho_n(s)=\sum_{i=0}^{\lfloor 2nT\rfloor-1}a_i\one_{s\in[a_i,a_{i+1})}.
\end{align*}
By construction, we have $a_0<1/n$, $1/4n< a_{i+1}-a_i<1/n$, whence $\sup_{[0,T]}|s-\rho_n(s)|\leq1/n$, and
\begin{align} 
\label{grho}
\int_{a_0}^TJ_\gamma(g_{\rho_n(s)})ds\leq3\int_0^TJ_\gamma(g_s)ds+3.
\end{align}
We end this subsection with the following lemma.

\begin{lemme}
For any $v,v_*\in\mathbb R^3$, we have (recall (\ref{b}), (\ref{cboltz}) and (\ref{dboltz})) 
\begin{align} \label{ccarre}
\int_0^\infty\int_0^{2\pi}|c(v,v_*,z,\varphi)|^2dzd\varphi=k|v-v_*|^{\gamma+2},
\end{align}
and
\begin{align} \label{c-d}	
\int_0^\infty\int_0^{2\pi}|c(v,v_*,z,\varphi)-d(v,v_*,z,\varphi)|^2dzd\varphi\leq\int_0^\pi\theta^4\beta(\theta)d\theta|v-v_*|^{\gamma+2}.
\end{align}
\end{lemme}

\textbf{Proof.} 
We have for any $v,v_*\in\mathbb R^3$, $z\in[0,\infty)$ and $\varphi\in[0,2\pi)$ (recall (\ref{a(v)}),
\begin{align*}
|c(v,v_*,z,\varphi)|^2&=\Big|\frac{1-\cos G\big(\frac{z}{|v-v_*|^\gamma}\big)}{2}(v-v_*)-\frac{\sin G\big(\frac{z}{|v-v_*|^\gamma}\big)}{2}\Gamma(v-v_*,\varphi)\Big|^2\\
&=\frac{\Big(1-\cos G\big(\frac{z}{|v-v_*|^\gamma}\big)\Big)^2+\sin^2G\big(\frac{z}{|v-v_*|^\gamma}\big)}{4}|v-v_*|^2\\
&=\frac{1-\cos G\big(\frac{z}{|v-v_*|^\gamma}\big)}{2}|v-v_*|^2,
\end{align*}
since for any $X\in \mathbb R^3$, the vectors $X$ and $\Gamma(X,\varphi)$ are orthogonal, and $|\Gamma(X,\varphi)|=|X|$. Using the substitution $\theta=G\big(\frac{z}{|v-v_*|^\gamma}\big)$, we get
\begin{align*}
\int_0^\infty\int_0^{2\pi}|c(v,v_*,z,\varphi)|^2dzd\varphi=\pi\int_0^\pi(1-\cos\theta)\beta(\theta)d\theta|v-v_*|^{\gamma+2},
\end{align*}
and (\ref{ccarre}) follows. Using the same arguments, we have
\begin{align*}	
\int_0^\infty&\int_0^{2\pi}|c(v,v_*,z,\varphi)-d(v,v_*,z,\varphi)|^2dzd\varphi\\
&=\int_0^\pi\int_0^{2\pi}|v-v_*|^\gamma\Big|\frac{\cos\theta-1}{2}(v-v_*)+\frac{\sin\theta-\theta}{2}\Gamma(v-v_*,\varphi)\Big|^2\beta(\theta)d\varphi d\theta\\
&=2\pi\int_0^\pi\frac{(\cos\theta-1)^2+(\sin\theta-\theta)^2}{4}\beta(\theta)d\theta|v-v_*|^{\gamma+2}.
\end{align*}
It suffices to observe that for $\theta\in[0,\pi]$
\begin{align*}
2\pi\frac{(\cos\theta-1)^2+(\sin\theta-\theta)^2}{4}\leq\theta^4
\end{align*}
to conclude the proof. \hfill$\square$

\subsection{The proof}
We start with a preliminary lemma.
\begin{lemme} \label{prelim}
(i) There exists a constant $C$ depending on $m_2(f_0)$ and additionally on $\sup_{s\in[0,T]} J_{\gamma+1}(f_s)$ if $\gamma\in(-3,-1)$ such that for $0\leq t'\leq t\leq T$ with $t-t'<1$,
\begin{align*}
\mathbb E\Big[|V_t-V_{t'}|^2\Big]\leq C(t-t').
\end{align*}
The same bound holds for $\mathbb E\Big[|Y_t-Y_{t'}|^2\Big]$ and $\mathbb E_\alpha\Big[|\tilde Y_t-\tilde Y_{t'}|^2\Big]$ with $C$ depending on $m_2(g_0)$ and additionally on $\sup_{s\in[0,T]} J_{\gamma+1}(g_s)$ if $\gamma\in(-3,-1)$.\\
(ii) For all $t\in[0,T]$, we have 
\begin{align*}
\mathbb E\Big[|V_t-V_{\rho_n(t)}|^2\Big]+\mathbb E\Big[|Y_t-Y_{\rho_n(t)}|^2\Big]+\mathbb E_\alpha\Big[|\tilde Y_t-\tilde Y_{\rho_n(t)}|^2\Big]\leq\frac{C}{n}.
\end{align*}
\end{lemme}

\textbf{Proof.}
Recalling that $t-\rho_n(t)\leq1/n$, we observe that $(ii)$ immediately follows from $(i)$ taking $t'=\rho_n(t)$. Let's prove $(i)$. Observing that
\begin{align*}
V_t-V_{t'}=&\int_{t'}^t\int_0^\infty\int_0^{2\pi}\int_0^1 c(V_{s-},\tilde V_s(\alpha),z,\varphi)\tilde N(ds,dz,d\varphi,d\alpha)\\
&-k\int_{t'}^t\int_0^1|V_s-\tilde V_s(\alpha)|^\gamma\big(V_s-\tilde V_s(\alpha)\big)dsd\alpha,
\end{align*}
and using (\ref{ccarre}), we get
\begin{align*}
\mathbb E\Big[|V_t-V_{t'}|^2\Big]&\leq 2\int_{t'}^t\int_0^\infty\int_0^{2\pi}\int_0^1 \mathbb E\Big[|c(V_{s},\tilde V_s(\alpha),z,\varphi)|^2\Big]dsdzd\varphi d\alpha\\
&\quad+2k^2\mathbb E\Big[\Big|\int_{t'}^t\int_0^1|V_s-\tilde V_s(\alpha)|^\gamma\big(V_s-\tilde V_s(\alpha)\big)dsd\alpha\Big|^2\Big]\\
&\leq 2k\int_{t'}^t\mathbb E\Big[\mathbb E_\alpha[|V_s-\tilde V_s|^{\gamma+2}]\Big]ds+2k^2\mathbb E\Big[\Big(\int_{t'}^t\mathbb E_\alpha[|V_s-\tilde V_s|^{\gamma+1}]ds\Big)^2\Big]\\
&=:A+B.
\end{align*}
We first deal with $A$. If $\gamma\in[-2,0)$, using that $|a|^{\gamma+2}\leq1+|a|^2$ and recalling that $\mathbb E(|V_s|^2)= \mathbb E_\alpha(|\tilde V_s|^2)=m_2(f_0)$, we have
\begin{align*}
A\leq4k\int_{t'}^t\mathbb E\Big[\mathbb E_\alpha[1+|V_s|^2+|\tilde V_s|^2]\Big]ds\leq4k(1+2m_2(f_0))(t-t'),
\end{align*}
and if $\gamma\in(-3,-2)$, then a.s., $\mathbb E_\alpha[|V_s-\tilde V_s|^{\gamma+2}]\leq1+\mathbb E_\alpha[|V_s-\tilde V_s|^{\gamma+1}]=1+\int_{\mathbb R^3}|V_s-v_*|^{\gamma+1}f_s(dv_*)\leq1+J_{\gamma+1}(f_s)$ (recall (\ref{Jalphadef})), so that
\begin{align*}
A\leq2k\int_{t'}^t(1+J_{\gamma+1}(f_s))ds\leq C(t-t'),
\end{align*}
where $C$ depends on $\sup_{s\in[0,T]} J_{\gamma+1}(f_s)$. We now deal with $B$. If $\gamma\in[-1,0)$, using first the Cauchy-Schwarz  inequality and then that $|a|^{2\gamma+2}\leq1+|a|^2$, we get
\begin{align*}
B&\leq 2k^2\mathbb E\Big[(t-t')\int_{t'}^t\mathbb E_\alpha[|V_s-\tilde V_s|^{2\gamma+2}]ds\Big]\\
&\leq4k^2(t-t')\int_{t'}^t\mathbb E\Big[\mathbb E_\alpha[1+|V_s|^2+|\tilde V_s|^2]\Big]ds\leq4k^2(1+2m_2(f_0))(t-t')^2,
\end{align*}
and if $\gamma\in(-3,-1)$, as previously, we have
\begin{align*}
B\leq2k^2\Big(\int_{t'}^tJ_{\gamma+1}(f_s)ds\Big)^2\leq C(t-t')^2.
\end{align*}
This finally gives
\begin{align*}
\mathbb E\Big[|V_t-V_{t'}|^2\Big]\leq C(t-t'),
\end{align*}
where $C$ depends on $m_2(f_0)$ and on $\sup_{s\in[0,T]} J_{\gamma+1}(f_s)$. The computation of $\mathbb E\Big[|Y_t-Y_{t'}|^2\Big]$ is very similar and we leave it for the reader. Since $(\mathcal L_\alpha\big(\tilde Y_t)_{t\geq0}\big)=(\mathcal L(Y_t)_{t\geq0}\big)$, we have $\mathbb E_\alpha\Big[|\tilde Y_t-\tilde Y_{t'}|^2\Big]=\mathbb E\Big[|Y_t-Y_{t'}|^2\Big]$ and that concludes the proof. \hfill$\square$
\vskip0.5cm

The following lemma states as follows.
\begin{lemme} \label{lem1}
There exists a constant $C$ depending on $m_2(f_0)$, $m_2(g_0)$, $\int_0^T J_\gamma(f_s+g_s)ds$ and additionally on $\sup_{[0,T]}J_{\gamma+1}(f_s+g_s)$ if $\gamma\in(-3,-1)$, such that, if $t\in[a_0,T]$,
\begin{align*}
\mathbb E\Big[|V_t-V_t^{n}|^2\Big]\leq C\Big(\frac{1}{n}+\int_{a_0}^tJ_\gamma(f_s+g_{\rho_n(s)})\Big(\mathbb{E}[|V_s-Y_s|^2]+ \mathbb E_\alpha[|\tilde V_s-\tilde Y_s|^2]\Big)ds\Big).
\end{align*}
\end{lemme}

\textbf{Proof.}
We have
\begin{align*}
V_t-V_t^n=&\int_0^{a_0}\int_0^\infty\int_0^{2\pi}\int_0^1 c(V_{s-},\tilde V_s(\alpha),z,\varphi)\tilde N(ds,dz,d\varphi,d\alpha)\\
&+\int_{a_0}^t\int_0^\infty\int_0^{2\pi}\int_0^1 \Big[c(V_{s-},\tilde V_s(\alpha),z,\varphi)\\
&\qquad\qquad\qquad-c(Y_{\rho_n(s)},\tilde Y_{\rho_n(s)}(\alpha),z,\varphi+\Phi_n(s,\alpha))\Big]\tilde N(ds,dz,d\varphi,d\alpha).
\end{align*}
Consequently,
\begin{align*}
\mathbb E\Big[|V_t-V_t^n|^2\Big]&\leq2\int_0^{a_0}\int_0^\infty\int_0^{2\pi}\int_0^1 \mathbb E\Big[|c(V_{s},\tilde V_s(\alpha),z,\varphi)|^2\Big]dsdzd\varphi d\alpha\\
&\quad+2\int_{a_0}^t\int_0^\infty\int_0^{2\pi}\int_0^1 \mathbb E\Big[\Big|c(V_{s},\tilde V_s(\alpha),z,\varphi)\\
&\qquad\qquad-c\Big(Y_{\rho_n(s)},\tilde Y_{\rho_n(s)}(\alpha),z,\varphi+\Phi_n(s,\alpha)\Big)\Big|^2\Big]dsdzd\varphi d\alpha.
\end{align*}
So using (\ref{ccarre}) and Fournier-Gu\'erin \cite[Lemma 2.3]{FOUGUE1}, we have
\begin{align*}
\mathbb E\Big[|V_t-V_t^n|^2\Big]&\leq C\int_0^{a_0}\int_0^1\mathbb E[|V_s-\tilde V_s(\alpha)|^{\gamma+2}]dsd\alpha\\
&\quad+C\int_{a_0}^t\int_0^1\mathbb E\Big[\big(|V_s-Y_{\rho_n(s)}|^2+|\tilde V_s(\alpha)-\tilde Y_{\rho_n(s)}(\alpha)|^2\big)\\
&\quad\quad\quad\quad\quad \big(|V_s-\tilde V_s(\alpha)|^\gamma+|Y_{\rho_n(s)}-\tilde Y_{\rho_n(s)}(\alpha)|^\gamma\big)\Big]dsd\alpha.
\end{align*}
Since for any $x\geq0$, $x^{\gamma+2}\leq C_\gamma(1+x^2)$ if $\gamma\in[-2,0)$ and $\mathbb E_\alpha[|V_s-\tilde V_s|^{\gamma+2}]\leq1+\mathbb E_\alpha[|V_s-\tilde V_s|^{\gamma+1}]\leq1+\int_{\mathbb R^3}|V_s-v|^{\gamma+1}f_s(dv)\leq1+J_{\gamma+1}(f_s)$ a.s. if $\gamma\in(-3,-2)$, we get (recall that $a_0<1/n$)
\begin{align*} 
	\int_0^{a_0}\int_0^1\mathbb E[|V_s-\tilde V_s(\alpha)|^{\gamma+2}]dsd\alpha=\int_0^{a_0}\mathbb E_\alpha[\mathbb E[|V_s-\tilde V_s|^{\gamma+2}]]ds\leq\frac{C}{n}.
\end{align*}
We thus have
\begin{align*}
\mathbb E\Big[|V_t-V_t^n|^2\Big]&\leq \frac{C}{n}+C\int_{a_0}^t \mathbb E\Big[|V_s-Y_{\rho_n(s)}|^2\mathbb E_\alpha\big(|V_s-\tilde V_s|^\gamma+|Y_{\rho_n(s)}-\tilde Y_{\rho_n(s)}|^\gamma\big)\Big]ds\\
&\quad+C\int_{a_0}^t \mathbb E_\alpha\Big[|\tilde V_s-\tilde Y_{\rho_n(s)}|^2\mathbb E\big(|V_s-\tilde V_s|^\gamma+|Y_{\rho_n(s)}-\tilde Y_{\rho_n(s)}|^\gamma\big)\Big]ds\\
&\leq \frac{C}{n}+C\int_{a_0}^t \mathbb E[|V_s-Y_{\rho_n(s)}|^2]J_\gamma(f_s+g_{\rho_n(s)})ds\\
&\quad+C\int_{a_0}^t\mathbb E_\alpha[|\tilde V_s-\tilde Y_{\rho_n(s)}|^2]J_\gamma(f_s+g_{\rho_n(s)})ds.
\end{align*}
Using first that $\mathbb E[|V_s-Y_{\rho_n(s)}|^2]\leq2\mathbb E[|V_s-Y_s|^2]+2\mathbb E[|Y_s-Y_{\rho_n(s)}|^2]$, $\mathbb E_\alpha[|\tilde V_s-\tilde Y_{\rho_n(s)}|^2]\leq2\mathbb E_\alpha[|\tilde V_s-\tilde Y_s|^2]+2\mathbb E_\alpha[|\tilde Y_s-\tilde Y_{\rho_n(s)}|^2]$, next Lemma \ref{prelim} and (\ref{grho}) concludes the proof.  \hfill$\square$
\vskip0.5cm

We next estimate $V_t^n-I_t^n$.
\begin{lemme} \label{lem2}
There exists a constant $C$ depending on $T$, $\gamma$, $\int_0^TJ_\gamma(f_s)ds$, $m_2(f_0)$, $m_2(g_0)$ and additionally on $\sup_{[0,T]}J_{\gamma+1}(f_s+g_s)$ if $\gamma\in(-3,-1)$ such that, if $t\in[a_0,T]$,
\begin{align*}
\mathbb E\big(|V_t^n-I_t^n|^2\big)\leq C\int_0^\pi\theta^4\beta(\theta)d\theta.
\end{align*}
\end{lemme}

\textbf{Proof.}
We have (recall (\ref{blandau}))
\begin{align*}
V_t^n-I_t^n=&\int_{a_0}^t\int_0^\infty\int_0^{2\pi}\int_0^1 \Big[c(Y_{\rho_n(s)},\tilde Y_{\rho_n(s)}(\alpha),z,\varphi+\Phi_n(s,\alpha))\\
&\qquad\qquad\qquad-d(Y_{\rho_n(s)},\tilde Y_{\rho_n(s)}(\alpha),z,\varphi+\Phi_n(s,\alpha))\Big]\tilde N(ds,dz,d\varphi,d\alpha)\\
&-(k-2)\int_0^t\int_0^1|V_s-\tilde V_s(\alpha)|^\gamma\big(V_s-\tilde V_s(\alpha)\big)dsd\alpha.
\end{align*}
Recalling (\ref{b}) and that $\int_0^\pi\theta^2\beta(\theta)d\theta=\frac{4}{\pi}$, we first observe that
\begin{align*}
	|k-2|=\pi\Big|\int_0^\pi(1-\cos\theta-\frac{\theta^2}{2})\beta(\theta)d\theta\Big|\leq \frac{\pi}{24}\int_0^\pi\theta^4\beta(\theta)d\theta.
\end{align*}
So recalling (\ref{c-d}), we get (recall that $\int_0^\pi\theta^4\beta(\theta)d\theta\leq1$)
\begin{align*}
\mathbb E[|V_t^n-I_t^n|^2]&\leq2\int_{a_0}^t\int_0^\infty\int_0^{2\pi}\int_0^1 \mathbb E\Big[|c(Y_{\rho_n(s)},\tilde Y_{\rho_n(s)}(\alpha),z,\varphi+\Phi_n(s,\alpha))\\
&\qquad\qquad\qquad\qquad-d(Y_{\rho_n(s)},\tilde Y_{\rho_n(s)}(\alpha),z,\varphi+\Phi_n(s,\alpha))|^2\Big]dsdzd\varphi d\alpha\\
&\quad+2(k-2)^2\mathbb E\Big[\Big(\int_0^t\int_0^1 |V_s-\tilde V_s(\alpha)|^{\gamma+1}dsd\alpha\Big)^2\Big]\\
&\leq C\int_0^\pi\theta^4\beta(\theta)d\theta\Big(\int_{a_0}^t \mathbb E\Big[\mathbb E_\alpha[|Y_{\rho_n(s)}-\tilde Y_{\rho_n(s)}|^{\gamma+2}] \Big]ds\\&
\qquad\qquad\qquad\qquad+\mathbb E\Big[\Big(\int_0^t\mathbb E_\alpha[|V_s-\tilde V_s|^{\gamma+1}]ds\Big)^2\Big]\Big).
\end{align*}
We conclude using the same arguments as in the proof of Lemma \ref{prelim} (recall that $Y_{\rho_n(s)}\sim g_{\rho_n(s)}$). \hfill $\square$
\vskip0.5cm

The following lemma is the key point of the proof of Theorem \ref{dist}. 
\begin{lemme} \label{lem3}
Assume that $m_{p+2}(g_0)<\infty$ for some $p\geq5$. We can couple the Poisson measure $N$ and the white noise $W$ in such a way that there exists a constant $C$ depending on $\gamma,T$, $m_{p+2}(g_0)$, $H(g_0)$, and additionally on $\sup_{[0,T]}J_{\gamma+1}(f_s+g_s)$ if $\gamma\in(-3,-2)$ such that for any $M>\sqrt{2m_2(g_0)}$, any $\eta\in(0,\pi)$, any $t\in[a_0,T]$, 
\begin{align*}
\mathbb E[|I_t^n-J_t^n|^2]\leq C\Big[\eta^2M^2n\Big(\log^2(r_\eta)+\log^2(n\eta^2)+M\Big)+\int_\eta^\pi\theta^2\beta(\theta)d\theta+\frac{1}{M^p}\Big].
\end{align*}
\end{lemme}

It is because of this lemma that we do not have an optimal rate of convergence (recall that we obtain here a bound in $\epsilon^{1/2-}$ for $\mathcal W_2(f_t^\epsilon,g_t)$ while we get a bound in $\epsilon$ for the Kac equation, see \cite{FOUGOD}). More precisely, it is due to the fact that we need to partition the interval $[0,T]$ in order to use Proposition \ref{wasserstein}. 
\vskip0.5cm

\textbf{Proof.}
We fix $\eta\in(0,\pi)$ and $M>\sqrt{2m_2(g_0)}$ for the whole proof, which we divide in several steps.

\underline{Step 1}: For $0<u<u'$ and $y$ fixed, we set
\begin{align*}
\mu_u^{u'}(y):=\mathcal L\Big(\int_u^{u'}\int_0^\infty\int_0^{2\pi}\int_0^1d\big(y,\tilde Y_u(\alpha)&,z,\varphi\big)\one_{\{G(z/|y-\tilde Y_u(\alpha)|^\gamma)\leq\eta\}}\\
&\one_{\{|\tilde Y_u(\alpha)|<M\}}\tilde N(ds,dz,d\varphi,d\alpha)\Big),
\end{align*}
and
\begin{align*}
\nu_u^{u'}(y):=\mathcal L\Big(\sqrt {r_\eta}\int_u^{u'}\int_0^1 \sigma\big(y-\tilde Y_{u}(\alpha)\big)\one_{\{|\tilde Y_{u}(\alpha)|<M\}}W(ds,d\alpha)\Big).
\end{align*}
We have
\begin{align*}
	Cov\big(\nu_u^{u'}(y)\big)&=r_\eta\int_u^{u'}\int_0^1 \sigma\big(y-\tilde Y_{u}(\alpha)\big)\sigma^*\big(y-\tilde Y_{u}(\alpha)\big)\one_{\{|\tilde Y_{u}(\alpha)|<M\}}dsd\alpha\\
	&=r_\eta(u'-u)\int_0^1l\big(y-\tilde Y_{u}(\alpha)\big)\one_{\{|\tilde Y_{u}(\alpha)|<M\}}d\alpha\\
	&=(u'-u)\zeta_u(y),
\end{align*}
where
\begin{align*}
\zeta_u(y)=r_\eta\int_0^1l\big(y-\tilde Y_{u}(\alpha)\big)\one_{\{|\tilde Y_{u}(\alpha)|<M\}}d\alpha.
\end{align*}
We thus observe that $\nu_u^{u'}(y)=\mathcal N\big(0,(u'-u)\zeta_u(y)\big)$. Now in order to compute $Cov\big(\mu_u^{u'}(y)\big)$, we first observe that for $v\in\mathbb R^3$ (recall (\ref{a(v)}))
\begin{align*}
\int_0^{2\pi}\Gamma(v,\varphi)\Gamma^*(v,\varphi)d\varphi&=\int_0^{2\pi}\big(\cos\varphi I(v)+\sin\varphi J(v)\big)\big(\cos\varphi I(v)^*+\sin\varphi J(v)^*\big)d\varphi\\
&=\pi(I(v)I(v)^*+J(v)J(v)^*)\\
&=\pi(|v|^2I_3-vv^*)=\pi|v|^{-\gamma}l(v).
\end{align*} 
Using that $\int_0^\infty G^2(z/x)\one_{\{G(z/x)\leq\eta\}}dz=x\int_0^\eta\theta^2\beta(\theta)d\theta=\frac{4x}{\pi}r_\eta$, we have (recall (\ref{dboltz}) and (\ref{alandau}))
\begin{align*}
&Cov\big(\mu_u^{u'}(y)\big)\\
&=\int_u^{u'}\int_0^\infty\int_0^{2\pi}\int_0^1d\big(y,\tilde Y_u(\alpha),z,\varphi\big)d^*\big(y,\tilde Y_u(\alpha),z,\varphi\big)\one_{\{G(z/|y-\tilde Y_u(\alpha)|^\gamma)\leq\eta\}}\\
&\qquad\qquad\qquad\qquad\qquad\qquad\qquad\qquad\one_{\{|\tilde Y_u(\alpha)|<M\}}d\alpha d\varphi dzds\\
&=\frac{1}{4}\int_u^{u'}\int_0^\infty\int_0^{2\pi}\int_0^1G^2(z/|y-\tilde Y_u(\alpha)|^\gamma)\Gamma(y-\tilde Y_u(\alpha),\varphi)\Gamma^*(y-\tilde Y_u(\alpha),\varphi)\\
&\qquad\qquad\qquad\qquad\one_{\{G(z/|y-\tilde Y_u(\alpha)|^\gamma)\leq\eta\}}\one_{\{|\tilde Y_u(\alpha)|<M\}}d\alpha d\varphi dzds\\
&=\frac{r_\eta}{\pi}(u'-u)\int_0^1|y-\tilde Y_u(\alpha)|^\gamma\int_0^{2\pi}\Gamma(y-\tilde Y_u(\alpha),\varphi)\Gamma^*(y-\tilde Y_u(\alpha),\varphi)d\varphi\\
&\qquad\qquad\qquad\qquad\qquad\qquad\qquad\qquad\one_{\{|\tilde Y_u(\alpha)|<M\}}d\alpha\\
&=(u'-u)r_\eta\int_0^1l\big(y-\tilde Y_{u}(\alpha)\big)\one_{\{|\tilde Y_{u}(\alpha)|<M\}}d\alpha\\
&=(u'-u)\zeta_u(y).
\end{align*}
\vskip0.2cm

\underline{Step 2}: the aim of this step is to prove that
\begin{align}
\label{Xab}
&\mathcal W_2^2\Big(\mu_u^{u'}(y),\nu_u^{u'}(y)\big)\Big)\leq C\eta^2M^2(1+|y|^7)\Big(\log^2\frac{(u'-u)r_\eta}{\eta^2}+M\Big).
\end{align}
We set   
\begin{align*}
	\kappa_u(y):=\sup_{\alpha,z,\varphi}|\zeta_u^{-1/2}(y)d\big(y,\tilde Y_u(\alpha),z,\varphi&\big)\one_{\{G(z/|y-\tilde Y_u(\alpha)|^\gamma)\leq\eta\}}\one_{\{|\tilde Y_u(\alpha)|<M\}}|,
\end{align*}
where the supremum is taken over all $\alpha\in[0,1]$, $z\in[0,\infty)$ and $\varphi\in [0,2\pi]$. By Step 1 and Proposition \ref{wasserstein}, we have
\begin{align} \label{Xab2}
	\mathcal W_2^2\Big(\mu_u^{u'}(y),\nu_u^{u'}(y)\big)\Big)&\leq C\kappa_u^2(y)|\zeta_u(y)|\max\Big(1,\log\big(\frac{u'-u}{\kappa_u^2(y)}\big)\Big)^2\\
	\nonumber
	&\leq C|\zeta_u(y)|\psi\big(\kappa_u^2(y)\big),
\end{align}
where $\psi(x)=x\big(1+\log^2 \frac{u'-u}{x}\big)$ for any $x>0$. Let's first deal with $\zeta_u(y)$. Setting $\bar l^h(v)=\int_{\mathbb R^3}l(v-v_*)h(v_*)dv_*$ for a nonnegative function $h$, we observe that 
\begin{align*}
\zeta_u(y)=\lambda_{M,u}r_\eta\bar l^{g_{M,u}}(y),
\end{align*}
with $\lambda_{M,u}=\int_{\mathbb R^3}g_u(v)\one_{\{|v|<M\}}dv$ and $g_{M,u}(v)=\lambda_{M,u}^{-1}g_u(v)\one_{\{|v|<M\}}$ (recall that $\mathcal L_\alpha(\tilde Y_u)=g_u$). Observing that $\lambda_{M,u}\geq1-m_2(g_u)/M^2=1-m_2(g_0)/M^2$, we have $\lambda_{M,u}>1/2$ for any $u\in[0,T]$ since $M>\sqrt{2m_2(g_0)}$ by assumption. We  thus have
\begin{align*}
m_2(g_{M,u})=\lambda_{M,u}^{-1}\int_{\mathbb R^3}|v|^2g_u(v)\one_{\{|v|<M\}}dv\leq2\int_{\mathbb R^3}|v|^2g_0(v)dv=:E_0,
\end{align*}
and
\begin{align*}
H(g_{M,u})&=\lambda_{M,u}^{-1}\int_{\mathbb R^3}g_u(v)\one_{\{|v|<M\}}\log\big(\lambda_{M,u}^{-1}g_u(v)\big)dv\\
&=\lambda_{M,u}^{-1}\int_{\mathbb R^3}g_u(v)\one_{\{|v|<M\}}\Big(\log\big(\lambda_{M,u}^{-1}\big)+\log\big(g_u(v)\big)\Big)dv\\
&\leq \log\big(\lambda_{M,u}^{-1}\big)+2\int_{\mathbb R^3}g_u(v)\log\big(g_u(v)\big)\one_{\{|v|<M\}}dv\\
&\leq \log(2)+2\int_{\mathbb R^3}g_u(v)|\log g_u(v)|dv\\
&\leq \log(2)+2H(g_u)+C(1+m_2(g_u))\\
&\leq \log(2)+2H(g_0)+C(1+m_2(g_0))=:H_0.
\end{align*}
We first used that classically $\int_{\mathbb R^3} g(v)|\log g(v)|dv\leq H(g)+C(1+m_2(g))$ for any $g\in\mathcal P_2(\mathbb R^3)$ and we then used (\ref{conservationl})-(\ref{entropie}). So using Proposition \ref{ellipticity}, there is $c=c(\gamma,E_0,H_0)$ such that for all $u\in[0,T]$, all $\xi\in\mathbb R^3$,
\begin{align*}
	(\bar l^{g_{M,u}}(y)\xi).\xi\geq c(1+|y|)^\gamma|\xi|^2,
\end{align*}
and thus
\begin{align*}
\big(\zeta_u(y)\xi\big).\xi\geq cr_\eta\big(1+|y|\big)^\gamma|\xi|^2.
\end{align*}
This gives
\begin{align*}
|\zeta_u(y)^{-1/2}|^2\leq Cr_\eta^{-1}\big(1+|y|\big)^{|\gamma|},
\end{align*}
and we thus get (recall that $d(y,\tilde Y_u(\alpha),z,\varphi)=\frac{1}{2}G\big(z/|y-\tilde Y_u(\alpha)|^{\gamma}\big)\Gamma(y-\tilde Y_u(\alpha),\varphi)$ and that $|\Gamma(X,\varphi)|=|X|$ for any $X\in\mathbb R^3$) 
\begin{align} \label{kappa2}
\kappa_u^2(y)&\leq Cr_\eta^{-1}\big(1+|y|\big)^{|\gamma|}\eta^2\sup_{\alpha\in[0,1]}|y-\tilde Y_u(\alpha)|^2\one_{\{|\tilde Y_u(\alpha)|<M\}}\\
\nonumber
&\leq Cr_\eta^{-1}\big(1+|y|\big)^{|\gamma|}\eta^2(|y|^2+M^2).
\end{align}
We also have
\begin{align}
\label{normzeta}
|\zeta_u(y)|&\leq r_\eta\int_0^1|y-\tilde Y_u(\alpha)|^{\gamma+2}d\alpha\\
\nonumber
&=r_\eta\int_{\mathbb R^3}|y-v|^{\gamma+2}g_u(dv)\\
\nonumber
&\leq Cr_\eta\big(|y|^{\gamma+2}+m_2(g_u)\big)\one_{\gamma\in[-2,0)}+Cr_\eta J_{\gamma+2}(g_u)\one_{\gamma\in(-3,-2)}\\
\nonumber
&\leq Cr_\eta(1+|y|^{\gamma+2}\one_{\{\gamma\in[-2,0)\}}),
\end{align}
where $C$ depends on $\sup_{[0,T]}J_{\gamma+1}(g_s)$ if $\gamma\in(-3,-2)$ (of course, $J_{\gamma+2}(g_u)$ is controlled by $J_{\gamma+1}(g_u)$ since $\gamma+1<\gamma+2<0$) or on $m_2(g_0)$ if $\gamma\in[-2,0)$. Coming back to (\ref{Xab2}), observing that $\psi$ is an increasing function of $x$ and using (\ref{kappa2}) and (\ref{normzeta}), we get
\begin{align*}
	\mathcal W_2^2\Big(\mu_u^{u'}&(y),\nu_u^{u'}(y)\big)\Big)\leq C(1+|y|^{\gamma+2}\one_{\{\gamma\in[-2,0)\}})\big(1+|y|\big)^{|\gamma|}\eta^2(|y|^2+M^2)\\
	&\qquad\qquad\qquad\qquad\qquad\qquad\qquad\Big(1+\log^2\frac{(u'-u)r_\eta}{\big(1+|y|\big)^{|\gamma|}\eta^2(|y|^2+M^2)}\Big)\\
	&\leq C\eta^2(1+|y|^{\gamma+2}\one_{\{\gamma\in[-2,0)\}})\big(1+|y|\big)^{|\gamma|}(|y|^2+M^2)\Big(1+\log^2\frac{(u'-u)r_\eta}{\eta^2}\\
	&\qquad\qquad\qquad\qquad\qquad\qquad\qquad\qquad+\log^2\big[\big(1+|y|\big)^{|\gamma|}(|y|^2+M^2)\big]\Big),
\end{align*}
since $\log^2(a/b)\leq2\log^2(a)+2\log^2(b)$. Observing that $x\log^2x\leq C(1+x^{1.1})$ for any $x\geq0$, we have
\begin{align*}
 \big(1+|y|\big)^{|\gamma|}(|y|^2+M^2)\log^2\big[\big(1+|y|\big)^{|\gamma|}&(|y|^2+M^2)\big]\\
 &\leq C(1+\big(1+|y|\big)^{1.1|\gamma|}(|y|^2+M^2)^{1.1}).
\end{align*}
Using that 
\begin{align*}
(1+|y|^{\gamma+2}\one_{\{\gamma\in[-2,0)\}})\big(1+|y|\big)^{|\gamma|}\leq C(1+|y|^3)
\end{align*}
and
\begin{align*}(
1+|y|^{\gamma+2}\one_{\{\gamma\in[-2,0)\}})\big(1+|y|\big)^{1.1|\gamma|}\leq C(1+|y|^4)
\end{align*}
(recall that $\gamma\in(-3,0)$), we finally get 
\begin{align*}
	\mathcal W_2^2\Big(\mu_u^{u'}(y),\nu_u^{u'}(y)\big)\Big)&\leq C\eta^2\Big[(1+|y|^3)(|y|^2+M^2)\Big(1+\log^2\frac{(u'-u)r_\eta}{\eta^2}\Big)\\
	&\qquad\qquad+(1+|y|^4)(|y|^2+M^2)^{1.1}\Big]\\
	&\leq C\eta^2\Big[M^2(1+|y|^3)(1+|y|^2)\Big(1+\log^2\frac{(u'-u)r_\eta}{\eta^2}\Big)\\
	&\qquad\qquad+M^3(1+|y|^4)(1+|y|^3)\Big]\\
	&\leq C\eta^2M^2(1+|y|^7)\Big(\log^2\frac{(u'-u)r_\eta}{\eta^2}+M\Big).
\end{align*}
\vskip0.2cm

\underline{Step 3}: recall that the white noise $W$ is fixed. In this step we want to build the Poisson measure $N$ in order to have $\mathbb{E}[|I_t^n-J_t^n|^2]$ as small as possible.  

For any $i\in\{0,...,\lfloor 2nT\rfloor-1\}$, we build a $(\mathcal F_t)_{t\in[0,T]}$-Poisson measure $N^{*,i}$ on $[a_i,a_{i+1})\times[0,\infty)\times[0,2\pi]\times[0,1]$ with intensity measure $dsdzd\varphi d\alpha$ such that a.s.
\begin{align}
\label{couplage}
&\mathcal W_2^2\Big(\mu_{a_i}^{a_{i+1}}(Y_{a_i}),\nu_{a_i}^{a_{i+1}}(Y_{a_i})\Big)\\
\nonumber
&=\mathbb E\Big[\Big|\int_{a_i}^{a_{i+1}}\int_0^\infty\int_0^{2\pi}\int_0^1 d(Y_{a_i},\tilde Y_{a_i}(\alpha),z,\varphi)\\
\nonumber
&\qquad\qquad\qquad\qquad\one_{\{G(z/|Y_{a_i}-\tilde Y_{a_i}(\alpha)|^\gamma)<\eta\}}\one_{\{|\tilde Y_{a_i}(\alpha)|<M\}}\tilde N^{*,i}(ds,dz,d\varphi,d\alpha)\\
\nonumber
&\qquad-\sqrt {r_\eta}\int_{a_i}^{a_{i+1}}\int_0^1 \sigma\big(Y_{a_i}-\tilde Y_{a_i}(\alpha)\big)\one_{\{|\tilde Y_{a_i}(\alpha)|<M\}}W(ds,d\alpha)\Big|^2|\mathcal F_{a_i}\Big].
\end{align}
We are able to do this because $Y_{a_i}$ is $\mathcal F_{a_i}$-measurable. We now consider a $(\mathcal F_t)_{t\in[0,T]}$-Poisson measure $N_{ini}^{*}$ on $[0,a_{0})\times[0,\infty)\times[0,2\pi]\times[0,1]$ with intensity measure $dsdzd\varphi d\alpha$. For any $[u,u']\subset[0,T]$ and $A\subset[0,\infty)\times[0,2\pi]\times[0,1]$, we set $N^*([u,u']\times A):=N_{ini}^{*}(([u,u']\cap[0,a_{0})\times A)+\sum_{i=0}^{\lfloor 2nT\rfloor-1}N^{*,i}(([u,u']\cap[a_i,a_{i+1}))\times A)$ (observe that $N^*$ is a $(\mathcal F_t)_{t\in[0,T]}$-Poisson measure on $[0,T]\times[0,\infty)\times[0,2\pi]\times[0,1]$ with intensity measure $dsdzd\varphi d\alpha$).

Recalling that $\Phi_n(s,\alpha)\in[0,2\pi)$, we set ($\varphi-\Phi_n(s,\alpha)$ should be interpreted modulo $2\pi$)
\begin{align*}
\psi:[u,u']\times[0,\infty)\times[0,2\pi]\times[0,1]&\rightarrow [u,u']\times[0,\infty)\times[0,2\pi]\times[0,1]\\
 (t,z,\varphi,\alpha)&\mapsto(t,z,\varphi-\Phi_n(s,\alpha),\alpha).
\end{align*}
We consider the image $N$ of $N^*$ by $\psi$. Using a remark of Tanaka \cite{TAN}, since $\Phi_n(s,\alpha)=\varphi_0(V_s-\tilde V_s(\alpha),Y_{\rho_n(s)}-\tilde Y_{\rho_n(s)}(\alpha))$ is predictable, we get that $N$ is also a $(\mathcal F_t)_{t\in[0,T]}$-Poisson measure on $[0,T]\times[0,\infty)\times[0,2\pi]\times[0,1]$ with intensity measure $dsdzd\varphi d\alpha$.
\vskip0.5cm

\underline{Step 4}: we set
\begin{align*}
A=\mathbb E\Big[\Big|&\int_{a_0}^t\int_0^\infty\int_0^{2\pi}\int_0^1 d(Y_{\rho_n(s)},\tilde Y_{\rho_n(s)}(\alpha),z,\varphi+\Phi_n(s,\alpha))\\
&\qquad\qquad\qquad\one_{\{G(z/|Y_{\rho_n(s)}-\tilde Y_{\rho_n(s)}(\alpha)|^\gamma)\leq\eta\}}\one_{\{|\tilde Y_{\rho_n(s)}(\alpha)|<M\}}\tilde N(ds,dz,d\varphi,d\alpha)\\
&-\sqrt {r_\eta}\int_{a_0}^t\int_0^1 \sigma\big(Y_{\rho_n(s)}-\tilde Y_{\rho_n(s)}(\alpha)\big)\one_{\{|\tilde Y_{\rho_n(s)}(\alpha)|<M\}}W(ds,d\alpha)\Big|^2\Big].
\end{align*}
The aim of this step is to show that
\begin{align}
\label{A}
A\leq C\eta^2M^2n\Big(\log^2(r_\eta)+\log^2(n\eta^2)+M\Big),
\end{align}
where $C$ depends on $\gamma$, $T$, $m_7(g_0)$ and additionally on $\sup_{[0,T]}J_{\gamma+1}(g_s)$ if $\gamma\in(-3,-2)$. By construction of $N$, we have
\begin{align*}
A=\mathbb E\Big[\Big|&\int_{a_0}^t\int_0^\infty\int_0^{2\pi}\int_0^1 d(Y_{\rho_n(s)},\tilde Y_{\rho_n(s)}(\alpha),z,\varphi)\\
&\qquad\qquad\qquad\one_{\{G(z/|Y_{\rho_n(s)}-\tilde Y_{\rho_n(s)}(\alpha)|^\gamma)\leq\eta\}}\one_{\{|\tilde Y_{\rho_n(s)}(\alpha)|<M\}}\tilde N^*(ds,dz,d\varphi,d\alpha)\\
&-\sqrt {r_\eta}\int_{a_0}^t\int_0^1 \sigma\big(Y_{\rho_n(s)}-\tilde Y_{\rho_n(s)}(\alpha)\big)\one_{\{|\tilde Y_{\rho_n(s)}(\alpha)|<M\}}W(ds,d\alpha)\Big|^2\Big],
\end{align*}
and setting for any $0<u<u'<T$ and $y\in\mathbb R^3$,
\begin{align*}
X_u^{u'}(y):=\int_u^{u'}\int_0^\infty\int_0^{2\pi}\int_0^1d\big(y,\tilde Y_u(\alpha),z,\varphi&\big)\one_{\{G(z/|y-\tilde Y_u(\alpha)|^\gamma)\leq\eta\}}\\
&\one_{\{|\tilde Y_u(\alpha)|<M\}}\tilde N^*(ds,dz,d\varphi,d\alpha),
\end{align*}
we have
\begin{align*}
A&\leq \mathbb E\Big[\Big|\sum_{i=0}^{\lfloor 2nT\rfloor-1}\Big(X_{a_i}^{a_{i+1}}(Y_{a_i})\\
&\qquad\qquad-\sqrt {r_\eta}\int_{a_i}^{a_{i+1}}\int_0^1 \sigma\big(Y_{a_i}-\tilde Y_{a_i}(\alpha)\big)\one_{\{|\tilde Y_{a_i}(\alpha)|<M\}}W(ds,d\alpha)\Big)\Big|^2\Big]\\
&=\sum_{i=0}^{\lfloor 2nT\rfloor-1}\mathbb E\Big[\Big|X_{a_i}^{a_{i+1}}(Y_{a_i})\\
&\qquad\qquad-\sqrt {r_\eta}\int_{a_i}^{a_{i+1}}\int_0^1 \sigma\big(Y_{a_i}-\tilde Y_{a_i}(\alpha)\big)\one_{\{|\tilde Y_{a_i}(\alpha)|<M\}}W(ds,d\alpha)\Big|^2\Big],
\end{align*}
since for $i\neq j$, $\Big(N^*_{|[a_i,a_{i+1})}$, $W_{|[a_i,a_{i+1})}\Big)$ and $\Big(N^*_{|[a_j,a_{j+1})}$, $W_{|[a_j,a_{j+1})}\Big)$ are independent, which gives
\begin{align*}
\mathbb E\Big[&\Big(X_{a_i}^{a_{i+1}}(Y_{a_i})-\sqrt {r_\eta}\int_{a_i}^{a_{i+1}}\int_0^1 \sigma\big(Y_{a_i}-\tilde Y_{a_i}(\alpha)\big)\one_{\{|\tilde Y_{a_i}(\alpha)|<M\}}W(ds,d\alpha)\Big).\\
&\Big(X_{a_j}^{a_{j+1}}(Y_{a_j})-\sqrt {r_\eta}\int_{a_j}^{a_{j+1}}\int_0^1 \sigma\big(Y_{a_j}-\tilde Y_{a_j}(\alpha)\big)\one_{\{|\tilde Y_{a_j}(\alpha)|<M\}}W(ds,d\alpha)\Big)\Big]=0.
\end{align*}
First taking the conditional expectation with respect to $\mathcal F_{a_i}$ for each term of the sum, and then using (\ref{couplage}) and (\ref{Xab}), we have
\begin{align*}
A&\leq\sum_{i=0}^{\lfloor 2nT\rfloor-1}\mathbb E\Bigg[\mathbb E\Big[\Big|X_{a_i}^{a_{i+1}}(Y_{a_i})\\
&\qquad\qquad-\sqrt {r_\eta}\int_{a_i}^{a_{i+1}}\int_0^1 \sigma\big(Y_{a_i}-\tilde Y_{a_i}(\alpha)\big)\one_{\{|\tilde Y_{a_i}(\alpha)|<M\}}W(ds,d\alpha)\Big|^2|\mathcal F_{a_i}\Big]\Bigg]\\
&=\sum_{i=0}^{\lfloor 2nT\rfloor-1}\mathbb E\Big[\mathcal W_2^2\Big(\mu_{a_i}^{a_{i+1}}(Y_{a_i}),\nu_{a_i}^{a_{i+1}}(Y_{a_i})\Big)\Big]\\
&\leq C\sum_{i=0}^{\lfloor 2nT\rfloor-1}\eta^2M^2\Big(1+\mathbb E[|Y_{a_i}|^7]\Big)\Big(\log^2\big(\frac{r_\eta(a_{i+1}-a_i)}{\eta^2}\big)+M\Big)\\
&\leq C\eta^2M^2n\Big(\log^2\big(\frac{r_\eta}{n\eta^2}\big)+M\Big)\leq C\eta^2M^2n\Big(\log^2(r_\eta)+\log^2(n\eta^2)+M\Big),
\end{align*}
where we used that $1/4n<a_{i+1}-a_i<1/n$ by construction (recall Proposition \ref{subdivision}) and that $\mathbb E[|Y_{a_i}|^7]\leq Cm_7(g_0)$.
\vskip 0.5cm

\underline{Step 5}: we finally compute $\mathbb E[|I_t^n-J_t^n|^2]$. We have
\begin{align*}
I_t^n-J_t^n=&\int_{a_0}^t\int_0^\infty\int_0^{2\pi}\int_0^1 d(Y_{\rho_n(s)},\tilde Y_{\rho_n(s)}(\alpha),z,\varphi+\Phi_n(s,\alpha))\tilde N(ds,dz,d\varphi,d\alpha)\\
&-\int_{a_0}^t\int_0^1 \sigma\big(Y_{\rho_n(s)}-\tilde Y_{\rho_n(s)}(\alpha)\big)W(ds,d\alpha).
\end{align*}
This gives $\mathbb E[|I_t^n-J_t^n|^2]\leq 4(A+B+D)$ with

\begin{align*}
B=(\sqrt {r_\eta}-1)^2\mathbb E\Big[\Big|\int_{a_0}^t\int_0^1 \sigma\big(Y_{\rho_n(s)}-\tilde Y_{\rho_n(s)}(\alpha)\big)\one_{\{|\tilde Y_{\rho_n(s)}(\alpha)|<M\}}W(ds,d\alpha)\Big|^2\Big],
\end{align*}
and
\begin{align*}
D=\mathbb E\Big[\Big|&\int_{a_0}^t\int_0^\infty\int_0^{2\pi}\int_0^1 d(Y_{\rho_n(s)},\tilde Y_{\rho_n(s)}(\alpha),z,\varphi+\Phi_n(s,\alpha))\\
&\qquad\qquad\qquad\one_{\{G(z/|Y_{\rho_n(s)}-\tilde Y_{\rho_n(s)}(\alpha)|^\gamma)>\eta\}\cup\{|\tilde Y_{\rho_n(s)}(\alpha)|>M\}}\tilde N(ds,dz,d\varphi,d\alpha)\\
&-\int_{a_0}^t\int_0^1 \sigma\big(Y_{\rho_n(s)}-\tilde Y_{\rho_n(s)}(\alpha)\big)\one_{\{|\tilde Y_{\rho_n(s)}(\alpha)|>M\}}W(ds,d\alpha)\Big|^2\Big].
\end{align*}
Using that $\sum_{i,k=1}^3\sigma_{ik}^2(Y_{\rho_n(s)}-\tilde Y_{\rho_n(s)}(\alpha))=2|Y_{\rho_n(s)}-\tilde Y_{\rho_n(s)}(\alpha)|^{\gamma+2}$ (recall (\ref{sigmalandau})), 
\begin{align}
\label{B}
B&=(\sqrt {r_\eta}-1)^2\int_{a_0}^t\int_0^1\mathbb E\Big[\sum_{i,k=1}^3\sigma_{ik}^2(Y_{\rho_n(s)}-\tilde Y_{\rho_n(s)}(\alpha))\Big]\one_{\{|\tilde Y_{\rho_n(s)}(\alpha)|<M\}}dsd\alpha\\
\nonumber
& \leq2(\sqrt {r_\eta}-1)^2\int_{a_0}^t\int_0^1\mathbb E\Big[|Y_{\rho_n(s)}-\tilde Y_{\rho_n(s)}(\alpha)|^{\gamma+2}\Big]dsd\alpha\\
\nonumber
&\leq C\Big(\int_\eta^\pi\theta^2\beta(\theta)d\theta\Big)^2\Big(t\one_{\{\gamma\in[-2,0)\}}+\int_{a_0}^tJ_{\gamma+2}(g_{\rho_n(s)})ds\one_{\{\gamma\in(-3,-2)\}}\Big)\\
\nonumber
&\leq Ct\int_\eta^\pi\theta^2\beta(\theta)d\theta,
\end{align}
where $C$ depends on $m_2(g_0)$ and on $\sup_{[0,T]}J_{\gamma+2}(g_s)$ if $\gamma\in(-3,-2)$ (which is controlled by $\sup_{[0,T]}J_{\gamma+1}(g_s)$). We used that $|\sqrt {r_\eta}-1|\leq C|r_\eta-1|$ and that $r_\eta-1=-\frac{\pi}{4}\int_\eta^\pi\theta^2\beta(\theta)d\theta$ by (\ref{A2}). We removed the square of $\int_\eta^\pi\theta^2\beta(\theta)d\theta$ because it will appear without square in the computation of $D$. Using first that $|a-b|^2\leq2|a|^2+2|b|^2$, and then the substitution $\theta=G(z/|Y_{\rho_n(s)}-\tilde Y_{\rho_n(s)}(\alpha)|^\gamma)$ for which $dz=|Y_{\rho_n(s)}-\tilde Y_{\rho_n(s)}(\alpha)|^\gamma\beta(\theta)d\theta$ (recall (\ref{dboltz})) for the Poisson integral, we get
\begin{align}
\nonumber
D&\leq C\int_{a_0}^t\int_0^\pi\int_0^1\theta^2\beta(\theta)\mathbb E[|Y_{\rho_n(s)}-\tilde Y_{\rho_n(s)}(\alpha)|^{\gamma+2}]\one_{\{\theta>\eta\}\cup\{|\tilde Y_{\rho_n(s)}(\alpha)|>M\}}d\alpha d\theta ds\\
\nonumber
&\quad+C\int_{a_0}^t\int_0^1 \mathbb E[|Y_{\rho_n(s)}-\tilde Y_{\rho_n(s)}(\alpha)|^{\gamma+2}]\one_{\{|\tilde Y_{\rho_n(s)}(\alpha)|>M\}}d\alpha ds.
\end{align}
If $\gamma\in[-2,0)$, we have
\begin{align}
\nonumber
D&\leq C\int_\eta^\pi\theta^2\beta(\theta)d\theta\int_{a_0}^t \Big(1+\mathbb E[|Y_{\rho_n(s)}|^{2}]+\mathbb E_\alpha[|\tilde Y_{\rho_n(s)}|^2]\Big)ds\\
\nonumber
&\quad+C\int_{a_0}^t \mathbb E_\alpha\Big[\Big(1+\mathbb E[|Y_{\rho_n(s)}|^{2}]+|\tilde Y_{\rho_n(s)}(\alpha)|^2\Big)\one_{\{|\tilde Y_{\rho_n(s)}(\alpha)|>M\}}\Big]ds\\
\nonumber
&\leq Ct\int_\eta^\pi\theta^2\beta(\theta)d\theta+C\int_{a_0}^t\frac{1+\mathbb E_{\alpha}[|\tilde Y_{\rho_n(s)}|^p]+\mathbb E_\alpha[|\tilde Y_{\rho_n(s)}|^{2+p}]}{M^p}\\
\label{D1}
&\leq Ct\int_\eta^\pi\theta^2\beta(\theta)d\theta+\frac{Ct}{M^p},
\end{align}
where $C$ depends on $m_{p+2}(g_0)$. If $\gamma\in(-3,-2)$
\begin{align}
\label{D2}
D&\leq C\int_\eta^\pi\theta^2\beta(\theta)d\theta\int_{a_0}^tJ_{\gamma+2}(g_{\rho_n(s)})ds+C\int_{a_0}^tJ_{\gamma+2}(g_{\rho_n(s)})\mathbb E_\alpha[\one_{\{|\tilde Y_{\rho_n(s)}(\alpha)|>M\}}]ds\\
\nonumber
&\leq C\int_\eta^\pi\theta^2\beta(\theta)d\theta\int_{a_0}^tJ_{\gamma+2}(g_{\rho_n(s)})ds+\frac{C}{M^p}\int_{a_0}^tJ_{\gamma+2}(g_{\rho_n(s)})ds\\
\nonumber
&\leq Ct\int_\eta^\pi\theta^2\beta(\theta)d\theta+\frac{Ct}{M^p},
\end{align}
where $C$ depends on $m_p(g_0)$ and on $\sup_{[0,T]}J_{\gamma+1}(g_s)$. It suffices to use (\ref{A}), (\ref{B}), (\ref{D1}) and (\ref{D2}) to conclude. \hfill$\square$
\vskip 0.5cm

We finally state the last lemma needed to conclude the proof of Theorem \ref{dist}.

\begin{lemme} \label{lem4}
There exists a constant $C$ depending on $\gamma$, $T$, $\int_0^TJ_\gamma(f_s+g_s)ds$, $m_2(g_0)$ and additionnally on $\sup_{[0,T]}J_{\gamma+1}(g_s)$ if $\gamma\in(-3,-1)$ such that, if $t\in[a_0,T]$,  
\begin{align*}
\mathbb E[|J_t^n-Y_t|^2]\leq& C\Big(\mathbb E[|V_0-Y_0|^2]+\frac{1}{n}\\
&+\int_0^t\Big(\mathbb E[|V_s-Y_s|^2]+\mathbb E_\alpha[|\tilde V_s-\tilde Y_s|^2]\Big)J_\gamma(f_s+g_s)ds\Big).
\end{align*}
\end{lemme}

\textbf{Proof.}
We have
\begin{align*}
J_t^n-Y_t=&V_0-Y_0-\int_0^{a_0}\int_0^1\sigma\big(Y_s-\tilde Y_s(\alpha)\big)W(ds,d\alpha)\\
&+\int_{a_0}^t\int_0^1 \Big[\sigma\big(Y_{\rho_n(s)}-\tilde Y_{\rho_n(s)}(\alpha)\big)-\sigma\big(Y_s-\tilde Y_{s}(\alpha)\big)\Big]W(ds,d\alpha)\\
&+\int_0^t\int_0^1\Big[b\big(V_s-\tilde V_s(\alpha)\big)-b\big(Y_s-\tilde Y_s(\alpha)\big)\Big]dsd\alpha.
\end{align*}
Using It\^o's formula and taking expectations, we get
\begin{align*}
\mathbb E[|J_t^n-Y_t|^2]&=\mathbb E[|V_0-Y_0|^2]+\int_0^{a_0}\int_0^1\mathbb E\Big[\sum_{i,k=1}^3\sigma_{ik}^2(Y_s-\tilde Y_s(\alpha))\Big]dsd\alpha\\
&\quad+\int_{a_0}^t\int_0^1 \sum_{i,k=1}^3\mathbb E\Big[\Big(\sigma_{ik}\big(Y_{\rho_n(s)}-\tilde Y_{\rho_n(s)}(\alpha)\big)\\
&\qquad\qquad\qquad\qquad-\sigma_{ik}\big(Y_s-\tilde Y_{s}(\alpha)\big)\Big)^2\Big]dsd\alpha\\
&\quad+2\int_0^t\int_0^1\mathbb E\Big[\Big(b\big(V_s-\tilde V_s(\alpha)\big)-b\big(Y_s-\tilde Y_s(\alpha)\big)\Big).\Big(J_s^n-Y_s\Big)\Big]dsd\alpha\\
&=:\mathbb E[|V_0-Y_0|^2]+2\int_0^{a_0}\int_0^1\mathbb E[|Y_s-\tilde Y_s(\alpha)|^{\gamma+2}]dsd\alpha+A+B,
\end{align*}
since $\sum_{i,k=1}^3\sigma_{ik}^2(Y_s-\tilde Y_s(\alpha))=2|Y_s-\tilde Y_s(\alpha)|^{\gamma+2}$ (recall (\ref{sigmalandau})). Using that $|a|^{\gamma+2}\leq1+|a|^2$ if $\gamma\in[-2,0)$ (recall also that $\mathbb E(|Y_s|^2)=\mathbb E_\alpha(|\tilde Y_s|^2)=m_2(g_0)$) and that $\mathbb E[|Y_s-\tilde Y_s(\alpha)|^{\gamma+2}]\leq J_{\gamma+2}(g_s)\leq1+J_{\gamma+1}(g_s)$ if $\gamma\in(-3,-2)$, we have (recall that $a_0\leq1/n$)
\begin{align*}
\int_0^{a_0}\int_0^1\mathbb E[|Y_s-\tilde Y_s(\alpha)|^{\gamma+2}]dsd\alpha\leq\frac{C}{n}.
\end{align*}
Using Fournier-Gu\'erin \cite[Remark 2.2]{FOUGUE2}, we get
\begin{align*}
A&\leq C\int_{a_0}^t\int_0^1\mathbb E\Big[|Y_{\rho_n(s)}-Y_s+\tilde Y_s(\alpha)-\tilde Y_{\rho_n(s)}(\alpha)|^2\big(|Y_{\rho_n(s)}-\tilde Y_{\rho_n(s)}(\alpha)|^\gamma\\
&\qquad\qquad\qquad\qquad\qquad\qquad\qquad\qquad\qquad\qquad\qquad\quad+|Y_s-\tilde Y_s(\alpha)|^\gamma\big)\Big]dsd\alpha\\
&\leq C\int_{a_0}^t\Big(\mathbb E\big[|Y_{\rho_n(s)}-Y_s|^2\big]+\mathbb E_\alpha\big[|\tilde Y_s-\tilde Y_{\rho_n(s)}|^2\big]\Big)J_\gamma(g_{\rho_n(s)}+g_s)ds\\
&\leq \frac{C}{n}\Big(\int_0^TJ_\gamma(g_s)ds+1\Big),
\end{align*}
(recall Lemma \ref{prelim} and (\ref{grho})) and
\begin{align*}
B&\leq C\int_0^t\int_0^1\mathbb E\Big[|V_s-Y_s+\tilde Y_s(\alpha)-\tilde V_s(\alpha)|\big(|V_s-\tilde V_s(\alpha)|^\gamma\\
&\qquad\qquad\qquad\qquad\qquad\qquad\qquad\qquad\qquad+|Y_s-\tilde Y_s(\alpha)|^\gamma\big)|J_s^n-Y_s|\Big]dsd\alpha\\
&\leq C\int_0^t\mathbb E\Big[\big(|V_s-Y_s|^2+|J_s^n-Y_s|^2\big)\mathbb E_\alpha[|V_s-\tilde V_s(\alpha)|^\gamma+|Y_s-\tilde Y_s(\alpha)|^\gamma]\Big]ds\\
&\quad +C\int_0^t\mathbb E_\alpha\Big[|\tilde V_s-\tilde Y_s|^2\mathbb E[|V_s-\tilde V_s(\alpha)|^\gamma+|Y_s-\tilde Y_s(\alpha)|^\gamma]\Big]ds\\
&\leq C\int_0^t\Big(\mathbb E\big[|V_s-Y_s|^2\big]+\mathbb E\big[|J_s^n-Y_s|^2\big]+\mathbb E_\alpha[|\tilde V_s-\tilde Y_s|^2]\Big)J_\gamma(f_s+g_s)ds.
\end{align*}
We thus get
\begin{align*}
\mathbb E[|J_t^n-&Y_t|^2]\leq  \mathbb E[|V_0-Y_0|^2]+\frac{C}{n}\\
&+C\int_0^t\Big(\mathbb E\big[|V_s-Y_s|^2\big]+\mathbb E_\alpha\big[|\tilde V_s-\tilde Y_s|^2\big]+\mathbb E\big[|J_s^n-Y_s|^2\big]\Big)J_\gamma(f_s+g_s)ds,
\end{align*}
and we conclude by Gr\"onwall's lemma. \hfill$\square$
\vskip0.5cm

We can now prove Theorem \ref{dist}.

\textbf{Proof of Theorem \ref{dist}.} We couple the Poisson measure $N$ and the white noise $W$ as in Lemma \ref{lem3}. Recall that $\mathbb E_\alpha[|\tilde V_s-\tilde Y_s|^2]=\mathcal W_2^2(f_s,g_s)\leq\mathbb E[|V_s-Y_s|^2]=:u(s)$ and $\mathbb E[|V_0-Y_0|^2]=\mathcal W_2^2(f_0,g_0)$. We first observe that if $t<a_0$,
\begin{align*}
	u(t)\leq 4\mathbb E[|V_t-V_0|^2]+4\mathbb E[|V_0-Y_0|^2]+4\mathbb E[|Y_t-Y_0|^2]&\leq C\Big(\mathbb E[|V_0-Y_0|^2]+a_0\Big)\\
	&\leq C\Big(\mathcal W_2^2(f_0,g_0)+\frac{1}{n}\Big),
\end{align*}
by Lemma \ref{prelim} and the result is proved when $t<a_0$. Using Lemmas \ref{lem1}, \ref{lem2}, \ref{lem3}, \ref{lem4} and (\ref{grho}), we have, for $t\in[a_0,T]$, 
\begin{align*}
u(t)&\leq C\Big(\frac{1}{n}+\int_{a_0}^tJ_\gamma(f_s+g_{\rho_n(s)})\big(\mathbb E[|V_s-Y_s|^2]+\mathbb E_\alpha[|\tilde V_s-\tilde Y_s|^2]\big)ds\Big)\\
&\quad+C\int_0^\pi\theta^4\beta(\theta)d\theta\\
&\quad+C\Big[\eta^2M^2n\Big(\log^2(r_\eta)+\log^2(n\eta^2)+M\Big)+\int_\eta^\pi\theta^2\beta(\theta)d\theta+\frac{1}{M^p}\Big]\\
&\quad +C\Big(\mathbb E[|V_0-Y_0|^2]+\frac{1}{n}+\int_0^t\big(\mathbb E[|V_s-Y_s|^2]+\mathbb E_\alpha[|\tilde V_s-\tilde Y_s|^2]\big)J_\gamma(f_s+g_s)ds\Big)\\
&\leq C\Big(\mathcal W_2^2(f_0,g_0)+\int_{0}^tJ_\gamma(f_s+g_s+g_{\rho_n(s)}\one_{\{s\geq a_0\}})u(s)ds+\int_0^\pi\theta^4\beta(\theta)d\theta\\
&\quad+\frac{1}{n}+ \eta^2M^2n\Big(\log^2(r_\eta)+\log^2(n\eta^2)+M\Big)+\int_\eta^\pi\theta^2\beta(\theta)d\theta+\frac{1}{M^p}\Big),
\end{align*}
for all $n\in\mathbb N^*$, $\eta\in(0,\pi)$ and $M>\sqrt{2m_2(g_0)}$.
Using the generalized Gr\"onwall Lemma and (\ref{grho}), we get
\begin{align*}
u(t)\leq C\Big(\mathcal W_2^2(f_0,g_0)+\int_0^\pi\theta^4\beta(\theta)d\theta+\frac{1}{n}&+ \eta^2M^2n\Big(\log^2(r_\eta)+\log^2(n\eta^2)+M\Big)\\
&+\int_\eta^\pi\theta^2\beta(\theta)d\theta+\frac{1}{M^p}\Big).
\end{align*} 
This concludes the proof since $\mathcal W_2^2(f_t,g_t)\leq\mathbb E[|V_t-Y_t|^2]=u(t)$. \hfill $\square$

\section{The Coulomb Case}

This section is devoted to the proof of Theorem \ref{colrascoul}. We thus assume (\ref{AC}) and consider $f_0\in\mathcal P_p(\mathbb R^3)\cap L^\infty(\mathbb R^3)$ for some $p\geq7$. By Theorems \ref{uniboltzcoul} and \ref{unilandau} $(iii)$, we can consider $T>0$ and $(f_t^{\epsilon})_{t\in[0,T]}$, $(g_t)_{t\in[0,T]}$ solutions to (\ref{boltz}) and (\ref{landau}) respectively, both starting from $f_0$ and lying in $L^\infty([0,T],L^\infty(\mathbb R^3))\cap L^\infty([0,T],\mathcal P_2(\mathbb R^3))$ (uniformly in $\epsilon$ for $f^\epsilon$) with $g$ which additionally lies in $L^\infty([0,T],\mathcal{P}_p(\mathbb R^3))$.

\subsection{Some preliminary results}
The main difficulty of the Coulomb case is the fact that $\int_{|v|<1}|v|^{-3}dv$ is not finite. We will use the following lemma stated in the paper of Fournier \cite[Lemma 4]{FOU2} in order to deal with this difficulty.
\begin{lemme} \label{r3v3} 
Let $\alpha\in(-3,0]$. There is a constant $C_\alpha$ such that for all $h\in\mathcal P(\mathbb R^3)\cap L^\infty(\mathbb R^3)$, all $\epsilon\in(0,1]$,
\begin{align*}
&\sup_{v\in\mathbb R^3}\int_{\mathbb R^3}|v-v_*|^\alpha h(v_*)dv_*\leq 1+C_\alpha||h||_\infty,\\
&\int_{\mathbb R^3}\int_{\mathbb R^3}|v-v_*|^\alpha h(v)h(v_*)dvdv_*\leq1+C_\alpha||h||_\infty,\\
&\sup_{v,w\in\mathbb R^3}\int_{|v-v_*|\leq\epsilon}|w-v_*|^\alpha h(v_*)dv_*\leq C_\alpha||h||_\infty\epsilon^{3+\alpha}.
\end{align*}
There is a constant $C$ such that for all $h\in\mathcal P(\mathbb R^3)\cap L^\infty(\mathbb R^3)$, all $\epsilon\in(0,1]$,
\begin{align*}
\sup_{v\in\mathbb R^3}\int_{|v-v_*|\geq\epsilon} |v-v_*|^{-3}h(v_*)dv_*\leq 1+C||h||_\infty\log(1/\epsilon).
\end{align*}
\end{lemme}

We will need to use a generalisation of the Gr\"onwall Lemma. To this aim, we consider the increasing continuous function $\psi:[0,\infty)\rightarrow\mathbb R_+$ defined by
\begin{align} \label{functionpsi}
\psi(x)=x(1-\one_{x\leq1}\log x).
\end{align}
Setting $\tilde\psi(x):=x(1-\log x)\one_{x\in[0,1/2]}+(x\log2+1/2)\one_{x\geq1/2}$, we observe that $\psi(x)/2\leq\tilde\psi(x)\leq2\psi(x)$ for any $x\geq0$. Since the function $\tilde\psi:\mathbb R_+\rightarrow\mathbb R_+$ is concave increasing, this last observation will almost allow us to apply the Jensen inequality to the function $\psi$.

As mentioned before, we only need the parameter $h_\epsilon$ in order to have easily existence and uniqueness of solutions of (\ref{boltz}). In order to point out that we do not need this cutoff parameter in quite all calculus, we consider (recall (\ref{HetGcoul}) and (\ref{a(v)}))
\begin{align} \label{ccoul}
c_\epsilon(v,v_*,z,\varphi)=a(v,v_*,G_\epsilon(z/|v-v_*|^{-3}),\varphi),
\end{align}
and
\begin{align} \label{chcoul}
c_{h_\epsilon,\epsilon}(v,v_*,z,\varphi)=a(v,v_*,G_\epsilon(z/(|v-v_*|+h_\epsilon)^{-3}),\varphi).
\end{align}

\begin{lemme} \label{ccoulomb}
(i) For any $v,v_*\in\mathbb R^3$,
\begin{align*}
\int_0^\infty\int_0^{2\pi}|c_\epsilon(v,v_*,z,\varphi)|^2dzd\varphi=k_\epsilon|v-v_*|^{-1},
\end{align*}
where
\begin{align} \label{keps}
k_\epsilon=\pi\int_0^\pi(1-\cos\theta)\beta_\epsilon(\theta)d\theta.
\end{align}
We also have $k_\epsilon\leq2$.\\
(ii) For any $v,v_*\in\mathbb R^3$,
\begin{align*}
\int_0^\infty\int_0^{2\pi}|(c_{h_\epsilon,\epsilon}-c_\epsilon)(v,v_*,z,\varphi)|^2dzd\varphi\leq Ch_\epsilon|v-v_*|^{-2}.
\end{align*}
(iii) For any $v,v_*,\tilde v, \tilde v_*\in\mathbb R^3$,
\begin{align*}
\int_0^\infty\int_0^{2\pi}\Big|c_\epsilon(v,v_*,z&,\varphi)-c_\epsilon(\tilde v,\tilde v_*,z,\varphi+\varphi_0(v-v_*,\tilde v-\tilde v_*))\Big|^2dzd\varphi\\
&\leq C\min\big\{|v-v_*|^{-1}+|\tilde v-\tilde v_*|^{-1},\\
&\qquad\qquad(|v-\tilde v|^2+|v_*-\tilde v_*|^2)(|v-v_*|^{-3}+|\tilde v-\tilde v_*|^{-3})\\
&\qquad+\frac{1}{\log\frac{1}{\epsilon}}[|v-v_*|^{-2}+|\tilde v-\tilde v_*|^{-2}+|v-v_*|^{2}+|\tilde v-\tilde v_*|^{2}]\Big\}.
\end{align*}
\end{lemme}

\textbf{Proof.}
We easily get the first part of Point $(i)$ from (\ref{ccarre}). The fact that $k_\epsilon\leq2$ comes from (\ref{theta2}), just observing that $1-\cos\theta\leq\theta^2/2$.

We now prove $(ii)$. Recalling (\ref{a(v)}) and using that for any $X\in \mathbb R^3$, the vectors $X$ and $\Gamma(X,\varphi)$ are orthogonal, and $|\Gamma(X,\varphi)|=|X|$, we have
\begin{align*}
&|(c_{h_\epsilon,\epsilon}-c_\epsilon)(v,v_*,z,\varphi)|^2\\
&= \Big|\frac{1}{2}\Big[1-\cos G_\epsilon\Big(\frac{z}{\big(|v-v_*|+h_\epsilon\big)^{-3}}\Big)-\Big(1-\cos G_\epsilon\Big(\frac{z}{|v-v_*|^{-3}}\Big)\Big)\Big](v-v_*)\\
&\quad+\frac{1}{2}\Big[\sin G_\epsilon\Big(\frac{z}{\big(|v-v_*|+h_\epsilon\big)^{-3}}\Big)-\sin G_\epsilon\Big(\frac{z}{|v-v_*|^{-3}}\Big)\Big]\Gamma(v-v_*,\varphi)\Big|^2\\
&= \frac{1}{4}\Big[\cos G_\epsilon\Big(\frac{z}{\big(|v-v_*|+h_\epsilon\big)^{-3}}\Big)-\cos G_\epsilon\Big(\frac{z}{|v-v_*|^{-3}}\Big)\Big]^2|v-v_*|^2\\
&\quad+\frac{1}{4}\Big[\sin G_\epsilon\Big(\frac{z}{\big(|v-v_*|+h_\epsilon\big)^{-3}}\Big)-\sin G_\epsilon\Big(\frac{z}{|v-v_*|^{-3}}\Big)\Big]^2|v-v_*|^2\\
&\leq \frac{1}{2}\Big[G_\epsilon\Big(\frac{z}{\big(|v-v_*|+h_\epsilon\big)^{-3}}\Big)-G_\epsilon\Big(\frac{z}{|v-v_*|^{-3}}\Big)\Big]^2|v-v_*|^2,
\end{align*}
since $(\cos\theta-\cos\theta')^2+(\sin\theta-\sin\theta')^2\leq2(\theta-\theta')^2$, which gives
\begin{align*}
&\int_0^\infty\int_0^{2\pi}|(c_{h_\epsilon,\epsilon}-c_\epsilon)(v,v_*,z,\varphi)|^2dzd\varphi\\
&\leq\pi|v-v_*|^2\int_0^\infty\Big|G_\epsilon\Big(\frac{z}{\big(|v-v_*|+h_\epsilon\big)^{-3}}\Big)-G_\epsilon\Big(\frac{z}{|v-v_*|^{-3}}\Big)\Big|^2dz\\
&\leq\pi|v-v_*|^2\kappa_2\Big(\frac{\big((|v-v_*|+h_\epsilon)^{-3}-|v-v_*|^{-3}\big)^2}{(|v-v_*|+h_\epsilon)^{-3}+|v-v_*|^{-3}}\\
&\qquad\qquad\qquad\qquad+\frac{|v-v_*|^{-3}}{\log\frac{1}{\epsilon}}\log\frac{|v-v_*|^{-3}}{(|v-v_*|+h_\epsilon)^{-3}}\Big),
\end{align*}
by (\ref{A5}). Now using that for any $x,h>0$,
\begin{align*}
x^2\frac{\big((x+h)^{-3}-x^{-3}\big)^2}{(x+h)^{-3}+x^{-3}}&=x^2\frac{\big(x^3-(x+h)^3\big)^2}{(x+h)^6x^6}\frac{(x+h)^3x^3}{x^3+(x+h)^3}\\
&=\frac{h^2(3x^2+3xh+h^2)^2}{x(x+h)^3\big(x^3+(x+h)^3\big)}\\
&\leq C\frac{h}{x^2}\frac{h(x^4+x^2h^2+h^4)}{(x+h)^5}\\
&\leq C\frac{h}{x^2},
\end{align*}
and
\begin{align*}
x^{-1}\log\frac{x^{-3}}{(x+h)^{-3}}\leq3x^{-1}\log\Big(1+\frac{h}{x}\Big)\leq 3hx^{-2},
\end{align*}
we get
\begin{align*}
\int_0^\infty\int_0^{2\pi}|(c_{h_\epsilon,\epsilon}-c_\epsilon)(v,v_*,z,\varphi)|^2dzd\varphi&\leq C\Big(h_\epsilon+\frac{h_\epsilon}{\log\frac{1}{\epsilon}}\Big)|v-v_*|^{-2}\\
&\leq Ch_\epsilon|v-v_*|^{-2}.
\end{align*}
We finally prove $(iii)$. First, by Point $(i)$, we have
\begin{align*}
\int_0^\infty\int_0^{2\pi}\Big|c_\epsilon(v,v_*,z,\varphi)\Big|^2dzd\varphi&=\int_0^\infty\int_0^{2\pi}\Big|c_\epsilon(v,v_*,z,\varphi+\varphi_0(v-v_*,\tilde v-\tilde v_*))\Big|^2dzd\varphi\\
&=k_\epsilon|v-v_*|^{-1},
\end{align*}
with $k_\epsilon\leq2$ and we thus get the first bound in the $\min$. For the second bound, we set $\Delta:=\Big|c_\epsilon(v,v_*,z,\varphi)-c_\epsilon(\tilde v,\tilde v_*,z,\varphi+\varphi_0(v-v_*,\tilde v-\tilde v_*))\Big|^2$. Looking at the proof of Fournier-Gu\'erin \cite[Lemma 2.3]{FOUGUE1}, we get
\begin{align*}
\Delta\leq C\Big[&|(v-v_*)-(\tilde v-\tilde v_*)|^2\Big(G_\epsilon^2\Big(\frac{z}{|v-v_*|^{-3}}\Big)+G_\epsilon^2\Big(\frac{z}{|\tilde v-\tilde v_*|^{-3}}\Big)\Big)\\
&+\min(|v-v_*|^2,|\tilde v-\tilde v_*|^2)\Big|G_\epsilon\Big(\frac{z}{|v-v_*|^{-3}}\Big)-G_\epsilon\Big(\frac{z}{|\tilde v-\tilde v_*|^{-3}}\Big)\Big|^2\Big].
\end{align*}
Using the substitution $\theta=G_\epsilon(z/\Phi(|v-v_*|))$ or $\theta=G_\epsilon(z/\Phi(|\tilde v-\tilde v_*|))$, we have
\begin{align*}
&\int_0^\infty\int_0^{2\pi}\Big(G_\epsilon^2\Big(\frac{z}{|v-v_*|^{-3}}\Big)+G_\epsilon^2\Big(\frac{z}{|\tilde v-\tilde v_*|^{-3}}\Big)\Big)dzd\varphi\\
&=2\pi\int_0^\pi\theta^2\beta_\epsilon(\theta)d\theta\Big(|v-v_*|^{-3}+|\tilde v-\tilde v_*|^{-3}\Big)=8\Big(|v-v_*|^{-3}+|\tilde v-\tilde v_*|^{-3}\Big).
\end{align*}
We set $a=|v-v_*|$ and $b=|\tilde v-\tilde v_*|$. Using (\ref{A5}) (observe that $\log\frac{\max(x,y)}{\min(x,y)}\leq|\log x|+|\log y|$), we get
\begin{align*}
&\min(a^2,b^2)\int_0^\infty\int_0^{2\pi}\Big|G_\epsilon\Big(\frac{z}{a^{-3}}\Big)-G_\epsilon\Big(\frac{z}{b^{-3}}\Big)\Big|^2dzd\varphi\\
&\leq C\min(a,b)^2\Big[\frac{(a^{-3}-b^{-3})^2}{a^{-3}+b^{-3}}+\frac{1}{\log\frac{1}{\epsilon}}\max(a^{-3},b^{-3})[|\log a^{-3}|+|\log b^{-3}|]\Big]\\
&\leq C(a-b)^2(a^{-3}+b^{-3})+\frac{C}{\log\frac{1}{\epsilon}}\min(a,b)^{-1}[|\log a|+|\log b|]\\
&\leq C(a-b)^2(a^{-3}+b^{-3})+\frac{C}{\log\frac{1}{\epsilon}}[a^{-2}+b^{-2}+a^2+b^2],
\end{align*}
where we used that
\begin{align*}
\min(a,b)^2\frac{(a^{-3}-b^{-3})^2}{a^{-3}+b^{-3}}&\leq 9\min(a,b)^2(a-b)^2\frac{\min(a,b)^{-8}}{a^{-3}+b^{-3}}\\
&\leq 9(a-b)^2\min(a,b)^{-3}\leq 9(a-b)^2(a^{-3}+b^{-3}),
\end{align*}
and that (observe that $|\log a|\leq a^{-1}+a$)
\begin{align*}
\min(a,b)^{-1}[|\log a|+|\log b|]&\leq \frac{1}{2}[a^{-1}+b^{-1}]^2+\frac{1}{2}[|\log a|+|\log b|]^2\\
&\leq C(a^{-2}+b^{-2}+a^2+b^2).
\end{align*}
We thus have
\begin{align*}
\int_0^\infty\int_0^{2\pi}\Delta dzd\varphi\leq &C|(v-v_*)-(\tilde v-\tilde v_*)|^2\Big(|v-v_*|^{-3}-|\tilde v-\tilde v_*|^{-3}\Big)\\
&+\frac{C}{\log\frac{1}{\epsilon}}[|v-v_*|^{-2}+|\tilde v-\tilde v_*|^{-2}+|v-v_*|^{2}+|\tilde v-\tilde v_*|^{2}],
\end{align*}
which concludes the proof. \hfill $\square$

\subsection{Definition of the processes}
We consider a random variable $V_0$ with law $f_0$. We fix a white noise $W$ on $[0,T]\times[0,1]$ with covariance measure $dsd\alpha$ and we consider a process $(Y_t)_{t\in[0,T]}$ and an $\alpha$-process $(\tilde Y_t)_{t\in[0,T]}$ such that for any $t\in[0,T]$, $\mathcal L(Y_t)=\mathcal L_\alpha(\tilde Y_t)=g_t$, such that $\mathcal L_\alpha\big((\tilde Y_t)_{t\in[0,T]}\big)=\mathcal L\big((Y_t)_{t\in[0,T]}\big)$ and such that (\ref{EDSlandau}) is satisfied with $\gamma=-3$ (see Proposition \ref{probalandau}). 
For any $t\in[0,T]$, we consider an $\alpha$-random variable $\tilde V_t^\epsilon$ with law $f_t^{\epsilon}$ such that $\mathcal W_2^2(f_t^{\epsilon},g_t)=\mathbb E_\alpha[|\tilde V_t^\epsilon-\tilde Y_t|^2]$ and we consider the solution $(V_t^\epsilon)_{t\in[0,T]}$ to (\ref{Vt}) (with $\Phi(|v-v_*|)=(|v-v_*|+h_\epsilon)^{-3}$) for some $(\mathcal F_t)_{t\in[0,T]}$-Poisson measure $N$ as in Proposition \ref{probaboltz}. We will precise later the dependence of $N$ with the white noise $W$. We recall the equations satisfied by $(V_t^\epsilon)_{t\in[0,T]}$ and $(Y_t)_{t\in[0,T]}$, and we introduce some intermediate processes (here $n\in\mathbb N^*$ is fixed) 
\begin{align}
\nonumber
V_t^\epsilon=&V_0+\int_0^t\int_0^\infty\int_0^{2\pi}\int_0^1 c_{h_\epsilon,\epsilon}(V_{s-}^\epsilon,\tilde V_s^\epsilon(\alpha),z,\varphi)\tilde N(ds,dz,d\varphi,d\alpha)\\
\nonumber
&-k_\epsilon\int_0^t\int_0^1\Big(|V_s^\epsilon-\tilde V_s^\epsilon(\alpha)|+h_\epsilon\Big)^{-3}\big(V_s^\epsilon-\tilde V_s^\epsilon(\alpha)\big)dsd\alpha,\\
\nonumber
W_t^\epsilon=&V_0+\int_0^t\int_0^\infty\int_0^{2\pi}\int_0^1 c_\epsilon(V_{s-}^\epsilon,\tilde V_s^\epsilon(\alpha),z,\varphi)\tilde N(ds,dz,d\varphi,d\alpha)\\
\nonumber
&-k_\epsilon\int_0^t\int_0^1|V_s^\epsilon-\tilde V_s^\epsilon(\alpha)|^{-3}\big(V_s^\epsilon-\tilde V_s^\epsilon(\alpha)\big)dsd\alpha,\\
\nonumber
V_t^{n,\epsilon}=&V_0+\int_{0}^t\int_0^\infty\int_0^{2\pi}\int_0^1 c_\epsilon(Y_{\rho_n(s)},\tilde Y_{\rho_n(s)}(\alpha),z,\varphi+\Phi_n(s,\alpha))\tilde N(ds,dz,d\varphi,d\alpha)\\
\nonumber
&-k_\epsilon\int_0^t\int_0^1|V_s^\epsilon-\tilde V_s^\epsilon(\alpha)|^{-3}\big(V_s^\epsilon-\tilde V_s^\epsilon(\alpha)\big)dsd\alpha,\\
\nonumber
I_t^{n,\epsilon}=&V_0+\int_{0}^t\int_0^\infty\int_0^{2\pi}\int_0^1 d_\epsilon(Y_{\rho_n(s)},\tilde Y_{\rho_n(s)}(\alpha),z,\varphi+\Phi_n(s,\alpha))\tilde N(ds,dz,d\varphi,d\alpha)\\
\nonumber
&+\int_0^t\int_0^1b\big(V_s^\epsilon-\tilde V_s^\epsilon(\alpha)\big)dsd\alpha,\\
\nonumber
J_t^{n,\epsilon}=&V_0+\int_{0}^t\int_0^1 \sigma\big(Y_{\rho_n(s)}-\tilde Y_{\rho_n(s)}(\alpha)\big)W(ds,d\alpha)\\
\nonumber
&+\int_0^t\int_0^1b\big(V_s^\epsilon-\tilde V_s^\epsilon(\alpha)\big)dsd\alpha,\\
\nonumber
Y_t=&V_0+\int_0^t\int_0^1\sigma\big(Y_s-\tilde Y_s(\alpha)\big)W(ds,d\alpha)+\int_0^t\int_0^1b\big(Y_s-\tilde Y_s(\alpha)\big)dsd\alpha,
\end{align}
where (recall Lemma \ref{tanaka})
\begin{align}
\Phi_n(s,\alpha)&=\varphi_0(V_s^\epsilon- \tilde V_s^\epsilon(\alpha),Y_{\rho_n(s)}-\tilde Y_{\rho_n(s)}(\alpha)),
\end{align}
and $d_\epsilon$ is defined by replacing $\gamma$ by $-3$ and $G$ by $G_\epsilon$ in (\ref{dboltz}). Recall that $b(v)=-2|v|^{-3}v$ and that $k_\epsilon$ is defined in (\ref{keps}). Finally, $\rho_n$ is defined as follows.

We consider some subdivision $0=a_0^{n}<...<a_{\lfloor 2nT\rfloor-1}^{n}<a_{\lfloor 2nT\rfloor}^{n}=T$ of $[0,T]$ such that $1/4n<a_{i+1}^{n}-a_i^{n}<1/n$. In order to lighten notation, we write $a_i=a_i^{n}$. For $s\in[0,T]$, we set 
\begin{align*}
\rho_n(s)=\sum_{i=0}^{\lfloor 2nT\rfloor-1}a_i\one_{s\in[a_i,a_{i+1})}.
\end{align*}
Observe that by construction, we have $\sup_{[0,T]}|s-\rho_n(s)|\leq1/n$.

\subsection{The proof.}
The ideas will be the same as for Theorem \ref{dist}. The proofs will thus be very similar to those used for Lemmas \ref{prelim}, \ref{lem1}, \ref{lem2}, \ref{lem3} and \ref{lem4}. So instead of rewriting all the proofs, we will only point out the modifications that we have to handle.

We start by a lemma where we compute the error due to the parameter $h_\epsilon$ in the collision kernel. Observe that after this lemma, we will use a collision kernel which corresponds to the real Coulomb case (without the parameter $h_\epsilon$) for our computations. Furthermore, the errors that we will get after this lemma will not depend on $h_\epsilon$. This confirms the fact that the parameter $h_\epsilon$ is not useful to get a rate of convergence for the grazing collisions limit for the Coulomb potential. Here again, we recall that we only introduce this parameter in order to get easily existence and uniqueness of $(f_t^\epsilon)_{t\in[0,T]}$ (and of the process $(V_t^\epsilon)_{t\in[0,T]}$). 
\begin{lemme} \label{lemh}
There exists a constant $C$ depending on $\sup_{[0,T]}||f_s^{\epsilon}||_\infty$ such that for any $t\in[0,T]$, any $\epsilon\in(0,1)$,
\begin{align*}
\mathbb E[|V_t^\epsilon-W_t^\epsilon|^2]\leq Ch_\epsilon^{e^{-C}}.
\end{align*}
\end{lemme}

\textbf{Proof.}
We have
\begin{align*}
V_t^\epsilon-W_t^\epsilon=&\int_{0}^t\int_0^\infty\int_0^{2\pi}\int_0^1 (c_{h_\epsilon,\epsilon}-c_\epsilon)\Big(V_s^\epsilon,\tilde V_s^\epsilon(\alpha),z,\varphi\Big)\tilde N(ds,dz,d\varphi,d\alpha)\\
&-k_\epsilon\int_0^t\int_0^1\Big(\big(|V_s^\epsilon-\tilde V_s^\epsilon(\alpha)|+h_\epsilon\big)^{-3}\\
&\qquad\qquad\qquad-|V_s^\epsilon-\tilde V_s^\epsilon(\alpha)|^{-3}\Big)\big(V_s^\epsilon-\tilde V_s^\epsilon(\alpha)\big)dsd\alpha.
\end{align*}
Using It\^o's formula and taking expectations, we thus get
\begin{align*}
\mathbb E[|V_t^\epsilon-W_t^\epsilon|^2]&=\int_{0}^t\int_0^\infty\int_0^{2\pi}\int_0^1 \mathbb E\Big[\Big|(c_{h_\epsilon,\epsilon}-c_\epsilon)\Big(V_s^\epsilon,\tilde V_s^\epsilon(\alpha),z,\varphi\Big)\Big|^2\Big]dsdzd\varphi d\alpha\\
&\quad-2k_\epsilon\int_0^t\int_0^1\mathbb E\Big[\Big(\big(|V_s^\epsilon-\tilde V_s^\epsilon(\alpha)|+h_\epsilon\big)^{-3}-|V_s^\epsilon-\tilde V_s^\epsilon(\alpha)|^{-3}\Big)\\
&\qquad\qquad\qquad\qquad\qquad\big(V_s^\epsilon-\tilde V_s^\epsilon(\alpha)\big).\big(V_s^\epsilon-W_s^\epsilon\big)\Big]dsd\alpha\\
&=:A+B.
\end{align*}
Using Point $(ii)$ of Lemma \ref{ccoulomb}, we get
\begin{align*}
A&\leq Ch_\epsilon\int_{0}^t\mathbb E\Big[\mathbb E_\alpha[|V_s^\epsilon-\tilde V_s^\epsilon(\alpha)|^{-2}]\Big]ds\leq Ch_\epsilon\int_{0}^t(1+||f_s||_\infty)ds,
\end{align*}
by Lemma \ref{r3v3}. For $B$, we first observe that for any $x,h,y>0$
\begin{align*}
\big(x^{-3}-(x+h)^{-3}\big)xy&\leq \one_{y\geq1}x^{-3}xy+\one_{y\leq x} \big(x^{-3}-(x+h)^{-3})\big)x^2\\
&\quad+\one_{y^2\leq x\leq y<1}x^{-3}y^2+\one_{x<y^2<1}x^{-3}xy\\
&\leq \one_{y\geq1}x^{-2}y^2+3\one_{y\leq x}hx^{-2}\\
&\quad+\one_{y^2\leq x\leq y<1}x^{-3}y^2+\one_{x<y^2<1}x^{-2}.
\end{align*}
We thus get (recall that $k_\epsilon\leq2$)
\begin{align*}
B&\leq C\int_{0}^t\mathbb E\Big[\mathbb E_\alpha[\one_{|V_s^\epsilon-W_s^\epsilon|\geq1}|V_s^\epsilon-W_s^\epsilon|^2|V_s^\epsilon-\tilde V_s^\epsilon|^{-2}\\
&\qquad\qquad\quad+\one_{|V_s^\epsilon-W_s^\epsilon|\leq|V_s^\epsilon-\tilde V_s^\epsilon|}h_\epsilon|V_s^\epsilon-\tilde V_s^\epsilon|^{-2}\\
&\qquad\qquad\quad +\one_{|V_s^\epsilon-W_s^\epsilon|^2\leq|V_s^\epsilon-\tilde V_s^\epsilon|\leq|V_s^\epsilon-W_s^\epsilon|<1}|V_s^\epsilon-W_s^\epsilon|^2|V_s^\epsilon-\tilde V_s^\epsilon|^{-3}\\
&\qquad\qquad\quad +\one_{|V_s^\epsilon-\tilde V_s^\epsilon|<|V_s^\epsilon-W_s^\epsilon|^2<1}|V_s^\epsilon-\tilde V_s^\epsilon|^{-2}]\Big]ds.
\end{align*}
Using that $\mathcal L_\alpha(\tilde V_s^\epsilon)=f_s^\epsilon$ and Lemma \ref{r3v3}, we have
\begin{align*}
B&\leq C\int_{0}^t\mathbb E\Big[(1+||f_s^{\epsilon}||_\infty)|V_s^\epsilon-W_s^\epsilon|^2+(1+||f_s^{\epsilon}||_\infty) h_\epsilon\\
&\qquad\qquad\quad +\one_{|V_s^\epsilon-W_s^\epsilon|<1}|V_s^\epsilon-W_s^\epsilon|^2\big(1-||f_s^{\epsilon}||_\infty\log(|V_s^\epsilon-W_s^\epsilon|^2)\big)\\
&\qquad\qquad\quad +||f_s^{\epsilon}||_\infty|V_s^\epsilon-W_s^\epsilon|^2\Big]ds\\
&\leq C\int_{0}^t(1+||f_s^\epsilon||_\infty)\mathbb E\Big[|V_s^\epsilon-W_s^\epsilon|^2+h_\epsilon+\psi(|V_s^\epsilon-W_s^\epsilon|^2)\Big]ds,
\end{align*}
where $\psi$ was defined in (\ref{functionpsi}). Using that $x\leq\psi(x)$ for any $x\geq0$ and the approximate Jensen inequality (recall the paragraph just after (\ref{functionpsi})), we thus get
\begin{align*}
\mathbb E[|V_t^\epsilon-W_t^\epsilon|^2]\leq Ch_\epsilon+C\int_0^t\psi(\mathbb E[|V_s^\epsilon-W_s^\epsilon|^2])ds,
\end{align*}
where $C$ depends on $\sup_{[0,T]}||f_s^{\epsilon}||_\infty$. The conclusion follows by Lemma \ref{grongen}. \hfill$\square$

\begin{lemme} \label{cprelim}
(i) There exists a constant $C$ depending on $\sup_{s\in[0,T]} ||f_s^{\epsilon}||_\infty$ and on $m_2(f_0)$ such that for $0\leq t'\leq t\leq T$ with $t-t'<1$, for any $\epsilon\in(0,1)$,
\begin{align*}
\mathbb E\Big[|V_t^\epsilon-V_{t'}^\epsilon|^2\Big]\leq C(t-t').
\end{align*}
The same bound holds for $\mathbb E\Big[|Y_t-Y_{t'}|^2\Big]$ and $\mathbb E_\alpha\Big[|\tilde Y_t-\tilde Y_{t'}|^2\Big]$ with $C$ depending on $m_2(g_0)$ and on $\sup_{[0,T]} ||g_s||_\infty$.\\
(ii) For all $t\in[0,T]$, we have 
\begin{align*}
\mathbb E\Big[|V_t^\epsilon-V_{\rho_n(t)}^\epsilon|^2\Big]+\mathbb E\Big[|Y_t-Y_{\rho_n(t)}|^2\Big]+\mathbb E_\alpha\Big[|\tilde Y_t-\tilde Y_{\rho_n(t)}|^2\Big]\leq\frac{C}{n}.
\end{align*}
\end{lemme}

\textbf{Proof.} 
Since $t-\rho_n(t)\leq 1/n$, $(ii)$ immediately follows from $(i)$. To prove $(i)$ (for example for $(V_t^\epsilon)_{t\in[0,T]}$), we follow the line of the proof of Lemma \ref{prelim}, and we get (observe that $(a+h)^{-3}\leq a^{-3}$)
\begin{align*}
\mathbb E\Big[|V_t^\epsilon-V_{t'}^\epsilon|^2\Big]&\leq 2k_\epsilon\int_{t'}^t\mathbb E \Big[\mathbb E_\alpha[|V_s^\epsilon-\tilde V_s^\epsilon|^{-1}]\Big]ds\\
&\quad+2k_\epsilon^2\mathbb E \Big[\Big(\int_{t'}^t\mathbb E_\alpha[|V_s^\epsilon-\tilde V_s^\epsilon|^{-2}]ds\Big)^2\Big]\\
&\leq 2k_\epsilon\int_{t'}^t(1+||f_s^{\epsilon}||_\infty)ds+2k_\epsilon^2\Big(\int_{t'}^t(1+||f_s^{\epsilon}||_\infty)ds\Big)^2\\
&\leq C(t-t'),
\end{align*} 
by Lemma \ref{r3v3} and we conclude the proof as for Lemma \ref{prelim}. \hfill$\square$

\begin{lemme} \label{clem1}
There exists a constant $C$ depending on $\sup_{s\in[0,T]}||f_s^{\epsilon}+g_s||_\infty$ and on $m_2(f_0)$ such that, for any $n\geq2$, $\epsilon\in(0,1)$ and $t\in[0,T]$
\begin{align*}
\mathbb E[|W_t^\epsilon-V_t^{n,\epsilon}|^2]\leq &\frac{C}{\log\frac{1}{\epsilon}}+C\int_{0}^t\Big(\frac{\log n}{n}+\psi(\mathbb E[|V_s^\epsilon-Y_s|^2])+\psi(\mathbb E_\alpha[|\tilde V_s^\epsilon-\tilde Y_s|^2])\Big)ds.
\end{align*}
\end{lemme}

\textbf{Proof.}
Observing that
\begin{align*}
W_t^\epsilon-V_t^{n,\epsilon}=\int_0^t\int_0^\infty&\int_0^{2\pi}\int_0^1 \Big[c_\epsilon(V_{s-}^\epsilon,\tilde V_s^\epsilon(\alpha),z,\varphi)\\
&-c_\epsilon\Big(Y_{\rho_n(s)},\tilde Y_{\rho_n(s)}(\alpha),z,\varphi+\Phi_n(s,\alpha)\Big)\Big]\tilde N(ds,dz,d\varphi,d\alpha),
\end{align*}
we have
\begin{align*}
I:=\mathbb E\Big[|W_t^\epsilon-V_t^{n,\epsilon}|^2\Big]=\int_{0}^t&\int_0^\infty\int_0^{2\pi}\int_0^1 \mathbb E\Big[\Big|c_\epsilon(V_s^\epsilon,\tilde V_s^\epsilon(\alpha),z,\varphi)\\
&-c_\epsilon\Big(Y_{\rho_n(s)},\tilde Y_{\rho_n(s)}(\alpha),z,\varphi+\Phi_n(s,\alpha)\Big)\Big|^2\Big]dsdzd\varphi d\alpha.
\end{align*}
We set
\begin{align*}
\delta=\int_0^\infty\int_0^{2\pi}\Big|c_\epsilon(V_s^\epsilon,\tilde V_s^\epsilon(\alpha),z,\varphi)-c_\epsilon\Big(Y_{\rho_n(s)},\tilde Y_{\rho_n(s)}(\alpha),z,\varphi+\Phi_n(s,\alpha)\Big)\Big|^2dzd\varphi.
\end{align*}
Setting $a_s:=|V_s^\epsilon-Y_{\rho_n(s)}|+|\tilde V_s^\epsilon(\alpha)-\tilde Y_{\rho_n(s)}(\alpha)|$, $v_s:=|V_s^\epsilon-\tilde V_s^\epsilon(\alpha)|$, $y_s:=|Y_{\rho_n(s)}-\tilde Y_{\rho_n(s)}(\alpha)|$  and using Lemma \ref{ccoulomb}, we get
\begin{align*}
\delta&\leq C\one_{a_s\geq1}\Big(v_s^{-1}+y_s^{-1}\Big)\\
&\quad+C\one_{a_s\leq1} \one_{v_s\geq|V_s^\epsilon-Y_{\rho_n(s)}|^2,y_s\geq|V_s^\epsilon-Y_{\rho_n(s)}|^2}\Big[|V_s^\epsilon-Y_{\rho_n(s)}|^2\Big(v_s^{-3}+y_s^{-3}\Big)\\
&\qquad\qquad\qquad\qquad\qquad\qquad\qquad\qquad\qquad\qquad+\frac{1}{\log\frac{1}{\epsilon}}[v_s^{-2}+y_s^{-2}+v_s^2+y_s^2]\Big]\\
&\quad+C\one_{a_s\leq1} \one_{v_s\geq|\tilde V_s^\epsilon(\alpha)-\tilde Y_{\rho_n(s)}(\alpha)|^2,y_s\geq|\tilde V_s^\epsilon(\alpha)-\tilde Y_{\rho_n(s)}(\alpha)|^2}\\
&\qquad\Big[|\tilde V_s^\epsilon(\alpha)-\tilde Y_{\rho_n(s)}(\alpha)|^2\Big(v_s^{-3}+y_s^{-3}\Big)+\frac{1}{\log\frac{1}{\epsilon}}[v_s^{-2}+y_s^{-2}+v_s^2+y_s^2]\Big]\\
&\quad+C\one_{a_s\leq1}\one_{v_s\leq|V_s^\epsilon-Y_{\rho_n(s)}|^2}\Big(v_s^{-1}+y_s^{-1}\Big)\\
&\quad+C\one_{a_s\leq1}\one_{y_s\leq|V_s^\epsilon-Y_{\rho_n(s)}|^2}\Big(v_s^{-1}+y_s^{-1}\Big)\\
&\quad+C\one_{a_s\leq1}\one_{v_s\leq|\tilde V_s^\epsilon(\alpha)-\tilde Y_{\rho_n(s)}(\alpha)|^2}\Big(v_s^{-1}+y_s^{-1}\Big)\\
&\quad+C\one_{a_s\leq1}\one_{y_s\leq|\tilde V_s^\epsilon(\alpha)-\tilde Y_{\rho_n(s)}(\alpha)|^2}\Big(v_s^{-1}+y_s^{-1}\Big)\\
&=:C\sum_{i=1}^7\delta_i.
\end{align*}
We thus have $I\leq\sum_{i=1}^7I_i$ where $I_i=\int_{0}^t\int_0^1\mathbb E[\delta_i]dsd\alpha$. Using that $\one_{a_s\geq1}\leq a_s^2$, we have
\begin{align*}
I_1&\leq\int_{0}^t\int_0^1\mathbb E\Big[\Big(|V_s^\epsilon-Y_{\rho_n(s)}|+|\tilde V_s^\epsilon(\alpha)-\tilde Y_{\rho_n(s)}(\alpha)|\Big)^2\Big(|V_s^\epsilon-\tilde V_s^\epsilon(\alpha)|^{-1}\\
&\qquad\qquad\qquad\qquad\qquad\qquad\qquad\qquad\qquad\qquad+|Y_{\rho_n(s)}-\tilde Y_{\rho_n(s)}(\alpha)|^{-1}\Big)\Big]dsd\alpha\\
&\leq2\int_{0}^t\mathbb E\Bigg[|V_s^\epsilon-Y_{\rho_n(s)}|^2\mathbb E_\alpha\Big[|V_s^\epsilon-\tilde V_s^\epsilon|^{-1}+|Y_{\rho_n(s)}-\tilde Y_{\rho_n(s)}|^{-1}\Big]\Bigg]ds\\
&\quad+2\int_{0}^t\mathbb E_\alpha\Bigg[|\tilde V_s^\epsilon-\tilde Y_{\rho_n(s)}|^2\mathbb E\Big[|V_s^\epsilon-\tilde V_s^\epsilon|^{-1}+|Y_{\rho_n(s)}-\tilde Y_{\rho_n(s)}|^{-1}\Big]\Bigg]ds\\
&\leq C\int_{0}^t\Big(\mathbb E\Big[|V_s^\epsilon-Y_{\rho_n(s)}|^2\Big]+\mathbb E_\alpha\Big[|\tilde V_s^\epsilon-\tilde Y_{\rho_n(s)}|^2\Big]\Big)\Big(1+||f_s||_\infty+||g_{\rho_n(s)}||_\infty\Big)ds,
\end{align*}
by Lemma \ref{r3v3}. We thus get, using the triangular inequality and Lemma \ref{cprelim},
\begin{align} \label{I1}
I_1\leq C\int_{0}^t\Big(\frac{1}{n}+\mathbb E\Big[|V_s^\epsilon-Y_s|^2\Big]+\mathbb E_\alpha\Big[|\tilde V_s^\epsilon-\tilde Y_s|^2\Big]\Big)ds,
\end{align}
where $C$ depends on $\sup_{s\in[0,T]}||f_s^{\epsilon}+g_s||_\infty$. Using Lemma \ref{r3v3}, we have
\begin{align*}
I_2 &\leq\int_{0}^t\mathbb E\Big[|V_s^\epsilon-Y_{\rho_n(s)}|^2\one_{|V_s^\epsilon-Y_{\rho_n(s)}|\leq1}\\
&\qquad\qquad\qquad\qquad\qquad\Big(\mathbb E_\alpha[|V_s^\epsilon-\tilde V_s^\epsilon|^{-3}\one_{|V_s^\epsilon-\tilde V_s^\epsilon|\geq|V_s^\epsilon-Y_{\rho_n(s)}|^2}\\
&\qquad\qquad\qquad\qquad\qquad+|Y_{\rho_n(s)}-\tilde Y_{\rho_n(s)}|^{-3}\one_{|Y_{\rho_n(s)}-\tilde Y_{\rho_n(s)}|\geq|V_s^\epsilon-Y_{\rho_n(s)}|^2}]\Big)\Big]ds\\
&\quad+\frac{C}{\log\frac{1}{\epsilon}}\int_0^t\mathbb E\Big[\mathbb E_\alpha[|V_s^\epsilon-\tilde V_s^\epsilon|^{-2}+|Y_{\rho_n(s)}-\tilde Y_{\rho_n(s)}|^{-2}\\
&\qquad\qquad\qquad\qquad\qquad+|V_s^\epsilon-\tilde V_s^\epsilon|^{2}+|Y_{\rho_n(s)}-\tilde Y_{\rho_n(s)}|^{2}]\Big]ds\\
&\leq\int_{0}^t\mathbb E\Big[|V_s^\epsilon-Y_{\rho_n(s)}|^2\one_{|V_s^\epsilon-Y_{\rho_n(s)}|\leq1}\Big(1-\\
&\qquad\qquad\qquad\qquad\qquad\qquad\qquad\-C||f_s^\epsilon+g_{\rho_n(s)}||_\infty\log|V_s^\epsilon-Y_{\rho_n(s)}|^2\Big)\Big]ds\\
&\quad+\frac{C}{\log\frac{1}{\epsilon}}\int_0^t(1+||f_s^\epsilon+g_{\rho_n(s)}||_\infty+m_2(f_0))ds.
\end{align*}
Recalling (\ref{functionpsi}) and using the (approximate) Jensen inequality for the function $\psi$, the fact that $\psi(a+b)\leq\psi(a)+ \psi(b)$ and that the function $\psi$ is increasing, and the Lemma \ref{cprelim} we get
\begin{align}
\label{I2}
I_2&\leq C \int_{0}^t\mathbb E\Big[\psi(|V_s^\epsilon-Y_{\rho_n(s)}|^2)\Big]ds+\frac{C}{\log\frac{1}{\epsilon}}\\
\nonumber
&\leq C\int_{0}^t\Big(\psi(\frac{C}{n})+\psi(\mathbb E[|V_s^\epsilon-Y_s|^2])\Big)ds+\frac{C}{\log\frac{1}{\epsilon}}\\
\nonumber
&\leq C\int_{0}^t\Big(\frac{\log n}{n}+\psi(\mathbb E[|V_s^\epsilon-Y_s|^2])\Big)ds+\frac{C}{\log\frac{1}{\epsilon}},
\end{align}
where $C$ depends on $\sup_{s\in[0,T]}||f_s^{\epsilon}+g_s||_\infty$. Using the same ideas, we also have
\begin{align}
\label{I3}
I_3\leq C\int_{0}^t\Big(\frac{\log n}{n}+\psi(\mathbb E_\alpha[|\tilde V_s^\epsilon-\tilde Y_s|^2])\Big)ds+\frac{C}{\log\frac{1}{\epsilon}}.
\end{align}
We now deal with $I_4$.
\begin{align*}
I_4\leq \int_{0}^t\mathbb E\Big[\mathbb E_\alpha[\one_{a_s\leq1}\one_{[V_s^\epsilon-\tilde V_s^\epsilon|\leq|V_s^\epsilon-Y_{\rho_n(s)}|^2}(&|V_s^\epsilon-\tilde V_s^\epsilon|^{-1}\\
&+|Y_{\rho_n(s)}-\tilde Y_{\rho_n(s)}|^{-1})]\Big]ds.
\end{align*}
Using Lemma \ref{r3v3}, we first observe that
\begin{align*}
\mathbb E_\alpha\Big[\one_{a_s\leq1}\one_{[V_s^\epsilon-\tilde V_s^\epsilon|\leq|V_s^\epsilon-Y_{\rho_n(s)}|^2}&|V_s^\epsilon-\tilde V_s^\epsilon|^{-1}\Big]\\
&\leq C\one_{|V_s^\epsilon-Y_{\rho_n(s)}|\leq1}||f_s||_\infty|V_s^\epsilon-Y_{\rho_n(s)}|^4\\
&\leq C|V_s^\epsilon-Y_{\rho_n(s)}|^2.
\end{align*}
Next, using the H\"older inequality with $p=3$ and $q=3/2$, and then Lemma \ref{r3v3}, we get
\begin{align*}
\mathbb E_\alpha\Big[&\one_{[V_s^\epsilon-\tilde V_s^\epsilon|\leq|V_s^\epsilon-Y_{\rho_n(s)}|^2}|Y_{\rho_n(s)}-\tilde Y_{\rho_n(s)}|^{-1}\Big]\\
&\leq \mathbb E_\alpha\Big[\one_{[V_s^\epsilon-\tilde V_s^\epsilon|\leq|V_s^\epsilon-Y_{\rho_n(s)}|^2}\Big]^{\frac{1}{3}}\mathbb E_\alpha\Big[|Y_{\rho_n(s)}-\tilde Y_{\rho_n(s)}|^{\frac{-3}{2}}\Big]^{\frac{2}{3}}\\
&\leq \Big(C||f_s^{\epsilon}||_\infty|V_s^\epsilon-Y_{\rho_n(s)}|^6\Big)^{\frac{1}{3}}\Big(1+C||g_{\rho_n(s)}||_\infty\Big)^{\frac{2}{3}}\\
&\leq C(1+||f_s^{\epsilon}+g_{\rho_n(s)}||_\infty)|V_s^\epsilon-Y_{\rho_n(s)}|^2.
\end{align*}
We thus have
\begin{align}
\label{I4}
I_4&\leq C\int_{0}^t \mathbb E[|V_s^\epsilon-Y_{\rho_n(s)}|^2]ds\leq C\int_{0}^t \Big(\frac{1}{n}+\mathbb E[|V_s^\epsilon-Y_s|^2]\Big)ds
\end{align}
by Lemma \ref{cprelim}. With the same arguments, 
\begin{align}
\label{I5}
I_5\leq C\int_{0}^t \Big(\frac{1}{n}+\mathbb E[|V_s^\epsilon-Y_s|^2]\Big)ds,
\end{align}
and
\begin{align}
\label{I67}
I_6+I_7\leq C\int_{0}^t\Big(\frac{1}{n}+\mathbb E_\alpha[|\tilde V_s^\epsilon-\tilde Y_s|^2]\Big)ds.
\end{align}
It suffices to use (\ref{I1}), (\ref{I2}), (\ref{I3}), (\ref{I4}), (\ref{I5}), (\ref{I67}) and to observe that $x\leq \psi(x)$ for any $x\geq0$ to conclude the proof. \hfill $\square$

\begin{lemme} \label{clem2}
There exists a constant $C$ depending on $\sup_{s\in[0,T]}||f_s^{\epsilon}+g_s||_\infty$ and on $T$ such that for any $n\geq2$, $\epsilon\in(0,1)$ and $t\in[0,T]$,
\begin{align*}
\mathbb E[|V_t^{n,\epsilon}-I_t^{n,\epsilon}|^2]\leq C\int_0^\pi\theta^4\beta_\epsilon(\theta)d\theta.
\end{align*}
\end{lemme}

\textbf{Proof.}
As in the proof of Lemma \ref{lem2}, we have
\begin{align*}
\mathbb E[|V_t^{n,\epsilon}-I_t^{n,\epsilon}|^2]&\leq  C\int_0^\pi\theta^4\beta_\epsilon(\theta)d\theta\Big(\int_{0}^t \mathbb E\Big[\mathbb E_\alpha[|Y_{\rho_n(s)}-\tilde Y_{\rho_n(s)}|^{-1}] \Big]ds\\
&\qquad\qquad\qquad\qquad+\mathbb E\Big[\Big(\int_0^t\mathbb E_\alpha[|V_s^\epsilon-\tilde V_s^\epsilon|^{-2}]ds\Big)^2\Big]\Big)\\
&\leq C\int_0^\pi\theta^4\beta_\epsilon(\theta)d\theta\Big(\int_{0}^t(1+||g_{\rho_n(s)}||_\infty)ds\\
&\qquad\qquad\qquad\qquad+\mathbb E\Big[\Big(\int_0^t(1+||f_s^{\epsilon}||_\infty)ds\Big)^2\Big]\Big)\\
&\leq C\int_0^\pi\theta^4\beta_\epsilon(\theta)d\theta,
\end{align*}
by Lemma \ref{r3v3}. \hfill$\square$
\vskip0.5cm

The following lemma states as follows.
\begin{lemme} \label{clem3}
Assume that $m_{p+2}(f_0)<\infty$ for some $p\geq5$. We can couple the Poisson measure $N$ and the white noise $W$ in such a way that there exists a constant $C$ depending on $T$, $m_{p+2}(f_0)$, $H(f_0)$, and $\sup_{s\in[0,T]}||f_s^{\epsilon}+g_s||_\infty$ such that for any $M>\sqrt{2m_2(f_0)}$, $\eta\in[\epsilon,\pi]$, $n\geq2$ and $t\in[0,T]$, 
\begin{align*}
\mathbb E[|I_t^{n,\epsilon}-J_t^{n,\epsilon}|^2]\leq C\Big[\eta^2M^2n\Big(\log^2(r_\eta)+\log^2(n\eta^2)+M\Big)+\int_\eta^\pi\theta^2\beta_\epsilon(\theta)d\theta+\frac{1}{M^p}\Big],
\end{align*}
where
\begin{align*}
r_\eta=\frac{\pi}{4}\int_0^\eta\theta^2\beta_\epsilon(\theta)d\theta.
\end{align*}
\end{lemme}

\textbf{Proof.}
It suffices to follow the line of the proof of Lemma \ref{lem3}, recalling that
\begin{align*}
\mathbb E[|Y_{\rho_n(s)}-\tilde Y_{\rho_n(s)}|^{-1}]\leq (1+C||g_{\rho_n(s)}||_\infty)\leq C,
\end{align*}
by Lemma \ref{r3v3}. \hfill$\square$
\vskip0.5cm 

We now give the last lemma needed to prove Theorem \ref{colrascoul}.
\begin{lemme} \label{clem4}
There exists a constant $C$ depending on $\sup_{s\in[0,T]}||f_s^{\epsilon}+g_s||_\infty$ and on $T$ such that for any $n\geq2$, $\epsilon\in(0,1)$ and $t\in[0,T]$,
\begin{align*}
\mathbb E[|J_t^{n,\epsilon}-Y_t|^2]\leq C\frac{\log n}{n}+C\int_0^t \Big(\psi(\mathbb E[|V_s^\epsilon-Y_s|^2])&+\psi(\mathbb E_\alpha[|\tilde V_s^\epsilon-\tilde Y_s|^2])\\
&+\psi(\mathbb E[|J_s^{n,\epsilon}-Y_s|^2])\Big)ds.
\end{align*}
\end{lemme}

\textbf{Proof.}
The It\^o formula gives
\begin{align*}
\mathbb E[|J_t^{n,\epsilon}-Y_t|^2]&=\int_{0}^t\int_0^1\mathbb E\Big[|\sigma\big(Y_{\rho_n(s)}-\tilde Y_{\rho_n(s)}(\alpha)\big)-\sigma\big(Y_s-\tilde Y_s(\alpha)\big)|^2\Big]dsd\alpha\\
&\quad+\int_0^t\int_0^1\mathbb E\Big[\Big(b\big(V_s^\epsilon-\tilde V_s^\epsilon(\alpha)\big)-b\big(Y_s-\tilde Y_s(\alpha)\big)\Big).(J_s^{n,\epsilon}-Y_s)\Big]dsd\alpha\\
&=:A+B.
\end{align*}
Using Fournier \cite[Lemma 6]{FOU2}, we get
\begin{align*}
A&\leq 2\int_{0}^t\mathbb E_\alpha\Big[\mathbb E[|\sigma(Y_{\rho_n(s)}-\tilde Y_{\rho_n(s)})-\sigma(Y_{\rho_n(s)}-\tilde Y_s)|^2]\Big]ds\\
&\quad+2\int_{0}^t\mathbb E\Big[\mathbb E_\alpha[|\sigma(Y_{\rho_n(s)}-\tilde Y_s)-\sigma(Y_s-\tilde Y_s)|^2]\Big]ds\\
&\leq C\int_{0}^t(1+||g_{\rho_n(s)}||_\infty)\mathbb E_\alpha\Big[\psi(|\tilde Y_{\rho_n(s)}-\tilde Y_s|^2)\Big]ds\\
&\quad+C\int_{0}^t(1+||g_s||_\infty)\mathbb E\Big[\psi(|Y_{\rho_n(s)}-Y_s|^2)\Big]ds\\
&\leq C\int_{0}^t\psi(C/n)ds\\
&\leq C\frac{\log n}{n},
\end{align*}
where we used the (approximate) Jensen inequality for $\psi$, Lemma \ref{cprelim} and the fact that $\psi$ is increasing (recall (\ref{functionpsi})). For $B$, we first set $R=|\Big(b\big(V_s^\epsilon-\tilde V_s^\epsilon(\alpha)\big)-b\big(Y_s-\tilde Y_s(\alpha)\big)\Big).(J_s^{n,\epsilon}-Y_s)|$ and $E_s=\{|J_s^{n,\epsilon}-Y_s|\geq |V_s^\epsilon-Y_s|+|\tilde V_s^\epsilon(\alpha)-\tilde Y_s(\alpha)|\}$. Using \cite[Lemma 3]{FOU2}, we have
\begin{align*}
R&\leq \one_{E_s^c}(|V_s^\epsilon-Y_s|+|\tilde V_s^\epsilon(\alpha)-\tilde Y_s(\alpha)|)|b\big(V_s^\epsilon-\tilde V_s^\epsilon(\alpha)\big)-b\big(Y_s-\tilde Y_s(\alpha)\big)|\\
&\quad+ C\one_{E_s}\one_{|J_s^{n,\epsilon}-Y_s|\geq1}|J_s^{n,\epsilon}-Y_s|^2(|V_s^\epsilon-\tilde V_s^\epsilon(\alpha)|^{-2}+|Y_s-\tilde Y_s(\alpha)|^{-2})\\
&\quad+C\one_{E_s}\one_{|J_s^{n,\epsilon}-Y_s|\leq1}\one_{|V_s^\epsilon-\tilde V_s^\epsilon(\alpha)|>|J_s^{n,\epsilon}-Y_s|^4}\one_{|Y_s-\tilde Y_s(\alpha)|>|J_s^{n,\epsilon}-Y_s|^4}|J_s^{n,\epsilon}-Y_s|^2\\
&\qquad\qquad(|V_s^\epsilon-\tilde V_s^\epsilon(\alpha)|^{-3}+|Y_s-\tilde Y_s(\alpha)|^{-3})\\
&\quad+C\one_{E_s}\one_{|J_s^{n,\epsilon}-Y_s|\leq1}\one_{|V_s^\epsilon-\tilde V_s^\epsilon(\alpha)|\leq|J_s^{n,\epsilon}-Y_s|^4}(|V_s^\epsilon-\tilde V_s^\epsilon(\alpha)|^{-2}\\
&\qquad\qquad\qquad\qquad\qquad\qquad\qquad\qquad\qquad\qquad\qquad+|Y_s-\tilde Y_s(\alpha)|^{-2})\\
&\quad+C\one_{E_s}\one_{|J_s^{n,\epsilon}-Y_s|\leq1}\one_{|Y_s-\tilde Y_s(\alpha)|\leq|J_s^{n,\epsilon}-Y_s|^4}(|V_s^\epsilon-\tilde V_s^\epsilon(\alpha)|^{-2}+|Y_s-\tilde Y_s(\alpha)|^{-2})\\
&=:\sum_{i=1}^5R_i.
\end{align*}
We thus have $B\leq\sum_{i=1}^5B_i$ where $B_i:=\int_0^t\mathbb E\big[\mathbb E_\alpha[R_i]\big]ds$. Using \cite[Lemma 7]{FOU2}, we get
\begin{align*}
B_1\leq C\int_0^t(1+||f_s^{\epsilon}+g_s||_\infty)\Big(\psi(\mathbb E[|V_s^\epsilon-Y_s|^2])+\psi(\mathbb E_\alpha[|\tilde V_s^\epsilon-\tilde Y_s|^2])\Big)ds.
\end{align*}
For $B_2$, we easily get by Lemma \ref{r3v3},
\begin{align*}
B_2\leq C\int_0^t\mathbb E[|J_s^{n,\epsilon}-Y_s|^2](1+||f_s^{\epsilon}||_\infty+||g_s||_\infty)ds.
\end{align*}
Using Lemma \ref{r3v3}, we have 
\begin{align*}
B_3&\leq C\int_0^t\mathbb E\Big[|J_s^{n,\epsilon}-Y_s|^2\one_{|J_s^{n,\epsilon}-Y_s|<1}\mathbb E_\alpha[\one_{|V_s^\epsilon-\tilde V_s^\epsilon|>|J_s^{n,\epsilon}-Y_s|^4}|V_s^\epsilon-\tilde V_s^\epsilon(\alpha)|^{-3}]\\
&\qquad\qquad\qquad\qquad\qquad\qquad\qquad\qquad+\one_{|Y_s-\tilde Y_s|>|J_s^{n,\epsilon}-Y_s|^4}|Y_s-\tilde Y_s|^{-3}\Big]ds\\
&\leq C\int_0^t\mathbb E\Big[|J_s^{n,\epsilon}-Y_s|^2\one_{|J_s^{n,\epsilon}-Y_s|<1}(1+(||f_s^{\epsilon}||_\infty+||g_s||_\infty)\log\frac{1}{|J_s^{n,\epsilon}-Y_s|^4})\Big]ds.
\end{align*}
Recalling (\ref{functionpsi}), observing that $\log(x^4)=2\log(x^2)$ and using the (approximate) Jensen inequality for $\psi$, we get
\begin{align*}
B_3\leq C\int_0^t\psi(\mathbb E[|J_s^{n,\epsilon}-Y_s|^2])ds.
\end{align*}
We also have by Lemma \ref{r3v3},
\begin{align*}
B_4\leq &C\int_0^t ||f_s^{\epsilon}||_\infty\mathbb E\Big[|J_s^{n,\epsilon}-Y_s|^4\one_{|J_s^{n,\epsilon}-Y_s|\leq1}\Big]ds\\
&+C\int_0^t\mathbb E\Big[\one_{|J_s^{n,\epsilon}-Y_s|\leq1}\mathbb E_\alpha[\one_{|V_s^\epsilon-\tilde V_s|\leq|J_s^{n,\epsilon}-Y_s|^4}|Y_s-\tilde Y_s|^{-2}]\Big]ds.
\end{align*}
Using first the H\"older inequality with $p=5$ and $q=5/4$, and then Lemma \ref{r3v3}, we get
\begin{align*}
\mathbb E_\alpha[\one_{|V_s^\epsilon-\tilde V_s^\epsilon|\leq|J_s^{n,\epsilon}-Y_s|^4}&|Y_s-\tilde Y_s|^{-2}]\\
&\leq\mathbb E_\alpha[\one_{|V_s^\epsilon-\tilde V_s^\epsilon|\leq|J_s^{n,\epsilon}-Y_s|^4}]^{\frac{1}{5}}\mathbb E_\alpha[|Y_s-\tilde Y_s|^{\frac{-5}{2}}]^{\frac{4}{5}}\\
&\leq (C||f_s^{\epsilon}||_\infty|J_s^{n,\epsilon}-Y_s|^{12})^{\frac{1}{5}}(1+C||g_s||_\infty)^{\frac{4}{5}}\\
&\leq C(1+||f_s^{\epsilon}+g_s||_\infty)|J_s^{n,\epsilon}-Y_s|^{\frac{12}{5}}.
\end{align*}
We thus get
\begin{align*}
B_4&\leq C\int_0^t\mathbb E\Big[(|J_s^{n,\epsilon}-Y_s|^4+|J_s^{n,\epsilon}-Y_s|^{\frac{12}{5}})\one_{|J_s^{n,\epsilon}-Y_s|\leq1}\Big]ds\\
&\leq C\int_0^1 \mathbb E[|J_s^{n,\epsilon}-Y_s|^2]ds.
\end{align*}
We have the same bound for $B_5$ and thus (recalling that $x\leq\psi(x)$ for any $x\geq0$)
\begin{align*}
B\leq C\int_0^t \Big(\psi(\mathbb E[|V_s^\epsilon-Y_s|^2])+\psi(\mathbb E_\alpha[|\tilde V_s^\epsilon-\tilde Y_s|^2])+\psi(\mathbb E[|J_s^{n,\epsilon}-Y_s|^2])\Big)ds,
\end{align*}
which concludes the proof. \hfill$\square$

\subsection{Proof of Theorem \ref{colrascoul}}
We set $u(t):=\mathbb E[|V_t^\epsilon-Y_t|^2]$ and $v(t):=\mathbb E[|V_t^\epsilon-W_t^\epsilon|^2]+\mathbb E[|W_t^\epsilon-V_t^{n,\epsilon}|^2]+\mathbb E[|V_t^{n,\epsilon}-I_t^{n,\epsilon}|^2]+\mathbb E[|I_t^{n,\epsilon}-J_t^{n,\epsilon}|^2]+\mathbb E[|J_t^{n,\epsilon}-Y_t|^2]$. We have $u(t)\leq Cv(t)$ and using Lemmas \ref{lemh}, \ref{clem1}, \ref{clem2}, \ref{clem3} and \ref{clem4}, we get
\begin{align*}
v(t)\leq &Ch_\epsilon^{e^{-C}}+\frac{C}{\log\frac{1}{\epsilon}}\\
&+C\int_{0}^t\Big(\frac{\log n}{n}+\psi(\mathbb E[|V_s^\epsilon-Y_s|^2])+\psi(\mathbb E_\alpha[|\tilde V_s^\epsilon-\tilde Y_s|^2])\Big)ds\\
&+C\int_0^\pi\theta^4\beta_\epsilon(\theta)d\theta\\
&+C\Big[\eta^2M^2n\Big(\log^2(r_\eta)+\log^2(n\eta^2)+M\Big)+\int_\eta^\pi\theta^2\beta_\epsilon(\theta)d\theta+\frac{1}{M^p}\Big]\\
&+C\frac{\log n}{n}+C\int_0^t \Big(\psi(\mathbb E[|V_s^\epsilon-Y_s|^2])+\psi(\mathbb E_\alpha[|\tilde V_s^\epsilon-\tilde Y_s|^2])\\
&\qquad\qquad\qquad\qquad\qquad\qquad\qquad\qquad+\psi(\mathbb E[|J_s^{n,\epsilon}-Y_s|^2])\Big)ds.
\end{align*}
Since $\mathbb E[|J_s^{n,\epsilon}-Y_s|^2]\leq v(s)$ and $u(s)\leq Cv(s)$ for any $s\in(0,T]$, using  that the function $\psi$ (recall (\ref{functionpsi})) is increasing, we get (recall that $\mathbb E_\alpha[|\tilde V_s^\epsilon-\tilde Y_s|^2]=\mathcal W_2^2(f_s^{\epsilon},g_s)\leq u(s)$ for any $s\in[0,T]$)
\begin{align*}
v(t)\leq &Ch_\epsilon^{e^{-C}}+C\frac{\log n}{n}+\frac{C}{\log\frac{1}{\epsilon}}+C\int_0^\pi\theta^4\beta_\epsilon(\theta)d\theta+C\int_0^t\psi\big(v(s)\big)ds\\
&+C\Big[\eta^2M^2n\Big(\log^2(r_\eta)+\log^2(n\eta^2)+M\Big)+\int_\eta^\pi\theta^2\beta_\epsilon(\theta)d\theta+\frac{1}{M^p}\Big].
\end{align*}
Setting, for $\epsilon\in(0,1)$ fixed, $\eta=\frac{1}{\log\frac{1}{\epsilon}}$, $n \approx\Big(\log\frac{1}{\epsilon}\Big)^{\frac{2p}{2p+3}}$,  $M=\sqrt{2m_2(f_0)}\Big(\log\frac{1}{\epsilon}\Big)^{\frac{2}{2p+3}}$ and observing that $\int_0^{\pi}\theta^4\beta_\epsilon(\theta)d\theta\leq \frac{C}{\log\frac{1}{\epsilon}}$, $\int_\eta^{\pi}\theta^2\beta_\epsilon(\theta)d\theta\leq \frac{C\log\log\frac{1}{\epsilon}}{\log\frac{1}{\epsilon}}$, $\lim_{\epsilon\rightarrow0}r_\eta=1$ (whence $\log^2r_\eta$ is bounded for $\epsilon\in(0,1)$) and
\begin{align*}
&\eta^2M^2n\Big(\log^2(r_\eta)+\log^2(n\eta^2)+M\Big)\\
&\leq \frac{C}{\big(\log\frac{1}{\epsilon}\big)^{\frac{2p+2}{2p+3}}}\Big(1+\log^2\big(\log\frac{1}{\epsilon}\big)+\big(\log\frac{1}{\epsilon}\big)^{\frac{2}{2p+3}}\Big)\leq\frac{C}{\big(\log\frac{1}{\epsilon}\big)^{\frac{2p}{2p+3}}},
\end{align*}
we get
\begin{align*}
v(t)&\leq Ch_\epsilon^{e^{-C}}+\frac{C\log\log\frac{1}{\epsilon}}{\Big(\log\frac{1}{\epsilon}\Big)^{\frac{2p}{2p+3}}}+\frac{C}{\log\frac{1}{\epsilon}}+\frac{C}{\Big(\log\frac{1}{\epsilon}\Big)^{\frac{2p}{2p+3}}}\\
&\quad+\frac{C\log\log\frac{1}{\epsilon}}{\log\frac{1}{\epsilon}}+C\int_0^t\psi\big(v(s)\big)ds\\
&\leq Ch_\epsilon^{e^{-C}}+\frac{C}{\Big(\log\frac{1}{\epsilon}\Big)^{\frac{2p-1}{2p+3}}}+C\int_0^t\psi\big(v(s)\big)ds.
\end{align*} 
By Lemma \ref{grongen}, if $\epsilon$ is small enough (such that $Ch_\epsilon^{e^{-C}}+\frac{C}{\Big(\log\frac{1}{\epsilon}\Big)^{\frac{2p-1}{2p+3}}}\leq1$) we finally have
\begin{align*}
v(t)&\leq C\Big(h_\epsilon^{e^{-C}}+\frac{1}{\Big(\log\frac{1}{\epsilon}\Big)^{\frac{2p-1}{2p+3}}}\Big)^{e^{-C}}\leq Ch_\epsilon^{a}+\Big(\frac{C}{\log\frac{1}{\epsilon}}\Big)^{a},
\end{align*}
for some $a>0$. This concludes the proof since $W_2^2(f_t^{\epsilon},g_t)\leq\mathbb E[|V_s^\epsilon-Y_s|^2]=u(t)\leq Cv(t)$ and since for $\epsilon$ greater, we have $W_2^2(f_t^\epsilon,g_t)\leq2m_2(f_0)$. \hfill$\square$

\appendix

\section{Appendix}

\subsection{Distance between a compensated Poisson integral and a Gaussian variable}

We first recall a result of Zaitsev \cite{ZAI}. For $\tau\geq0$ and $d\in\mathbb N$, let $\mathcal A_d(\tau)$ be the class of probability distributions $F$ on $\mathbb R^d$ for which the function $\varphi(z)=\log\int_{\mathbb R^d}e^{z.x}F(dx)$ is analytic on $\{z\in\mathbb C^d, \ |z|\tau<1\}$ and $|d_ud_v^2\varphi(z)|\leq |u|\tau\mathbb Dv.v$ for all $u,v\in\mathbb R^d$ and $|z|\tau<1$, where $\mathbb D$ is the covariance matrix of $F$, and $d_u\varphi$ is the derivative of $\varphi$ in the direction $u$.

\begin{theo} \label{Zait} (Zaitsev \cite[Theorem 2]{ZAI}) Suppose that $\tau\geq1$ and that $\xi_1,...,\xi_n$ are independent random vectors with distributions $\mathcal L(\xi_k)\in\mathcal A_d(\tau)$, $\mathbb E(\xi_k)=0$, $Cov(\xi_k)=I_d$, $k=1,...,n$. Then one can build on some probability space a family of independent random vectors $X_1,...,X_n$ such that $\mathcal L(X_k)=\mathcal L(\xi_k)$ for any $k=1,...,n$ and a family of independent random vectors $Y_1,...,Y_n\sim\mathcal N(0,I_d)$ such that
\begin{align*}
\mathbb E\Big[\exp\Big(\frac{a\Delta_n(X,Y)}{\tau}\Big)\Big]\leq\exp\Big(b\max(1,\log n/\tau^2)\Big),
\end{align*}
where 
\begin{align*}
\Delta_n(X,Y)= \max_{1\leq k\leq n}\Big|\sum_{i=1}^kX_i-\sum_{i=1}^kY_i\Big|,
\end{align*}
and $a$, $b$ are positive quantities depending only on $d$.
\end{theo}

Using this result, we estimate the distance between a compensated Poisson integral and a Gaussian variable.

\begin{prop} \label{wasserstein}
Let $A$ be a measurable space endowed with a non negative $\sigma$-finite measure $\nu$ and $N$ be a Poisson measure on $[0,\infty)\times A$ with intensity measure $dt\nu(dz)$. We consider $h:A\rightarrow \mathbb R^d$ and we set $Z_t=\int_0^t\int_A h(z)\tilde N(ds,dz)$, $\mu_t=\mathcal L(Z_t)$ and $\Gamma=\int_A h(z)h^*(z)\nu(dz)$. If $\kappa:=\max_{z\in A} |\Gamma^{-1/2}h(z)|\in(0,\infty)$, then
\begin{align*}
\mathcal W_2^2(\mu_t, \mathcal N(0,t\Gamma))\leq C \kappa^2 |\Gamma|\Big[\max\Big(1,\log\frac{t}{\kappa^2}\Big)\Big]^2,
\end{align*}
where $C$ depends only on $d$ and where $\mathcal N(0,t\Gamma)$ is  the Gaussian distribution on $\mathbb R^d$ with mean 0 and covariance matrix $t\Gamma$.
\end{prop}

\textbf{Proof.}
For $n\in\mathbb N^*$ to be chosen later and $i\in\{1,...,n\}$, we consider
\begin{align*}
\xi_i=\sqrt{\frac{n}{t}}\Gamma^{-1/2}\int_{(i-1)t/n}^{it/n}\int_A h(z)\tilde N(ds,dz).
\end{align*}
We want to use Theorem \ref{Zait}. We first observe that the random variables $\xi_i$ are i.i.d., $\mathbb E(\xi_i)=0$ and $Cov(\xi_i)=I_d$. We now prove that $\xi_1\in\mathcal A_d(\tau)$ for some $\tau\geq1$. For $u\in\mathbb R^d$, we have $\mathbb E\Big(\exp( u.\xi_1)\Big)=\exp(\varphi(u))$, with
\begin{align*}
\varphi(u)=\frac{t}{n}\int_A \Big[\exp\Big(\sqrt{\frac{n}{t}} (\Gamma^{-1/2}h(z)).u\Big)-1-\sqrt{\frac{n}{t}} (\Gamma^{-1/2}h(z)).u\Big]\nu(dz).
\end{align*}
For $(x,y)\in \mathbb R^d\times\mathbb R^d$,
\begin{align*}
&d_xd_{y^2}^2\varphi(u)\\
&=\sqrt{\frac{n}{t}}\int_A \Big[\exp\Big(\sqrt{\frac{n}{t}} (\Gamma^{-1/2}h(z)).u\Big) [(\Gamma^{-1/2}h(z)).y]^2 (\Gamma^{-1/2}h(z)).x\Big]\nu(dz).
\end{align*}
We now search for $\tau>0$ such that $|d_xd_{y^2}^2\varphi(u)|\leq |x|\tau|y|^2$ for any $u$ satisfying $|u|<\frac{1}{\tau}$. We have, recalling that $\kappa:=\max_{z\in A} |\Gamma^{-1/2}h(z)|$,
\begin{align*}
|d_xd_{y^2}^2\varphi(u)|&\leq\sqrt{\frac{n}{t}}\int_A \exp\Big(\sqrt{\frac{n}{t}}|\Gamma^{-1/2}h(z)| |u|\Big) |\Gamma^{-1/2}h(z)|^2|y|^2|\Gamma^{-1/2}h(z)| |x|\nu(dz)\\
&\leq\sqrt{\frac{n}{t}}\exp\Big(\sqrt{\frac{n}{t}}\frac{\kappa}{\tau}\Big)|y|^2\kappa|x|\int_A |\Gamma^{-1/2}h(z)|^2\nu(dz),
\end{align*}
since $|u|<\frac{1}{\tau}$. We have, observing that $\Gamma$ is symetric,
\begin{align*}
\int_A |\Gamma^{-1/2}h(z)|^2\nu(dz)&=\int_A h^*(z)\Gamma^{-1}h(z)\nu(dz)\\
&=\sum_{i,j=1}^d\int_Ah_i(z)(\Gamma^{-1})_{ij}h_j(z)\nu(dz)\\
&=\sum_{i,j=1}^d(\Gamma^{-1})_{ij}\int_Ah_i(z)h_j(z)\nu(dz)\\
&=\sum_{i,j=1}^d(\Gamma^{-1})_{ij}\Gamma_{ij}=\sum_{i=1}^d(\sum_{j=1}^d(\Gamma^{-1})_{ij}\Gamma_{ji})=\sum_{i=1}^d(\Gamma^{-1}\Gamma)_{ii}=d.
\end{align*}
Setting $\tau=2d\kappa\sqrt{\frac{n}{t}}$, we thus have
\begin{align*}
|d_xd_{y^2}^2\varphi(u)|\leq |x||y|^2\sqrt{\frac{n}{t}}\exp\Big(\sqrt{\frac{n}{t}}\frac{\kappa}{\tau}\Big)d\kappa= |x||y|^2\frac{\tau}{2}\exp(\frac{1}{2d})\leq|x||y|^2\tau.
\end{align*}
So we have $\xi_i\in\mathcal A_d(\tau)$ with $\tau=2d\kappa\sqrt{\frac{n}{t}}$. Thus choosing $n\geq\frac{t}{4d^2\kappa^2}$ so that $\tau\geq1$, we can apply Theorem \ref{Zait}: one can construct on some probability space a sequence of independent random vectors $X_1,...,X_n$ such that $\mathcal L(X_k)=\mathcal L(\xi_k)$ for any $k=1,...,n$ and a sequence of independent random vectors $Y_1,...,Y_n\sim\mathcal N(0,I_d)$ such that
\begin{align*}
\mathbb E\Big[\exp\Big(\frac{a}{2d}\frac{\sqrt t}{\kappa}\frac{1}{\sqrt n}\Big|\sum_{i=1}^nX_i-\sum_{i=1}^nY_i\Big|\Big)\Big]\leq\exp\Big(b\max(1,\log\frac{t}{4d^2\kappa^2})\Big).
\end{align*}
Then setting $R_t:=\frac{1}{\sqrt n} \sum_{i=1}^n X_i$ (observe that $\mathcal L(R_t)=\mathcal L((t\Gamma)^{-1/2}Z_t)$) and $Y:=\frac{1}{\sqrt n}\sum_{i=1}^nY_i$ (observe that $\mathcal L(Y)=\mathcal N(0,I_d)$), we get
\begin{align*}
\mathbb E\Big[\exp\Big(\frac{a}{2d}\frac{\sqrt t}{\kappa}|R_t-Y|\Big)\Big]\leq\exp\Big(b\max(1,\log\frac{t}{\kappa^2})\Big).
\end{align*}
For $x\geq0$, we have
\begin{align*}
\mathbb P(|R_t-Y|^2\geq x)&=\mathbb P\Big(\exp\big(\frac{a}{2d}\frac{\sqrt t}{\kappa}|R_t-Y|\big)\geq \exp\big(\frac{a}{2d}\frac{\sqrt t}{\kappa}\sqrt x\big)\Big)\\
&\leq \exp\big(-\frac{a}{2d}\frac{\sqrt t}{\kappa}\sqrt x\big)\exp\big(b\max(1,\log\frac{t}{\kappa^2})\big).
\end{align*}
We consider $x_0$ verifying $\frac{a}{2d}\frac{\sqrt t}{\kappa}\sqrt x_0=b\max(1,\log\frac{t}{\kappa^2})$.
\begin{align*}
\mathbb E(|R_t-Y|^2)&=\int_0^\infty\mathbb P(|R_t-Y|^2\geq x)dx\\
&\leq x_0+\exp\big(b\max(1,\log\frac{t}{\kappa^2})\big)\int_{x_0}^{+\infty}\exp\big(-\frac{a}{2d}\frac{\sqrt t}{\kappa}\sqrt x\big)dx\\
&=x_0+\int_{x_0}^{+\infty}\exp\big(-\frac{a}{2d}\frac{\sqrt t}{\kappa}(\sqrt x-\sqrt{x_0})\big)dx\\
&=x_0+2\int_0^{+\infty}(y+\sqrt{x_0})\exp\big(-\frac{a}{2d}\frac{\sqrt t}{\kappa}y\big)dy\\
&=x_0+2\Big(\frac{4d^2\kappa^2}{a^2t}+\frac{2d\sqrt{x_0}\kappa}{a\sqrt t}\Big)\\
&\leq C\frac{\kappa^2}{t}\Big[\max\Big(1,\log\frac{t}{\kappa^2}\Big)\Big]^2.
\end{align*}
We thus have 
\begin{align*}
\mathcal W_2^2(R_t,\mathcal N(0,I_d))\leq C\frac{\kappa^2}{t}\Big[\max\Big(1,\log\frac{t}{\kappa^2}\Big)\Big]^2,
\end{align*}
and finally, since $Z_t$ has the same law as $\sqrt t\Gamma^{1/2}R_t$,
\begin{align*}
\mathcal W_2^2(Z_t, \mathcal N(0,t\Gamma))\leq C \kappa^2 |\Gamma|\Big[\max\Big(1,\log\frac{t}{\kappa^2}\Big)\Big]^2.
\end{align*}
\hfill$\square$

\subsection{Ellipticity of the diffusion matrix}
In this article, we need some ellipticity hypothesis for the diffusion matrix $l$, recall (\ref{alandau}). To this aim, we will extend some result stated in Desvillettes-Villani \cite{DESVIL} for $\gamma\geq0$.

\begin{prop}\label{ellipticity}
Let $\gamma\in[-3,0)$ and $E_0,H_0>0$ be two constants. Consider a nonnegative function $f$ such that $\int_{\mathbb R^3}f(v)dv=1$, $m_2(f)\leq E_0$ and $H(f)\leq H_0$. There exists a constant $c=c(\gamma,E_0,H_0)$ such that for any $v\in\mathbb R^3$ and any $\xi\in\mathbb R^3$,
\begin{align*}
(\bar l^f(v)\xi).\xi\geq c(1+|v|)^\gamma|\xi|^2,
\end{align*}
where $\bar l^f(v)=\int_{\mathbb R^3}l(v-v_*)f(v_*)dv_*$.
\end{prop}

\textbf{Proof.}
For $\gamma\in[-2,0)$, it is easy to check that in the proof of \cite[Proposition 4]{DESVIL}, they only use that $\gamma+2\geq0$. For $\gamma\in[-3,-2)$, we have to adapt a little bit their proof. In this case, estimate (44) of their proof still holds: for all $v\in\mathbb R^3$, $\theta\in(0,\pi/2)$ and $R_*>0$  
\begin{align*}
(\bar l^f(v)\xi).\xi&\geq \int_{\mathbb R^3\backslash D_{\theta,\xi}(v)}dv_*\one_{|v_*|\leq R_*}|v-v_*|^{\gamma+2}f(v_*)\sin^2\theta\\
&\geq (|v|+R_*)^{\gamma+2}\sin^2\theta\int_{\mathbb R^3\backslash D_{\theta,\xi}(v)}dv_*\one_{|v_*|\leq R_*}f(v_*),
\end{align*}
(recall that $\gamma+2<0$) where $D_{\theta,\xi}(v)=\Big\{v_*\in\mathbb R^3, \big|\frac{v-v_*}{|v-v_*|}.\xi\big|\geq\cos\theta\Big\}$ is the cone centred at $v$, of axis directed by $\xi$, and of angle $\theta$. Now following the scheme of their proof, we easily get that $(\bar l^f(v)\xi).\xi\geq K|v|^\gamma$ if $|v|\geq2R_*$ and that $(\bar l^f(v)\xi).\xi\geq K$ if $|v|<2R_*$ with $R_*=2\sqrt{E_0}$, which concludes the proof. \hfill$\square$

\subsection{Generalization of the Gr\"onwall Lemma}
In order to treat the Coulomb case, we need to use the following generalization of the Gr\"onwall lemma.
\begin{lemme} \label{grongen}
Let $T>0$ and $\gamma:[0,T]\rightarrow \mathbb R_+$ satisfy $\int_0^T\gamma(s)ds<\infty$. Let $\psi$ be defined by (\ref{functionpsi}). Consider a bounded function $\rho:[0,T]\rightarrow\mathbb R_+$ such that, for some $a\geq0$, for all $t\in[0,T]$, $\rho(t)\leq a+\int_0^t\gamma(s)\psi(\rho(s))ds$. We set $K:=\int_0^T\gamma(s)ds$. Then $\rho(t)\leq C(a^{e^{-K}}+a)$ for all $t\in[0,T]$, where $C$ only depends on $K$.
\end{lemme}

\textbf{Proof.}
From Chemin \cite[Lemme 5.2.1 p. 89]{CHE}, we get that $M(a)-M(\rho(t))\leq\int_0^t \gamma(s)ds$ for all $t\in[0,T]$, where $M(x):=\int_x^1(1/\psi(y))dy$ for $x>0$. 

Recalling that $\psi(y)=y(1-\one_{y\leq1}\log y)$, we get that $M(x)=\log(1-\log x)$ for $x\in[0,1]$ and $M(x)=-\log x$ for $x>1$. Let $t\in[0,T]$ be fixed.

If $a\leq1$ and $\rho(t)\leq1$, we have $\log\Big(\frac{1-\log a}{1-\log \rho(t)}\Big)\leq K$ which gives $\rho(t)\leq e^{1-e^{-K}}a^{e^{-K}}$.

If $a\leq1$ and $\rho(t)>1$, we have $\log\big((1-\log a)\rho(t)\big)\leq K$ which gives $\rho(t)\leq \frac{e^K}{1-\log a}$ and thus necessarily (since $\rho(t)>1$) $a>e^{1-e^{K}}$. Thus $\rho(t)\leq e^K\leq e^Ke^{e^K-1}a$.

If $a>1$ and $\rho(t)>1$, we have $\log\frac{\rho(t)}{a}\leq K$ which gives $\rho(t)\leq e^Ka$. 

If $a>1$ and $\rho(t)\leq1$, we have $\rho(t)\leq1<a$, which concludes the proof. \hfill $\square$

\subsection{Construction of a subdivision}
We end this paper with the following result.
\begin{prop} \label{subdivision}
For $T>0$ fixed, we consider $h\in \mathcal L^1([0,T])$ with $h(s)\geq0$ for any $s\in[0,T]$. For any $n\in\mathbb N^*$, there exist a subdivision $0<a_0^{n}<...<a_{\lfloor 2nT\rfloor-1}^{n}<a_{\lfloor 2nT\rfloor}^{n}=T$ such that $a_0^{n}<1/n$ and for any $i\in\{0,...,\lfloor 2nT\rfloor-1\}$, $1/4n<a_{i+1}^{n}-a_{i}^{n}<1/n$ and 
\begin{align*}
\sum_{i=0}^{\lfloor 2nT\rfloor-1}(a_{i+1}^{n}-a_{i}^{n})h(a_i^{n})\leq3\int_0^Th(s)ds+3.
\end{align*} 
\end{prop}

\textbf{Proof.} We take $a_i^{n}\in\big(\frac{i}{2n},\frac{2i+1}{4n}\big]$ such that $h(a_i^{n})\leq h(s)+1/T$ for any $s\in\big(\frac{i}{2n},\frac{2i+1}{4n}\big]$. We set $g(s)=\sum_{i=0}^{\lfloor 2nT\rfloor-1}h(a_i^{n})\one_{\big\{s\in\big(\frac{i}{2n},\frac{2i+1}{4n}\big]\big\}}$. We have $g(s)\leq h(s)+1/T$, $1/4n<a_{i+1}^n-a_i^n<3/4n$ and thus
\begin{align*}
\sum_{i=0}^{\lfloor 2nT\rfloor-1}(a_{i+1}^{n}-a_{i}^{n})h(a_i^{n})\leq\frac{3}{4n}\sum_{i=0}^{\lfloor 2nT\rfloor-1}h(a_i^{n})=3\int_0^Tg(s)ds\leq3\int_0^Th(s)ds+3,
\end{align*}
which concludes the proof. \hfill$\square$

\section*{Acknowledgements}
\addcontentsline{toc}{section}{Remerciements}
I would like to thank my Ph.D. advisor Nicolas Fournier for his insightful comments during the preparation of this work. I also want to point out that he provided Proposition \ref{wasserstein}.

\end{document}